# Unveiling the Potential of NOMA: A Journey to Next Generation Multiple Access

Adeel Ahmed, *Graduate Student Member, IEEE,* Wang Xingfu, *Member, IEEE,* Ammar Hawbani, *Member, IEEE,* Weijie Yuan, *Senior Member, IEEE,* Hina Tabassum, *Senior Member, IEEE,* Yuanwei Liu, *Fellow, IEEE,* Muhammad Umar Farooq Qaisar, *Member, IEEE,* Zhiguo Ding, *Fellow, IEEE,* Naofal Al-Dhahir, *Fellow, IEEE,* Arumugam Nallanathan, *Fellow, IEEE,* and Derrick Wing Kwan Ng, *Fellow, IEEE*

***Abstract*—The revolutionary sixth-generation wireless communications technologies and applications, notably digital twin networks (DTN), connected autonomous vehicles (CAVs), space-air-ground integrated networks (SAGINs), zero-touch networks, industry 5.0, healthcare 5.0, agriculture 5.0, and more, are driving the evolution of next-generation wireless networks (NGWNs). These innovative technologies and groundbreaking innovative applications will generate a sheer volume of data that requires the swift transmission of massive data across wireless networks and the capability to connect trillions of devices, thereby fueling the use of sophisticated next-generation multiple access (NGMA) schemes. In particular, NGMA strives to cater to the massive connectivity in the 6G era, enabling the smooth and optimized operations of NGWNs compared to existing multiple access (MA) schemes. This survey showcases non-orthogonal multiple access (NOMA) as the frontrunner for NGMA, spotlighting its novel contributions within the existing literature in terms of "*What has NOMA delivered?*", "*What is NOMA currently providing?*" and "*What lies ahead for NOMA?*". We present different variants of NOMA in this comprehensive survey, detailing their fundamental operations. In addition, this survey highlights NOMA's applicability in a broad range of wireless communications technologies such as multi-antenna systems, machine learning, reconfigurable intelligent surfaces (RIS), cognitive radio networks (CRN), integrated sensing and communications (ISAC), terahertz networks, unmanned aerial vehicles (UAVs), etc. This survey delves deeper by providing a comprehensive literature review of NOMA's interplay with various state-of-the-art wireless technologies. Furthermore, despite the numerous perks and advantages of NOMA, we also highlight several technical challenges of NOMA, which NOMA-assisted NGWNs may encounter. Next, we unveil the research trends of NOMA in the 6G era, offering reliable, robust, and swift communications. Finally, we offer design recommendations and insights along with the future perspectives of NOMA as the leading choice for NGMA within the realm of NGWNs.**

## I. INTRODUCTION

The world has been living under the charm of 5G for a few years. Operators have been tirelessly deploying and upgrading their networks, experimenting with different configurations and multi-band systems [1], to achieve maximum network performance. Behind the curtains, extensive efforts are already underway to shape the future of 6G. While 5G paved the way for massive Internet-of-Things (IoT) deployments, 6G holds the promise of delivering an immersive user experience, relying on advances in innovative technologies and applications such as smart healthcare [2], connected autonomous vehicles [3], Industry 5.0 [4], [5], smart agriculture [6], [7], digital twin networks (DTN) [8], wireless rechargeable sensor networks [9]–[11], wireless sensor networks [12]–[14] and more [15], [16]. 6G has been widely anticipated to usher in a radical shift in the mobile networking paradigm by achieving cutting-edge levels of network capabilities and unprecedented connectivity to satisfy the needs of the data-driven society [17]. Academics and industry scientists have forecast that the forthcoming 6G networks will possess superior capabilities in terms of intelligence, reliability, scalability, sustainability, and power efficiency, thereby surpassing the limitations of 5G [18].

Fig. 1 summarizes the ongoing 6G research and development key hotspots. A pioneering project is *6Genesis Flagship Program*[1], which is a research initiative aimed at developing, implementing, and testing integral technologies for 6G. *6G*

Adeel Ahmed and Wang Xingfu are with the School of Computer Science and Technology, University of Science and Technology of China, Hefei 230027, China. E-mail: adeelahmed@mail.ustc.edu.cn, wangxfu@ustc.edu.cn.

Ammar Hawbani is with the School of Computer Science, Shenyang Aerospace University, Shenyang 110136, China. Email: anmande@ustc.edu.cn.

Weijie Yuan is with the Department of Electrical and Electronics Engineering, Southern University of Science and Technology, Shenzhen 518055, China. Email: yuanwj@sustech.edu.cn.

Hina Tabassum is with the Department of Electrical Engineering and Computer Science, York University, Toronto, ON M3J 1P3, Canada. E-mail: Hina.Tabassum@lassonde.yorku.ca.

Yuanwei Liu is with the Department of Electrical and Electronic Engineering, The University of Hong Kong, Hong Kong. Email: yuanwei@hku.hk.

A. Nallanathan is with the School of Electronic Engineering and Computer Science, Queen Mary University of London, London and also with the Department of Electronic Engineering, Kyung Hee University, Yongin-si, Gyeonggi-do 17104, Korea. E-mail: a.nallanathan@qmul.ac.uk.

Muhammad Umar Farooq Qaisar is with the School of Computer Science, Northwestern Polytechnical University Xi'an, Shaanxi 710129, China. Email: muhammad@nwpu.edu.cn.

Zhiguo Ding is with the Department of Electrical Engineering and Computer Science, Khalifa University, Abu Dhabi, UAE. Email: zhiguo.ding@ieee.org.

Naofal Al-Dhahir is with the Department of Electrical and Computer Engineering, The University of Texas at Dallas, Richardson, TX 75080 USA. Email: aldhahir@utdallas.edu.

Derrick Wing Kwan Ng is with the School of Electrical Engineering and Telecommunications, University of New South Wales, Sydney, NSW 2052, Australia. Email: w.k.ng@unsw.edu.au.

Wang Xingfu and Ammar Hawbani are the corresponding authors.

[1]6Genesis Flagship Program– It is Finland's academic and industrial consortium aiming at developing key enabling technologies for 6G.

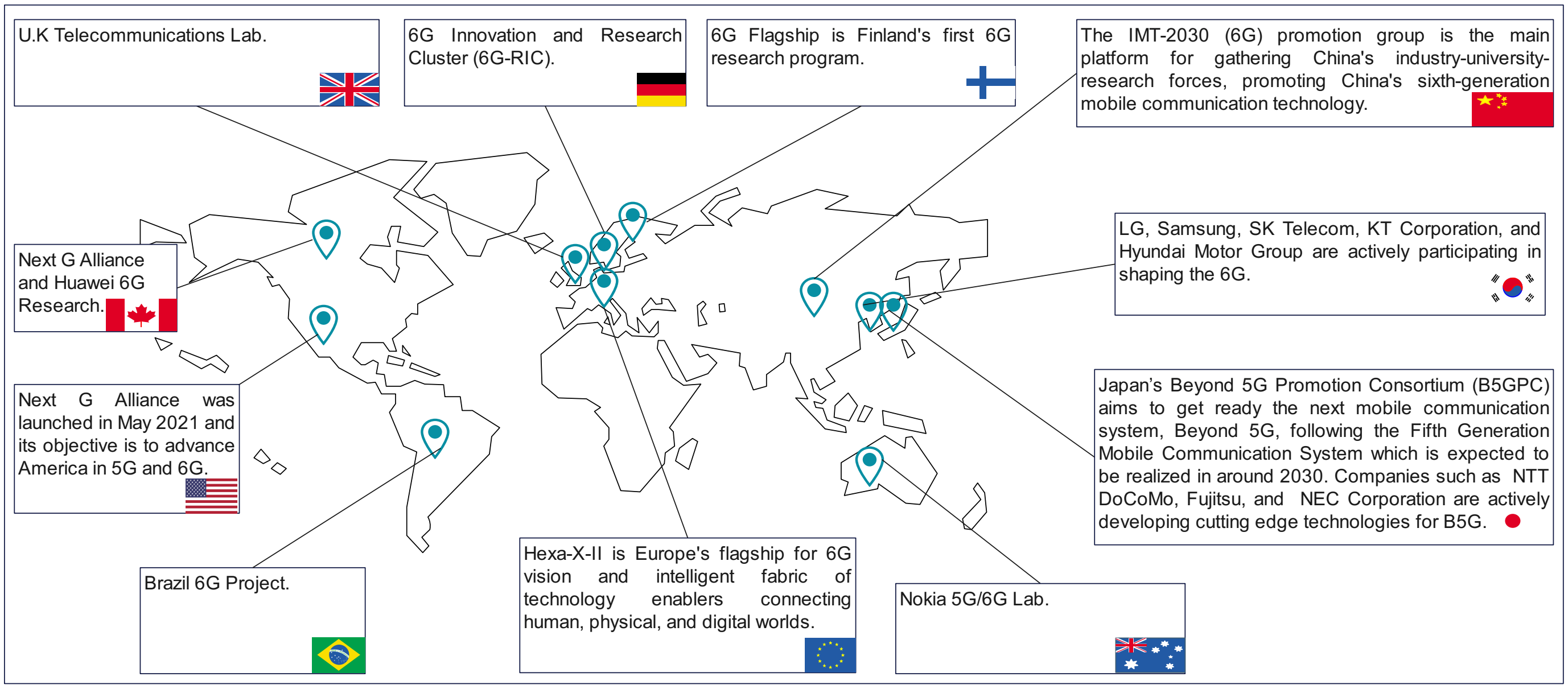

Fig. 1: A world map of 6G research and development hotspots.

*BRAINS*[2] is the H2020 research and innovation initiative, launched on January 1, 2021, and aims to set a precedent for future projects and worldwide standardization in industrial environments. It focuses on AI, multi-agent deep reinforcement learning, highly dynamic ultra-dense device-to-device (D2D) communications, cell-free networks, grant-free non-orthogonal multiple access (GF-NOMA), end-to-end (E2E) slicing, integrated access backhaul (IAB), industrial virtual assistant (IVA), simultaneous localization and mapping (SLAM), content distribution network (CDN), massive machine type communications (mMTC), and ultra-reliable low latency communications (URLLC). Another cutting-edge project is *DEDICAT 6G*[3], which aims to develop smart connectivity platforms using artificial intelligence and blockchain techniques that will enable 6G networks to combine the existing communications infrastructure with the novel distribution of intelligence (data, computation, and storage) at the edge to allow not only flexible but also an energy-efficient realization of the envisaged real-time experience. A number of other research initiatives are also currently underway with the intent of expanding the capabilities of wireless networks and paving the way for building the NGWN. These initiatives include *AI@EDGE*[4], *HEXA-XII*[5], *RISE-6G*[6], *MARSAL*[7], and *TeraFlow*[8]. Each of these projects explores the unique facets of 6G technologies, encompassing sophisticated AI applications to Terahertz communications and beyond.

6G aims to revolutionize connectivity by offering ultra-fast data rates of up to 1 Tb/s, achieving high spectral efficiency of 100 b/s/Hz, providing low latency of under 1 ms, and enabling the connection of up to 10 million devices per square kilometre. This ambitious vision supports the concept of the Internet-of-Everything (IoE) [19]. In the context of 6G, the surge in data demand and the advent of bandwidth-intensive applications like virtual reality (VR) and augmented reality (AR) are creating new challenges for multiple access. Additionally, the proliferation of Internet of Things (IoT) networks, the emergence of digital twins, the development of the metaverse, the integration of space-air-ground integrated networks (SAGINs), and the connectivity of autonomous vehicles are further driving the need for advanced multiple access techniques. More importantly, in contrast to the current wireless networks, which have mainly focused on providing communication services within terrestrial coverage, the next-generation wireless network's vision goes beyond this and is expected to include: [20] (i) Human-machine–things connections, (ii) Ubiquitous space-air-ground-sea coverage, (iii) Multi-functionality integration, and (iv) Native intelligent networks. From the aforementioned requirements and vision of 6G-based innovative and robust next-generation wireless networks, it is evident that such networks not only have to satisfy stringent communication requirements but also have to connect heterogeneous types of devices, provide ubiquitous coverage, integrate diverse functionalities, and support native intelligence. In line with this, communication-oriented MA schemes are expected to be replaced by advanced MA schemes, namely next-generation multiple access (NGMA) [20]. NGMA explores innovative approaches such as advanced waveforms, coding and modulation schemes, dynamic resource allocation, and intelligent interference management to cater to these emerging applications and ensure efficient connectivity in future wireless systems.

[2]6G BRAINS—It is the H2020 research and innovation action receiving supported by the European Commission Horizon 2020 Program.

[3]DEDICAT 6G— This project aims to achieve (i) more efficient use of resources; (ii) reduction of latency, response time, and energy consumption; (iii) reduction of operational and capital expenditures; and (iv) reinforcement of security, privacy, and trust.

[4]AI@EDGE— AI@EDGE targets significant breakthroughs in closed-loop network automation capable of supporting flexible and programmable pipelines for the creation, utilization, and adaptation of the secure, reusable, and trustworthy AI/ML models.

[5]HEXA-XII— Anticipating the conclusion of the Hexa-X project in June 2023, Nokia initiated a new project, which was launched as Hexa-X-II. This second European level 6G Flagship project started in January 2023 and will continue until June 2025.

[6]RISE-6G— RISE-6G project aims to investigate innovative solutions that capitalize on the latest advances in the emerging technology of reconfigurable intelligent surfaces, which offers dynamic and goal-oriented radio wave propagation control, enabling the concept of the wireless environment as a service.

[7]MARSAL— MARSAL project targets the development and evaluation of a complete framework for the management and orchestration of network resources in 5G and beyond by utilizing a converged optical-wireless network infrastructure in the access and fronthaul/midhaul segments.

[8]TeraFlow— TeraFlow will create a novel cloud-native SDN controller for beyond 5G networks. This new SDN controller shall be able to integrate with current NFV and MEC frameworks as well as to provide revolutionary features for flow aggregation, management (service layer), network equipment integration (infrastructure layer), and AI/ML-based security and forensic evidence for multi-tenancy.

TABLE I: LIST OF ACRONYMS.

| Acronym | Description | Acronym | Description |
|---|---|---|---|
| 3GPP | 3rd Generation Partnership Project | ADMM | Alternative Direction Method of Multipliers |
| AF | Amplify-and-Forward | AR | Augmented Reality |
| ATSC | Advanced Television Systems Committee | AWGN | Additive White Gaussian Noise |
| BC | Broadcast Channel | B5G | Beyond 5G |
| CDMA | Code Division Multiple Access | CF-mMIMO | Cell-Free Massive Multiple Input Multiple Output |
| CRAN | Cloud Radio Access Network | CSI | Channel State Information |
| CD-NOMA | Code Domain Non-Orthogonal Multiple Acces | CAV | Connected and Autonomous Vehicles |
| CRN | Cognitive Radio Networks | CoMP | Coordinated Multi-Point |
| DDPG | Deep Deterministic Policy Gradient | DF | Decode-and-Forward |
| DRL | Deep Reinforcement Learning | DoF | Degrees of Freedom |
| DPC | Dirty-Paper Coding | DL | Deep Learning |
| DSP | Digital Signal Processing | DML | Distributed Machine Learning |
| D2D | Device-2-Device | DTN | Digital Twin Networks |
| (F)eMBB | (Further) enhanced Mobile Broadband | FDMA | Frequency Division Multiple Access |
| FOVs | Fields of Views | GOCA | Group-Orthogonal Coded Access |
| GF-NOMA | Grant Free-Non Orthogonal Multiple Acces | HetNets | Heterogeneous Networks |
| ICT | Information and communications Technology | IDMA | Interleave Division Multiple Access |
| IFWT | Inverse Fast Fourier Transform | IoE | Internet-of-Everything |
| IoT | Internet-of-Things | IoRT | Internet of Robotic Things |
| ISAC | Integrated Sensing and Communication | LDPC | Low-Density Parity-Check |
| LDM | Layered Division Multiple Access | LTE | Long-Term Evolution |
| LoS | Line-of-Sight | LMMSE | Linear Minimum Mean Square Error |
| M2M | Machine to Machine | MA | Multiple Access |
| MEC | Mobile Edge Computing | MINLP | Mixed Integer Non-Linear Programming |
| MIMO | Multiple-Input Multiple-Output | MUSA | Multi-User Shared Access |
| MP | Message Passing | mmWave | millimeter-Wave |
| (u)mMTC | (ultra) massive Machine-Type communications | MU-MIMO | Multi-User Multiple-Input Multiple-Output |
| NCMA | Non-Orthogonal Coded Multiple Access | NCC | Non-Orthogonal Cover Codes |
| NOCA | Non-Orthogonal Coded Access | NOMA | Non-Orthogonal Multiple Acces |
| NGWN | Next Generation Wireless Network | NGMA | Next Generation Multiple Access |
| OFDM | Orthogonal Frequency Division Multiplexing | OFDMA | Orthogonal Frequency Division Multiple Access |
| PDMA | Pattern Division Multiple Access | PD-NOMA | Power Domain-Non-Orthogonal Multiple Acces |
| PA | Power Allocation | PS | Parameter Server |
| QoS | Quality of Service | RSMA | Rate Split Multiple Access |
| RDMA | Repetition Division Multiple Access | RA | Resource Allocation |
| RIS | Reconfigurable Intelligent Surface | SDMA | Space Division Multiple Access |
| TCMA | Trellis Coded Multiple Access | TDMA | Time Division Multiple Access |
| UAV | Unmanned Aerial Vehicles | URLLC | Ultra-Reliable and Low Latency communications |
| VR | Virtual Reality | V2X | Vehicle-to-Everything |
| WSMA | Welch-Bound Equality Spread Multiple Access | WSR | Weighted Sum Rate |

### *A. Towards Next Generation Multiple Access for 6G*

NGWNs are essentially multiuser systems where multiple access techniques play the critical role of utilizing the resources efficiently. The multiple access techniques can be broadly classified into two types, namely, orthogonal multiple access (OMA) and non-orthogonal multiple access (NOMA). Specifically, OMA techniques, including frequency division multiple access (FDMA), time division multiple access (TDMA), code division multiple access (CDMA), and orthogonal frequency division multiple access (OFDMA), allocate orthogonal resources to the individual users [21], [22]. On the other hand, innovative NOMA was first proposed for 5G, which allows multiple users to share the same bandwidth resource blocks by improving capacity, spectral efficiency, and low latency. NOMA yields several advantages over OMA techniques [23].

The advantages of NOMA are as follows [24], [25]:

- **Improved spectral efficiency and cell-edge throughput:** NOMA enables multiple users to share the same frequency, time, or code resources through power domain multiplexing. In NOMA, users are allocated the same resource block but with different power levels, which can significantly improve connectivity over the limited orthogonal resource blocks. In contrast, OMA techniques allocate orthogonal resources to each user, which limits the number of users that can access the network simul-taneously.

- **Massive connectivity:** Non-orthogonal resource allocation in NOMA indicates that the number of supportable users or devices is not strictly limited by the number of orthogonal resources available. Therefore, NOMA is capable of significantly increasing the number of simultaneous connections in rank-deficient scenarios; hence, it has the potential to support massive connectivity.
- **Low transmission latency and signalling cost:** In conventional OMA relying on access-grant requests, a user first has to send a scheduling request to the base station (BS). Then, upon receiving this request, the BS schedules the user's uplink transmission by responding with a clear-to-send signal in the downlink channel. Thus, a high transmission latency and a high signalling overhead will be imposed, which becomes unacceptable in the case of massive 5G-style connectivity. Specifically, the access grant procedure in LTE takes about 15.5 ms before the data is transmitted [120]. Hence, the stringent requirement of maintaining a user delay below 1 ms cannot be readily satisfied [112]. By contrast, dynamic scheduling is not required in some of the uplink NOMA schemes. To elaborate, in the uplink of a SCMA system, grant-free multiple access can be realized for users associated with pre-configured resources defined in the time- and frequency-domain, such as the codebooks, as well as the pilots. By contrast, at the receiver blind detection and compressive sensing (CS) techniques can be used for performing joint activity and data detection [91]. Hence, beneficial grant-free uplink transmission can be realized in NOMA, which is capable of significantly reducing both the transmission latency and the signalling overhead. Note that in some NOMA schemes using SIC receivers, the SIC process may impose extra latency. Therefore, the number of users relying on SIC should not be excessive, and advanced MIMO techniques can be invoked to serve more users. (iv) Relaxed channel feedback: The requirement of channel feedback will be relaxed in power-domain NOMA because the CSI feedback is only used for power allocation. Hence, there is no need for accurate instantaneous CSI knowledge. Therefore, regardless of whether fixed or mobile users are supported, having a limited-accuracy outdated channel feedback associated with a certain maximum inaccuracy and delay will not severely impair the attainable system performance as long as the channel does not change rapidly.

After briefly discussing the potential benefits of NOMA, it is important to analyze the NOMA's performance gain through the lens of information theory [24], [26], [27]. In fact, the concept of downlink NOMA (DL-NOMA) may also be interpreted as a special case of superposition coding (SC) in the downlink broadcast channel (BC). In [28], the capacity region of a realistic imperfect discrete memoryless BC was established by using SC. Bergmans found the Gaussian BC capacity region of single-antenna scenarios [29]. Inspired by [28] and [29], several researchers explored the potential performance gain of NOMA from an information theoretic perspective. In [30], the authors developed a new evaluation criterion for quantifying the performance gain of NOMA over OMA. More specifically, considering a simple two-user single-antenna scenario in conjunction with the Gaussian BC, the comparison of TDMA and NOMA in terms of their capacity regions was provided. The analytical results showed that NOMA is capable of outperforming TDMA in terms of both the individual user rates and the sum rate. In [31], the authors focused their attention on examining the capacity region of downlink NOMA by systematically designing practical schemes and investigated the gains of NOMA over OMA by designing practical encoders and decoders. Also, in [32], the authors examined the achievable capacity region of the RBC by invoking both decode-and-forward (DF) relaying and compress-and-forward (CF) relaying with/without dirty-paper coding (DPC). Besides, in [33], the achievable capacity region of multiuser MIMO-NOMA systems was investigated by invoking iterative linear minimum mean square error (LMMSE) detection. Furthermore, in [34], the authors investigated the performance of a two-user DL-NOMA in fading channels to exploit the time-varying nature of multi-user channels.

Despite the aforementioned discussions, the development of NOMA for the forthcoming 6G era is still in its infancy. To unveil its potential and pave the way forward, the goal of this paper is to showcase NOMA as the top next-generation multiple access (NGMA) contestant for the forthcoming 6G era. Indeed, researchers have already initiated efforts to develop NOMA-centric NGMA for NGWNs, as referenced in [35]–[37]. To align with the demands of 6G applications, several novel design considerations for NGMA must be addressed, which are elaborated upon in the subsequent sections.

*1) Limitless Access:* The vision of 6G-based NGWNs is to provide unrestricted connectivity to all applications, enabling seamless connections for anyone and anything, anywhere and anytime. NGWNs have the potential to significantly transform wireless communications services. Their primary objective is to provide unrestricted access and seamless connectivity to satisfy the increasing demands of connected devices and services. This spike in connectivity necessitates the NGMA systems to effectively cater to a large number of users with different requirements. This calls for the development of innovative resource allocation techniques, enhanced capacity, and improved energy efficiency to optimize network operations.

In the realm of information theory, researchers have long been captivated by the intricacies of multiple access techniques. These techniques, which allow multiple users to transmit data simultaneously over a shared channel, have been the subject of extensive research. However, most existing research works in this field have concentrated on determining the capacity limitations for scenarios involving infinite coding block lengths and a limited number of users. Yet, as technology continues to advance at an astonishing pace, a new frontier has emerged: the realm of massive access channels. In these channels, the number of users grows in tandem with the block length, presenting a unique set of challenges and opportunities. It is in this uncharted territory that the need to re-evaluate the fundamental limits of multiple access techniques becomes imperative. The existing research works on multiple access techniques have undoubtedly laid a solid foundation for understanding the intricacies of these systems. However, as the number of users continues to surge, it is crucial to reassess our understanding of the capacity limitations in this context. For example, in [38], the authors introduced the concept of a Many-Access Channel (MnAC) to describe a channel with one receiver and multiple transmitters, where the number of transmitters grows with the block length. In particular, the study in [38] provided innovative insights into achievable message length and symmetric capacity, demonstrating that separate identification and decoding are capacity-achieving strategies. These novel findings serve as valuable upper bounds for evaluating the performance of NGMA systems in the era of numerous users [27], [37], [39].

*2) Short-Packet Transmission:* In addition to tackling significant challenges related to massive access, NGMA systems must also accommodate the particular needs of short-packet transmission in IoT applications. These applications prioritize low latency and involve coding block lengths that are both finite and short. The significance of short-packet transmission in NGWNs is particularly notable due to the anticipated nature of messages originating from IoT devices, which are expected to be short. In [40], the authors provided an estimation of the maximum channel coding rate for a specific block length and error probability, thereby examining the rate gap compared to the scenario with infinite coding block length. However, extending this result to the regime characterized by a large yet finite number of users and finite coding block length remains an open problem that requires further investigation. Furthermore, achieving the fundamental limits of performance using practical transmission and multiple access schemes necessitates dedicated research efforts in the future [27].

*3) Ultra-Reliable and Low Latency Communications:* With the increasing demand for real-time and critical applications and technologies, such as Healthcare 5.0, Industry 5.0, Agriculture 5.0, and CAVs, NGMA must be designed to support minimal latency and reliable communications. To provide URLLC, NGMA must guarantee low latency to ensure swift transmission and reception of signals. This is particularly crucial for real-time applications where any delay could have severe consequences. Additionally, NGMA must exhibit high reliability to ensure accurate and prompt delivery of messages, especially for critical applications such as emergency services. Furthermore, NGMA should offer different levels of service to cater to the specific requirements of different applications, thereby ensuring Quality of Service (QoS). To measure the effectiveness of URLLC, various novel metrics, such as the age of information (AoI), transmission delay, and robustness, have been developed [41]–[43].

*4) High Mobility Scenarios:* The emergence of 6G technology is expected to unlock a plethora of groundbreaking possibilities, enabling an extensive range of highly mobile applications [44]. These include the seamless integration of connected autonomous vehicles (CAVs), non-terrestrial networks, and the empowerment of mobile robots, among a series of other transformative advancements. In the realm of NGMA systems, the imperative to embrace and facilitate exceptional mobility emerges as a paramount factor in the design process. In order to cater to the demands of modern users, these applications require seamless connectivity with lightning-fast data rates and low latency, even when users are on the move at high speeds. Hence, it is imperative that the advancement of NGMA places utmost importance on its capacity to facilitate seamless mobility and establish unwavering connectivity in swiftly evolving surroundings. To accomplish these objectives, NGMA has the potential to harness a myriad of advanced antenna techniques, including beamforming, massive MIMO, and sophisticated signal processing techniques. Beamforming enables the establishment of pencil-like energy-focused radio waves steering towards moving objects that increase the received signal strength at the desired receivers. Regarding the revolutionary concept of massive MIMO, it possesses the remarkable ability to enhance the signal-to-noise ratio (SNR) and amplify the system's capacity through the utilization of an extensive array of antennas. Moreover, sophisticated signal processing techniques possess the capability to efficiently mitigate the detrimental effects of Doppler shifts and fast-time varying fading that occur when devices are in fast motion. Furthermore, it is imperative that the design of NGMA places utmost importance on the aspects of scalability, energy optimization, and cost-efficiency.

*5) Flexibility and Adaptability:* The NGMA systems are at the forefront of driving the exponential growth of cutting-edge applications and technologies, such as the IoT, the vast realm of massive Machine-Type communications (mMTC), the intelligent landscapes of smart cities, the revolutionary realm of CAVs, and an array of other groundbreaking innovations. To meet the demands of such innovative applications, it is imperative to have cutting-edge wireless networks that possess exceptional performance capabilities. These networks must possess the ability to seamlessly adjust to dynamic conditions and cater to a wide array of services and devices. Moreover, it is imperative for NGMA systems to exhibit the ability to manage heterogeneous traffic patterns, catering to the diverse requirements of users. This will facilitate the optimal utilization of network resources, thereby enhancing the spectral efficiency.

*6) Sustainability:* It is crucial for NGMA systems to give utmost importance to sustainability during their design process, taking into account the possible ecological consequences of the growing quantity of intelligent sensors, devices, and ICT infrastructure. NGMA systems have the potential to make significant contributions to sustainability initiatives by optimizing spectral resources, reducing energy consumption, and facilitating the development of sustainable applications and services. Ensuring a sustainable future for our planet is crucial, and one way to achieve this is by integrating sustainability as a fundamental aspect of NGMA system design. This approach becomes even more important as the demand for NGWNs continues to rise. NGMA systems will have a crucial impact on the development of future NGWNs by promoting sustainability and meeting societal demands with minimal ecological impact.

### *B. Candidate NGMA Techniques*

As previously elucidated, the NGMA is currently undergoing a notable transformation from a state of orthogonality to one of non-orthogonality. This transition entails a departure from the conventional practice of allocating dedicated orthogonal resources to individual users or devices and instead embracing a system wherein multiple users or devices partake in the utilization of the same bandwidth resources. Several factors drive this transition. First and foremost, it has been observed that non-orthogonal transmission schemes exhibit superior performance compared to conventional orthogonal schemes in terms of efficiency and capacity. Furthermore, in light of the exponential proliferation of users and devices in forthcoming NGWNs, it is imperative to acknowledge that orthogonal schemes possess the capacity to accommodate merely a restricted number of users or devices, given the finite nature of the available orthogonal resources. In response to these challenges, researchers have embarked upon the task of formulating NOMA techniques for NGMA to effectively meet the demanding prerequisites of 6G networks [27], [37], [45], [46]. These efforts aim to utilize the principle of non-orthogonality to augment the efficacy and expandability of NGMA systems. Various NOMA-based candidates for NGMA have been proposed, supported by their technical foundations, as mentioned below [27].

*1) Non-orthogonal Multiple Access (NOMA):* Different NOMA variants can be broadly classified into two categories such as power-domain NOMA (PD-NOMA) and code-domain NOMA (CD-NOMA). PD-NOMA stands out as a distinguished wireless communications technique that utilizes the power domain to serve multiple users at the same time, even when they share identical time/frequency/code resources [45]. The fundamental concept of PD-NOMA is to differentiate users by allocating different power levels. This approach has been shown to be capacity-achieving in both the single-antenna broadcast channel (BC) and multiple access channel (MAC), owing to the use of superposition coding (SC) and successive interference cancellation (SIC) as the key enabling technologies [27]. PD-NOMA can be integrated with OFDMA for broadband communications over frequency-selective fading channels, where the channel coherence bandwidth is smaller than the system bandwidth. By assigning multiple users to each OFDMA subcarrier and serving them with PD-NOMA, multiuser superposition transmission (MUST) [47] has already been integrated into 3GPP Release 15 to simultaneously support two users on the same OFDMA subcarrier. (The fundamental operations of PD-NOMA will be presented in Section II.)

To assess the efficiency of NOMA, NTT DOCOMO carried out various field trials and performed performance evaluations and transmission experiments using prototype equipment [48]. The authors developed a two-user NOMA test bed to evaluate NOMA performance with the SIC receiver. Furthermore, in [48], LTE Release 8 frame structure was adopted, and channel estimation was based on CRS (cell-specific reference signal). For MIMO transmission, LTE TM3 was utilized for open-loop 2-by-2 single-user MIMO transmission. On the transmitter side, turbo encoding, modulation, and multiplication by precoding vector were applied for each user's data. The precoded signals of the two users were superposed according to a predefined power ratio. Results showed that NOMA could achieve a throughput gain of approximately 80% over OFDMA when a 2-by-2 SU-MIMO was adopted.

Inspired by its potential, numerous research studies have focused on PD-NOMA, including investigations into its performance, optimization, and implementation [49], [50]. Furthermore, researchers have explored the integration of PD-NOMA with other state-of-the-art wireless communications technologies, such as mMIMO, CoMP transmission, and MEC, to enhance its performance in various scenarios. For instance, non-cooperative NOMA, user-cooperative NOMA, and dedicated-relay cooperative NOMA have also been implemented and analyzed by using a software-defined radio (SDR) based testbed in [51]. Their results demonstrated that instead of channel quality difference for power allocation in NOMA, users can also be segregated based on the QoS requirement.

CD-NOMA was proposed to serve multiple users using sparse or non-orthogonal cross-correlation sequences with low cross-correlation [24]. At the receiver, multiuser detection (MUD) is usually carried out in an iterative manner using message passing (MP)-based algorithms. (Details of CD-NOMA variants will be provided in Section II.) The family of CD-NOMA schemes has many members, such as Low-Density Signature (LDS)-CDMA [52], LDS-OFDM [53], Sparse Code Multiple Access (SCMA) [54], and Pattern Division Multiple Access (PDMA) [55].

We noted that several researchers have integrated PD-NOMA and CD-NOMA together to optimize the performance of NGWN [56]–[58]. For instance, authors in [56] proposed a unified framework that covers both PD-NOMA and CD-NOMA for the satellite communications system. Moreover, in [56], authors jointly optimized the power allocation, carrier assignment, and beam scheduling to minimize the gap between user traffic demand and achieved capacity. Furthermore, in [57], authors proposed a hybrid-NOMA scheme for a joint system considering PD-NOMA and SCMA. The proposed hybrid-NOMA scheme in [57] exploits both the power difference and sparsity among the users and orthogonal resources for transmission. Furthermore, the authors offered some design guidelines for the deep learning-assisted hybrid-NOMA scheme for the massive connectivity regime. In a similar manner, authors in [58] proposed a hybrid NOMA scheme to enable massive connectivity in the uplink. The proposed hybrid NOMA scheme combines the CD-NOMA and PD-NOMA schemes by clustering the users in small path loss (strong) and large path loss (weak) groups. In the same vein, authors in [59], [60] proposed efficient hybrid NOMA schemes based on the PD-NOMA and CD-NOMA, which highlights the significance of designing innovative and robust joint PD-NOMA and CD-NOMA-based hybrid multiple access schemes to meet the stringent requirements of NGMA and NGWNs.

The major advantage of NOMA includes its potential to improve SE in severely overloaded scenarios by serving users with closely aligned channels and diverse channel strengths in the same time-frequency resources, Massive connectivity, and low transmission latency and signalling cost. However, On the other side, imperfect CSI can affect the performance of NOMA as CSI is crucial for the optimal resource allocation in NOMA. Therefore, a proper mechanism for CSI feedback and a suitable channel estimation scheme with proper reference signal design are important for achieving robust performance. Additionally, NOMA gets back the data from users depending on a technique called SIC. This technique causes an increase in the number of errors because of errors in single-user decoding using SIC.

NOMA has gained tremendous interest from both industry and academia. Moreover, recently, NOMA has been included in the ITU-IMT framework for 2030 because of its perks and compatibility with innovative wireless communications technologies [61]. In addition, it is worth mentioning that there has been a significant amount of research conducted on NOMA and its variants. Numerous online surveys have been conducted, covering a wide range of topics related to NOMA. These include general aspects of NOMA, different variants of NOMA, the use of deep learning to enhance NOMA's performance, and the information-theoretic limits and analytical foundations of NOMA. There remains a significant gap in the research community concerning the study of NOMA, especially when it comes to exploring its potential alongside enabling technologies from the perspective of NGMA. The absence of a comprehensive survey on NOMA's potential in the forthcoming era of massive connectivity in next-generation wireless networks, the foundations of NOMA, adoption of NOMA in the emerging technologies and the challenges it presents, and key insights into its performance have driven us to design this survey. Our aim is to showcase to the research community the unique features of NOMA that improve spectrum efficiency and enable massive connectivity, and low-latency communications in the upcoming 6G-enabled NGWNs.

*2) Rate Splitting Multiple Access (RSMA):* The realization of massive connectivity during the 6G era will be facilitated by the utilization of NOMA and its variants (RSMA, PD-NOMA, D-OMA, and so on) [62], [63]. RSMA, built upon the concept of rate-splitting (RS), has been recognized as a promising PHY-layer transmission paradigm for non-orthogonal transmission, interference management and MA strategies in 6G. The main idea behind RSMA is to split user messages into common and private parts, and enable the capability of partially decoding the interference and partially treating the interference as noise, which contrasts with the extreme interference management strategies used in SDMA and NOMA. The flexible nature of RSMA allows it to perform well for all levels of interference. When the interference is weak or strong, RSMA automatically reduces to SDMA or NOMA by tuning the powers and contents of the common and private streams [64]. The major disadvantages of RSMA include higher encoding complexity, higher signalling burden, and higher optimization burden.

There has been a significant amount of research work directed towards investigating the incorporation of RSMA within the framework of NGMA [65]–[69]. Authors in [69] investigated the weighted sum-rate maximization problem in order to optimize power and rate allocations in the hybrid RSMA-NOMA network. Moreover, in [69], power and rate allocation were considered to optimize the weighted system sum rate, while practical constraints, including QoS and SIC constraints, and the proportion of the maximum power budget for each multiple access scheme were considered. The hybrid RSMA-NOMA scheme enhances the system's sum rate and proportional fairness by leveraging the high data rate delivery capability of RSMA and the improved fairness provided by NOMA.

*3) Space Division Multiple Access (SDMA):* SDMA has gained significant ground in today's communications systems, including today's 5G standard and upcoming wireless networks, due to the utilization of multiple antennas, also known as MIMO technology. Multiple antennas at transmitters or receivers can be used to provide simultaneous service to multiple users or devices in the same time, frequency, or code domain while distinguishing them spatially. The most commonly used method in SDMA is linear precoding (LP) because of its low complexity. Spatial degrees of freedom (DoFs) are employed by LP to create suitable transmit and receive beamformers, which effectively minimize interference among users. It is worth mentioning that SDMA is typically applicable in situations where there is a low or critical load and where the spatial degrees of freedom are utilized to minimize interference between users. When there is a high number of users in relation to the available spatial DoFs, SDMA may become less effective as it may not completely eliminate interference between users [70].

One of the key advantages of SDMA is when the CSI is perfect, and the network is underloaded, SDMA can successfully suppress multi-user interference if the user channels are not aligned, and it achieves the maximum DoF of the underloaded multi-antenna BC. Furthermore, the transmitter and receiver complexities of SDMA are low as the transmitter employs linear precoding, and each receiver directly decodes the intended message by fully treating interference as noise. Despite its advantages, SDMA is sensitive to the network load. It is only suitable for underloaded systems, and the performance drops significantly when the network becomes overloaded. Also, SDMA is sensitive to the user deployment (including the angles and strengths of the user channels), which therefore imposes a stricter requirement on the scheduler. Moreover, SDMA is sensitive to CSI inaccuracy. In contrast to its good performance in the perfect CSI setting, SDMA cannot achieve the maximal DoFs when CSI is imperfect.

*4) Location Division Multiple Access (LDMA):* The concept of LDMA has been recently introduced as a potential technique to improve spectrum efficiency, presenting some benefits in comparison to traditional techniques. The LDMA system utilizes additional spatial resources in the distance domain to accommodate users situated at distinct locations within the near field. This is accomplished by analyzing angles and distances to determine the precise user locations that can be served [71]. The authors in [71] introduced a concept that involves leveraging the energy-focusing characteristic of near-field beams in order to cater to various users situated at different angles and locations. This approach aims to maximize the utilization of additional resources in the distance domain, thus enhancing the overall performance of the system. In this manner, near-field beams can effectively cater to users located at different angles and locations without inducing significant interference. The LDMA has numerous advantages. In contrast to classical SDMA, which necessitates a greater number of antennas to enhance the angular focusing capability of beams for interference mitigation, the concept of LDMA capitalizes on the additional focusing characteristic of near-field beams in the distance domain. Thus, users located at different angles and distances can be effectively served by near-field beams with minimal interference. The introduction of LDMA enhances spectrum efficiency in comparison to traditional SDMA techniques.

LDMA was initially proposed for near-field communications since it has many benefits to offer, such as optimized energy efficiency because spherical waves in near-field require less energy to transmit data over short distances compared to planar waves used in far-field communications [71], [72]. Additionally, due to the spherical nature of the wavefront of the transmitted signal, the receiver can be localized more precisely based on the phase and amplitude of the received signal. This makes NF communications an attractive technology for applications of NGMA-assisted NGWNs, such as indoor navigation and tracking, IoRT, and IIoT. This also revolutionizes spatial multiplexing, sets new standards for efficiency and performance, and showcases LDMA's innovation in enabling efficient near-field communications in the NGMA forthcoming era. The advantages of LDMA include enhanced spectrum efficiency, LDMA benefits from the asymptotic orthogonality of near-field beam focusing vectors in the distance domain, and also LDMA allows for precise focusing of signal energy on specific user locations, which can improve the accuracy and quality. Despite its advantages, its limitations include the requirement of an extremely large-scale antenna array, possibly thousands of antennas, to exploit the near-field properties effectively, which may not be feasible or cost-effective in all scenarios.

*5) Fluid-Antenna Multiple Access (FAMA):* Through a position-adaptable fluid antenna at each user, multiple access becomes possible by utilizing the natural interference null in a small area, referring to fluid antenna multiple access (FAMA) [73]. FAMA achieves a suitable channel condition for the desired signal without the need for complex signal processing by taking advantage of spatial moments of deep fade suffered by the interference. In particular, a fast FAMA (f-FAMA) [74] system can be used to exploit the sum-interference signal null, requiring each fluid antenna to switch to the best port on a symbol-by-symbol basis. However, a slow FAMA (s-FAMA) [75] scheme that switches to the sum-interference plus noise power null is more practical. For example, in [76], authors proposed a low-complexity port selection scheme for two user FAMA, where users can be served on the same channel by having each BS antenna transmit to one of the users, and each user has a fluid antenna to scan the entire fading envelope in its space and switch to the position (i.e., port) where the interference signal disappears naturally due to multipath fading. FAMA reduces multiple access to a simple port searching task on the user side without the need for precoding, and CSI at the BS [76]. FAMA is also a new multiple access technique and it brings a lot of benefits which can enhance the performance of the NGWN, such as, high density and multiplexing gains, enhanced sum rates, and reduced interference. However, implementing a robust system that can reliably change antenna positions in real-time to optimize reception and mitigate interference poses significant engineering challenges.

### *C. Research Methodology*

Despite the functional and performance advantages of NOMA as a key candidate for NGMA, there are still several important questions that require attention. Our survey focuses on addressing the following research questions:

**Research Questions**

- **Fundamentals of NOMA and Its Variants:** What are the fundamental operations of NOMA and its variants? (The answer to this question will be covered in Section II.)
- **Artificial Intelligence:** What role can artificial intelligence/machine learning play in the forthcoming NOMA-assisted NGMA? (The answer to this question will be discussed in the Subsection III-B and partly in the Section VII.)
- **Key Enabling Technologies:** What are the most promising key enabling technologies for NOMA, and what are NOMA's contributions to those technologies? (The answer to this question will be summarized in Section III.)
- **Challenges:** What are the main challenges associated with implementing NOMA in 5G/6G networks, and what techniques can be employed to deal with them? (This question will be answered in Section V.)
- **Trends:** What lies ahead of NOMA for the forthcoming innovative technologies and applications in the 6G era? (This question will be answered in Section IV.)
- **Design recommendations and Insights:** What are the design guidelines for NOMA towards NGMA? (This question will be answered in Section VI.)
- **Future Perspectives:** What are the exciting applications and use cases for the NOMA-driven NGWN in the forthcoming 6G era? (This question will be answered in Section VII.)

**Criteria for Selection:** The primary objective of this survey is to present a comprehensive overview of NOMA's performance in the context of NGMA. Our survey focuses on recent and innovative research contributions from both academia and industry, drawing insights from a wide range of literature sources. By carefully reviewing the existing literature, our survey aims to showcase the advancements in NOMA and its potential to enhance the performance of future game-changing applications and technologies. It also provides valuable design recommendations and key insights for strengthening NOMA for the implementation of upcoming NGWNs.

**Literature Classification:** We have identified and reviewed approximately three hundred papers for this survey. The selection criteria include papers that meet any of the following criteria: (i) Explicit usage of the term "NOMA" in the title; (ii) Explicit usage of the term "resource allocation" in NOMA;

(iii) Papers exploring the interplay between NOMA and wireless communications technologies; (iv) Papers on learning-enabled NOMA; (v) Papers on multi-antenna techniques for NOMA; (vi) Real-time implementations of NOMA; (vii) Papers on the use of blockchain, age of information and digital twin networks for NOMA; (viii) Papers on the security and privacy issues in NOMA; (ix) Papers based on the cross-layer design approaches; (x) Papers on the applicability of NOMA in URLLC; (xi) Papers on interference management; (xii) Papers on single and multi-carrier NOMA networks; (xiii) Papers on NOMA's standardization; (xiv) Papers on the acquisition of channel state information; and (xv) Papers on the successive interference cancellation.

Within each category, existing works were further classified based on specialized topics to highlight the specific applications and contexts in which NOMA is employed within the field of 5G and 6G wireless communications technologies. The categorization process allows us to clearly define the involvement of NOMA in various areas of research. The first category ensures the paper's prime focus on the NOMA, which streamlined the search and identification of the related topic. The related papers from the second category focus on the optimal resource allocation strategies, as NOMA's performance depends on the efficient management and allocation of radio resources. The third category includes research papers based on the integration of NOMA with state-of-the-art wireless communications technologies such as MIMO, mMIMO, RIS, HeTNets, and more. Then, in the fourth category, we included the research papers on the optimization of NOMA networks using machine learning-based techniques, which also makes the research area valuable for future research. Additionally, in the fifth category, we focused on NOMA-based multi-antenna networks, which improve spectrum efficiency, and it is also a trending research area. Then, in the sixth category, we reviewed papers on the real-time implementation of NOMA because this category provides insights into the real-world challenges of deploying NOMA, informing its feasibility and potential impact.

As the blockchain, age of information, and digital twins networks are emerging technologies, in the seventh category, we focused on the applicability of NOMA in the aforementioned technologies. The eighth category comprises research papers that address the unique security and privacy issues of NOMA due to its non-orthogonal nature. Furthermore, in the ninth category, we focused on the cross-layer design approaches in NOMA. In the tenth category, we focused on research papers on the NOMA-enabled IoV and URLLC for the 5G and 6G as NOMA's potential makes it crucial to further explore the mentioned area of research. As poor interference management leads to degraded performance of NOMA so, in the eleventh category, we focused on research papers based on interference management techniques for NOMA networks. In the twelfth category, we primarily focused on the multi-antenna architectures and techniques for NOMA, as such architectures and techniques are indispensable components of NGMA and offer huge potential that can be further explored. In the thirteenth category, we researched the standardization efforts around NOMA which helps in accessing its practical adoption and integration into wireless communication technologies. As accurate channel state information is crucial to the performance of NOMA, so in the fourteenth category we focused on reviewing the papers on the acquisition of channel state information in the NOMA networks. Along with the CSI acquisitions and efficient resource allocation, successive interference cancellation plays a vital role in signal decoding; that is why in the fifteenth category, we reviewed papers on the effects of successive interference cancellation and also how to design the decoding order to execute the successive interference cancellation technique in NOMA networks.

**Timeline:** Papers published from 2012 to 2024.

### D. Prior Works

The comprehensive comparisons of our survey and the existing surveys are presented in Tables II and III. The related surveys and magazine articles covering different aspects of NOMA are summarized below.

In [77], the authors provided a comprehensive survey that focuses on the progress of NOMA for 5G systems, state-of-the-art capacity analysis, power allocation strategies, user fairness, and user pairing schemes. In [78], the authors focused on the two-user single-carrier NOMA network to illustrate its basic principle. In [79], the authors provided an overview of fundamentals operations, research status, and applications of NOMA. In [21], the authors focused on the state-of-the-art modulation schemes for OMA and NOMA and compared their performance regarding spectral efficiency, out-of-band leakage, and bit error rate. In [80], the authors investigated the most recent developments in PD-NOMA principles and the interplay of NOMA with MIMO and cooperative communications. In addition to this, the authors presented a comparison between the various NOMA variations. In [24], the authors discussed the origins of NOMA and its evolution and applications in the context of 5G. Moreover, the authors presented the fundamental operations of NOMA variants.

In [81], the authors showcased a comprehensive overview of the promising NOMA schemes and the operations applied at the transmitter. Moreover, the authors also discussed multiuser detection algorithms corresponding to different NOMA schemes. The authors also focus on grant-free NOMA, which combines NOMA techniques with uplink uncoordinated access to meet 5G's massive connectivity requirements. In [82], the authors provided a comprehensive survey on the interplay of NOMA with wireless technologies such as massive MIMO, visible light communications, and physical layer security. In [83], the authors delivered a comprehensive review of different NOMA schemes from a grant-free perspective, along with potential challenges and prospective future directions. In [84], the authors surveyed rate optimization scenarios available in the literature when PD-NOMA is combined with one or more of the candidate schemes and technologies for 5G networks. In [85], the authors studied the recent trends in PD-NOMA-based cooperative communications networks along with performance evaluation of cooperative NOMA systems. In [86], the authors introduced innovative means of optimizing spectral and energy efficiency to ensure green communications.

TABLE II: A COMPARISON OF SURVEY PAPERS AND MAGAZINE ARTICLES IN CHRONOLOGICAL ORDER.

| | | | | NOMA Variants | | | | | | | | | | | | | | | | | | | | |
|---|---|---|---|---|---|---|---|---|---|---|---|---|---|---|---|---|---|---|---|---|---|---|---|---|
| **Type** | **Ref.** | **Year** | **Survey Contributions** | **A** | **B** | **C** | **D** | **E** | **F** | **G** | **H** | **I** | **J** | **K** | **L** | **M** | **N** | **O** | **P** | **Q** | **R** | **S** | **T** | **U** |
| Survey papers | [77] | 2016 | NOMA inter-cell interference mitigation and energy-spectral efficiency trade-offs. | ✓ | | | | | | | | | | | | | | | | | ✓ | | | * |
| | [78] | 2016 | Performance of two-user single-carrier and multi-user multi-carrier NOMA. | * | * | | | | | | | | | | | | | | | | ✓ | | | |
| | [79] | 2017 | Single- and multi-carrier NOMA, interplay with MIMO, mmWave, and cooperative communications are discussed. | ✓ | * | * | | | | | | | | | | | | | | | ✓ | | | * |
| | [21] | 2017 | Several modulation techniques were discussed for OMA and NOMA for 5G. | * | * | | | | | | | | | | | | | | | | ✓ | | | * |
| | [80] | 2017 | NOMA's primary concepts, benefits, and PD-NOMA milestones. | ✓ | * | * | | | | | | | | | | | | | | | ✓ | | | * |
| | [24] | 2018 | NOMA's development and comparison to OMA along with several NOMA schemes. | * | * | * | * | * | | * | | * | | | | | | | * | | * | | | * |
| | [81] | 2018 | Transmitter operations of NOMA schemes along with multiuser detection algorithms. | * | * | * | * | * | * | * | * | * | * | * | * | * | * | | | | | | | |
| | [82] | 2019 | Interplay of NOMA with some wireless communications technologies. | * | | | | | | | | | | | | | | | | | * | | | |
| | [83] | 2020 | A comprehensive review of NOMA from a grant-free access perspective. | * | * | * | * | * | * | * | * | * | * | * | * | | | | | | | | | |
| | [84] | 2020 | Interplay of NOMA with wireless communications technologies. | ✓ | | | | | | | | | | | | * | | | | | | ✓ | | |
| | [85] | 2020 | Basic operations and recent research trends in PD-NOMA based cooperative networks. | * | | | | | | | | | | | | | | | | | | * | | |
| | [86] | 2020 | Innovative approaches for NOMA in terms of spectral and energy efficiency. | | ✓ | | | | | | | | | | | | | | | | | | | |
| | [87] | 2021 | Research issues and challenges of NOMA along with 5G emerging technologies. | ✓ | | | | | | | | | | | | | | | | | | * | | |
| | [88] | 2021 | Basic principles and variants of NOMA. | * | | | * | | | | | * | * | * | | * | * | * | * | * | * | * | * | |
| | [27] | 2022 | Evolution of NOMA towards NGMA and a unified framework for multi-antenna assisted NOMA. | ✓ | * | | | | | | | | | | | | | | | | | * | | |
| | [89] | 2022 | NOMA network architectures, channel models, and mobility management. | * | * | * | * | | | | | | | | | | | | | | | * | | |
| | [90] | 2022 | Deep Learning aided NOMA. system. | * | | | | | | | | | | | | | | | | | * | | | |
| | [91] | 2022 | NOMA error rate analysis. analysis. | ✓ | ✓ | | | | | | | | | | | | | | | | ✓ | | | |
| | [92] | 2023 | Deep learning based NOMA: State of the art, key aspects, open challenges and future trends. | * | | | | | | | | | | | | | | | | | * | | | |
| Magazine articles | [93] | 2015 | NOMA variants, features, and research directions for 5G multiple access technologies. | * | * | * | * | | | | | | | | | | | | | | * | | | |
| | [94] | 2016 | Candidate NOMA schemes for 5G. | * | * | * | * | | | | | | | | | | | | | | * | | | * |
| | [95] | 2018 | Operations, enabling techniques, and optimization characteristics for CRNs with NOMA. | * | | | | | | | | | | | | | | | | | * | | | |
| | [96] | 2021 | Game theoretical approaches for NOMA networks and UAV-based relay networks. | * | | | | | | | | | | | | | | | | | | | | |
| | [37] | 2022 | Applications of NOMA in 6G networks and applying machine learning in NOMA. | * | | | | | | | | | | | | | | | | | * | | | |
| | [97] | 2023 | Artificial intelligence for NOMA towards NGMA. | * | | | | | | | | | | | | | | | | | * | | | |
| Our survey | | | Perks and benefits of NOMA as the frontrunner of NGMA, fundamentals of NOMA and its variants, interplay of NOMA with a broad set of technologies along with comprehensive literature review, challenges, opportunities, design guidelines and lessons learned along with the future perspective. | ✓ | ✓ | ✓ | ✓ | ✓ | ✓ | ✓ | ✓ | ✓ | ✓ | ✓ | ✓ | ✓ | ✓ | ✓ | ✓ | ✓ | ✓ | ✓ | ✓ | ✓ |

Note: (1). The symbol ✓ indicates that this specific aspect is covered in detail. The symbol * indicates that this specific aspect is only introduced briefly. (2). Moreover, this table consists of different columns represented by single alphabets, each associated with a specific NOMA variant or other aspects of this comprehensive survey. Here, A: PD-NOMA, B: SCMA, C: PDMA, D: MUSA, E: LSSA, F: RSMA, G: IGMA, H: IDMA, I: NCMA, J: NOCA, K: GOCA, L: RDMA, M: LDS-SVE, N: SSMA, O: WSMA, P: LDS-CDMA, Q: LPMA, R: Fundamental Operations, S: Literature Review, T: Research Methodology, U: Design Guidelines.

In [87], the authors reviewed the challenges related to resource allocation, signalling, practical implementation, and security aspects of NOMA and its integration with 5G and beyond wireless technologies. In [88], the authors presented an in-depth survey of state-of-the-art NOMA variants such as PD-NOMA, IDMA, SCMA, RDMA, LCRS, and more. In [89], the authors concentrated on the NOMA's applicability in 5G and B5G networks. In [27], the authors explored the evolution of NGMA with a particular focus on NOMA, the capacity limits of NOMA, new design requirements, and advanced mathematical optimization tools for NOMA. In [37], the authors offered a short survey on the development of NOMA towards NGMA with a focus on applying NOMA in 6G networks in massive connectivity scenarios. In [95], the authors stressed the significance of NOMA in cognitive radio networks. In [98], the authors studied the combination of NOMA and full duplex operation as a promising solution to improve the capacity of next-generation wireless systems.

## *E. Motivations*

Tomorrow's 6G network aims to offer limitless connectivity, which will foster technologies like healthcare 5.0 to industry 6.0 and intelligent transportation systems through

TABLE III: A COMPARATIVE EVALUATION OF SURVEY PAPERS AND MAGAZINE ARTICLES ON NOMA AND ITS INTEGRATION WITH WIRELESS COMMUNICATIONS TECHNOLOGIES.

| | | | | Enabling Technologies | | | | | | | | | | | | | Opportunities | | | | | | | | | | | | |
|---|---|---|---|---|---|---|---|---|---|---|---|---|---|---|---|---|---|---|---|---|---|---|---|---|---|---|---|---|---|
| **Type** | **Ref.** | **Year** | **A** | **B** | **C** | **D** | **E** | **F** | **G** | **H** | **I** | **J** | **K** | **L** | **M** | **N** | **O** | **P** | **Q** | **R** | **S** | **T** | **U** | **V** | **W** | **X** | **Y** | **Z** | **ZZ** |
| Survey papers | [77] | 2016 | | | * | | * | | | * | | | | | | | | | | | | | | | | | | | |
| | [78] | 2016 | | | * | | * | | | | | | | | | | | | | | | | | | | | | | |
| | [79] | 2017 | | | * | | * | | | | * | | | | | | | | | | | | | | | | | | |
| | [21] | 2017 | | | * | | * | | | | * | | | | | | | | | | | | | | | | | | |
| | [80] | 2017 | | | | | * | | | | * | | | | | * | | | | | | | | | | | | | |
| | [24] | 2018 | | | | | * | | | | | | | | | | | | | | | | | | | | | | |
| | [81] | 2018 | | | | | | | | | | | | | | | | | | | | | | | | | | | |
| | [82] | 2019 | | | * | | * | * | | | * | | | | | | | | | | | | | | | | | | |
| | [83] | 2020 | | | * | | | | | | | | * | | | | | | | | | | | | | | | | |
| | [84] | 2020 | | | ✓ | | ✓ | ✓ | ✓ | ✓ | ✓ | ✓ | | | | | | | | | | | | | | | | | |
| | [85] | 2020 | | | * | | | | | | * | | | | | | | | | | | | | | | | | | |
| | [86] | 2020 | | * | | | | | | | | | * | | * | | | | | | | | | | | | | | |
| | [87] | 2021 | | | ✓ | | ✓ | * | * | * | ✓ | | * | | | ✓ | | | | | | | | | | | | | |
| | [88] | 2021 | | | | | | | | | | | | | | | | | | | | | | | | | | | |
| | [27] | 2022 | * | ✓ | ✓ | | | * | * | | | | * | * | * | | | | | | | | | | | | * | | |
| | [89] | 2022 | | | | | | | | * | * | | | | | | | | | | | | | | | | | | |
| | [90] | 2022 | | ✓ | | | | | | | | | | | | | | | | | | | | | | | | | |
| | [91] | 2022 | | | * | | | | | | | * | * | | | | | | | | | | | | | | | | |
| | [92] | 2023 | | ✓ | | | | | | | | | | | | | | | | | | | | | | | | | |
| Magazine articles | [93] | 2015 | | | | | | | | | | | | | | | | | | | | | | | | | | | |
| | [94] | 2016 | | | | | | | | | | | | | | | | | | | | | | | | | | | |
| | [95] | 2018 | | | | | | | | | * | | | | | | | | | | | | | | | | | | |
| | [96] | 2021 | | | | | | | * | | | | | | | | | | | | | | | | | | | | |
| | [37] | 2022 | * | * | | | | | | | | | * | | * | | | | | * | | * | | | | | * | | |
| | [97] | 2023 | * | ✓ | | | | | | | | | * | * | * | | | | | | | | | | | | | | |
| Our survey | | | ✓ | ✓ | ✓ | ✓ | ✓ | ✓ | ✓ | ✓ | ✓ | ✓ | ✓ | ✓ | ✓ | ✓ | ✓ | ✓ | ✓ | ✓ | ✓ | ✓ | ✓ | ✓ | ✓ | ✓ | ✓ | ✓ | ✓ |

Note: (1). The symbol ✓ indicates that this specific aspect is covered in detail. The symbol * indicates that this specific aspect is only introduced briefly. (2). Moreover, this table consists of different columns represented by different alphabets, each associated with a specific wireless communications technology and different aspects of this comprehensive survey. A: NGMA, B: Artificial intelligence, C: Multi-antenna techniques, D: Ultra-Reliable and Low-Latency Communications, E: Cooperative communications, F: Mobile edge computing, G: Unmanned aerial vehicles, H: Visible Light Communications, I: Cognitive Radio Networks, J: Backscatter Communications, K: Reconfigurable Intelligent Surfaces, L: Integrated Sensing and Communications, M: Terahertz, N: Heterogenous Networks, O: Synergy of NOMA and Green Communications in the 6G Era, P: Green Connected Autonomous Vehicles, Q: 6G Space-Air-Ground Integrated Networks, R: Internet of Robotic Things for Smart and Sustainable Cities, S: Industry 5.0, T: Adaptability of NOMA in e-Health, U: Tactile Internet, V: Zero-Touch Networks, W: Satellite Communications, X: Semantic Communications, Y: Orthogonal Time Frequency Space Modulation, Z: Digital Twin Networks, ZZ: Age of Information.

IoE. The IoE anticipates a future in which a massive number of smart devices can collaborate and interact, allowing for seamless data exchange and task automation. Swift operations of such advanced applications and technologies will rely heavily on efficient MA techniques. 6G-based NGWNs are set to revolutionize connectivity by moving far beyond traditional terrestrial communication. These networks envision a seamlessly connected world where humans, machines, and intelligent devices interact fluidly across space, air, land, and sea. Designed for futuristic applications, they aim to deliver ubiquitous coverage and integrate advanced multi-functional capabilities, transforming the scope of connectivity into a truly boundless and intelligent network ecosystem.

Initially proposed for 5G, NOMA has attracted significant attention from the industry and academia because of its non-orthogonal characteristics. NOMA allows multiple users to efficiently share the same RB, which ensures enhanced spectrum efficiency, energy efficiency, compatibility, flexibility, user fairness, and improved throughput. Although NOMA has already been thoroughly investigated in the 5G and beyond networks, previous research focused on static devices and the data rate of broadband users. This ignores several fundamental problems for NGMA, for example, the effect of mobility and the design tradeoffs in terms of connectivity, reliability and latency. Our goal in this article is to fill this gap by exploring NOMA's working principles, variants, and state-of-the-art literature review to showcase what has been done so far to optimize the performance of the NOMA-driven wireless networks and its potential of integration with the innovative applications of 6G enabled NGWNs.

The rationale for considering NOMA as the top contender for NGMA can be articulated as follows. On the one hand, the overloaded regime is anticipated to serve as a significant application for next-generation wireless networks, where NOMA emerges as a promising technology to facilitate massive connectivity. Conversely, existing research highlights that NOMA offers enhanced compatibility and flexibility. This enables the seamless integration of NOMA with a range of advanced and intelligent technologies, such as multi-antenna techniques, RIS, NTN, CAVs, healthcare 5.0, ISAC, OTFS, DTN, Tactile Internet, and zero-touch networks. To add to this, we noted that a broad spectrum of research is currently available, which focuses on design techniques and principles, optimizing resource allocation and connectivity for static devices. On the other hand, advanced MA techniques will be necessary for the NGWNs to combine all horizons (land, air, sea, and space) for ubiquitous and seamless connectivity, as well as for the advent of CAVs, extended reality, IIoT, IoRT, and healthcare 5.0. Given the advantages that NOMA offers, this thorough survey seeks to highlight NOMA's capabilities as a potential MA technique for the NGMA, addressing the requirements of the Versatile and Ultra-Connected NGWNs to guarantee seamless and reliable connectivity. Researchers from academia and industry are still actively working on NOMA and NGMA, and hence there is undoubtedly more to explore in this domain. Note that there are some relevant surveys focusing on the fundamentals of NOMA or the enablers of NOMA towards 5G. We examine the various key enabling technologies along with relevant recent literature, challenges and proposed solution techniques in the literature, and opportunities that NOMA can bring to the table as an NGMA technique, followed by design guidelines and the future perspective of NOMA.

### *F. Contributions*

The vast majority of the previously conducted surveys were conducted in the context of 5G and concentrated on just one or a few components of PD-NOMA or CD-NOMA, including fundamental operations, rate optimization, and enabling technologies. In contrast, our survey provides a comprehensive review of NOMA. Spectrum efficiency and reliable connectivity are becoming increasingly crucial to the forthcoming 6G network because of technology's rapid development and the IoE's expansion. In this context, we disclosed NOMA's efficacy and potential benefits to drive it towards NGMA in enhancing the functionality of the next generation of wireless communications regarded as NGWNs. After a brief introduction to NGMA design requirements and candidate techniques, we explained the two distinct NOMA types, namely PD-NOMA and CD-NOMA. In addition, we also provided a brief overview of NOMA variants to set a firmer foundation of NOMA for NGMA. We also focused on showcasing the applicability of NOMA in a broad array of wireless communications technologies. Besides the benefits of NOMA's interplay, we also presented the literature review of NOMA's integration with a range of wireless communications technologies. By doing this, we not only highlighted the synergies and advantages offered by NOMA but also exhibited a comprehensive understanding of how NOMA can be integrated with state-of-the-art wireless communications technologies in the forthcoming 6G era. Massive connectivity, despite its benefits, introduces new challenges. To this end, we revealed the research challenges and open problems that need to be addressed for implementing the NOMA in the NGWNs. We also explored the transformative potential of NOMA with trending and innovative wireless communications technologies, unlocking new opportunities for novel applications. We compiled design guidelines and valuable lessons learned from this extensive survey. By doing so, we strive to pave the way for future advancements in the field, ensuring continued progress and innovation. We completed this survey by providing the future perspectives of NOMA along with the groundbreaking applications and use cases in the exciting 6G era.

### *G. Survey Organization*

Fig. 2 shows the organization of this paper, and Table I outlines the acronyms used in this paper. Tables II and III compare this survey with the existing surveys. Section II presents the basic functionality of NOMA along with the comparative analysis of different NOMA variants; Section III ascribes pivotal key enabling wireless communications technologies for NOMA along with a literature review; Section IV delivers the set of the research trends for NOMA in 6G wireless communications technologies, Section V underlines the research challenges and open problems; Section VI presents the design recommendations and insights. Section VII discusses future perspectives of NOMA, and finally, we conclude this survey in Section VIII.

## II. Fundamentals of NOMA and Its Variants

The sensations of the wireless communications revolution are constantly evolving, set to take another significant stride with the debut of 6G. The potential of NOMA as the frontrunner of NGMA to transform connectivity, spectral efficiency, ultra-low latency, and energy efficiency in NGWNs is immense. The exponential growth of smart devices and their diverse applications and use cases resulted in a pressing need to develop efficient multiple access schemes. This section introduces the fundamentals of NOMA, including its various variants. Subsection II-A highlights the operations of PD-NOMA along with uplink and downlink scenarios. Subsection II-B reflects the fundamental operations of different CD-NOMA variants. Table IV depicts the analysis of different NOMA variants.

### *A. PD-NOMA*

In contrast to the various multiple access schemes that depend on the allocation of resources in the time, frequency, or code domains, NOMA can be implemented in the power domain, referred to as PD-NOMA.[9] [10] Multiple user's signals are superimposed in the transmitter using SC. This permits the simultaneous transmission of multiple signals over the same resource block, enhancing the system's spectral efficiency. To enhance the communication efficiency, multiple users share the same bandwidth resources, which are then detected at the receiver by MUD algorithms such as SIC [99], MMSE [100], and maximum likelihood detector [101] resulting in enhanced spectral efficiency. Non-orthogonal multiplexing with superposition coding at the transmitter and SIC at the receiver outperforms traditional orthogonal multiplexing and is optimal for attaining the capacity regions of the uplink and downlink broadcast channel [24], [27]. Fig. 3 shows the uplink and downlink of two user PD-NOMA systems.

PD-NOMA has gained significant attention from both academia and industry due to its perks and benefits across a vast array of wireless communications technologies such as NOMA-MIMO [102], Network-NOMA [103], NOMA-MEC [104], NOMA-RIS [105], STAR-RIS-assisted NOMA [106], full-duplex NOMA [107] and more. In [108], authors considered the multi-cell NOMA network, which focuses on the coordination among BSs, and investigates the downlink user coordination mode selection and resource allocation. Multi-criteria user mode selection and fuzzy logic were used to strike a balance among multiple criteria to achieve robustness against the combined effects of shadowing, fading, and ICI. Moreover, the authors proposed two resource allocation algorithms: first, serving channel gain-based subchannel allocation (SCG-SA); second, a low complexity fuzzy logic user ranking order-based joint resource allocation algorithm (FLURO-JRA). Both the proposed algorithms provided the near-optimal solution with a relatively small number of users. Authors in [109] jointly optimized beamforming, uplink transmit power, and downlink power allocation factors using successive convex optimization (SCA) and semi-definite programming (SDP) for full-duplex cell-free NOMA in SAGIN. The proposed algorithms outperform the frequency division duplex (FDD) based cell-free NOMA/OMA and FD cell-based NOMA systems and enhance the spectrum efficiency. In [110], a hybrid MA scheme was considered based on the TDMA and NOMA. In this work, a novel approach utilizing sequential convex approximation and second-order cone (SOC) was employed to develop an iterative algorithm aimed at optimizing energy efficiency.

[9]This survey focuses specifically on the PD-NOMA-based schemes. The State-of-the-Art literature review in Section III of this survey explores the various schemes of PD-NOMA and other wireless communications technologies.

[10]The survey primarily addresses narrowband communications, but we would like to clarify that certain concepts and principles discussed may also be relevant to wideband communications. However, specific equations for wideband communications were not included in this survey as they fall outside the scope of our work.

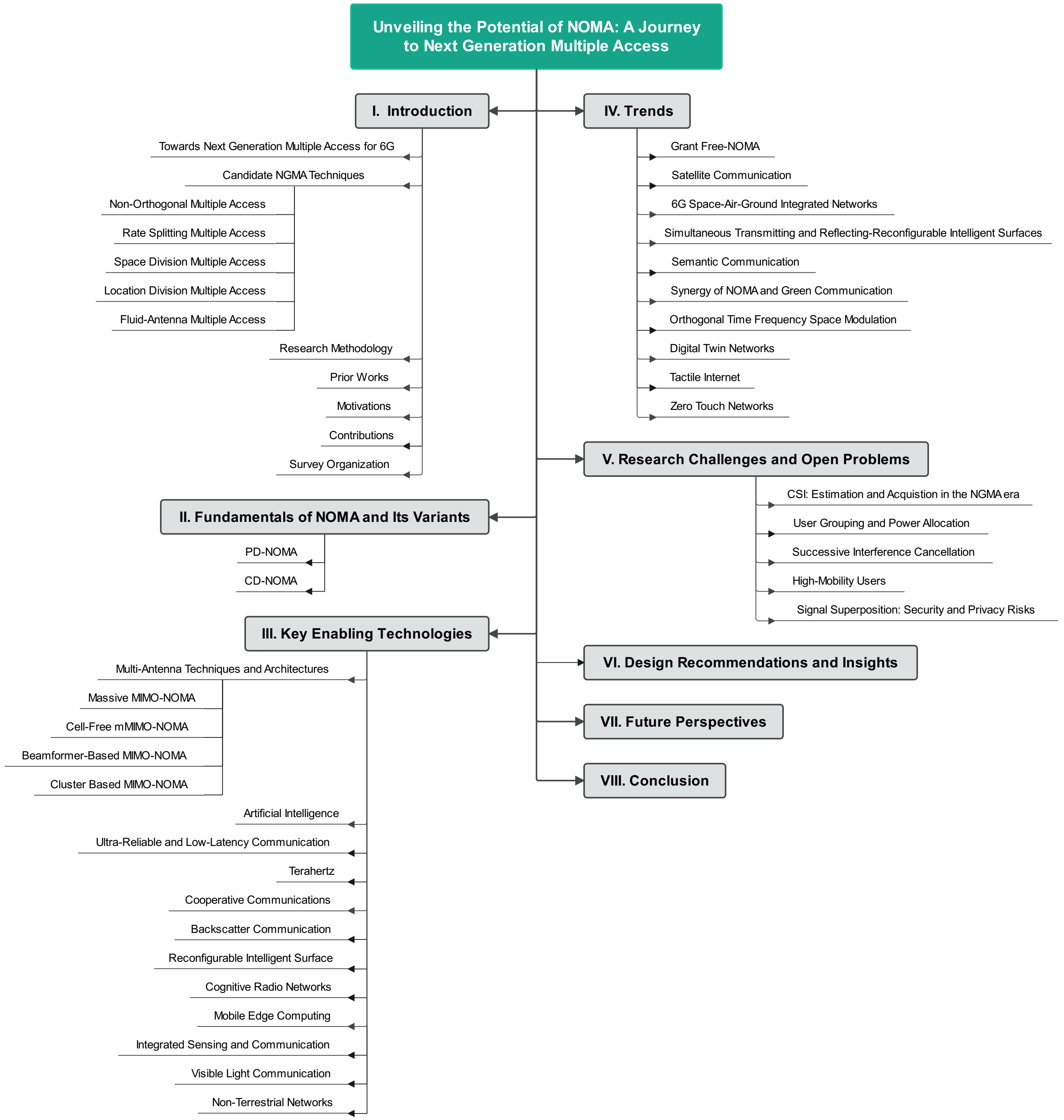

Fig. 2: Paper outline diagram with hierarchical sections and sub-sections.

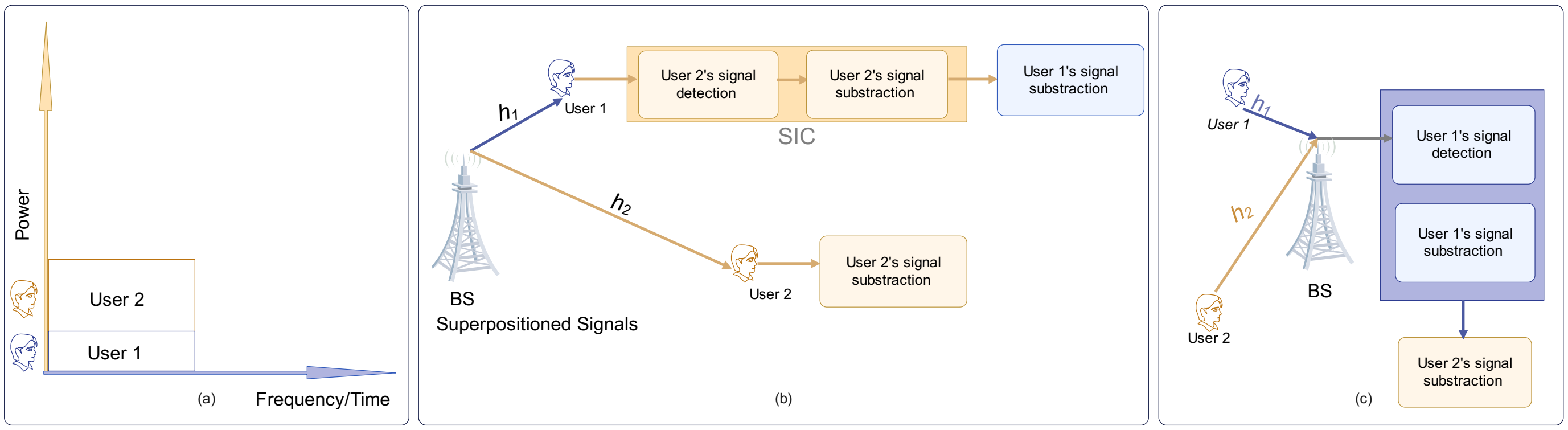

Fig. 3: Illustration of NOMA transmission; (a) Power domain multiplexing of two users's signals in same time/frequency resource; (b) Downlink two user PD-NOMA transmission; (c) Uplink two user PD-NOMA transmission.

*1) Downlink NOMA:* In the DL-NOMA system, the BS adopts the superposition coding technique to multiple signals of multiple users that have been assigned different power allocation coefficients. At the receiver, successive interference cancellation is exploited by receivers to efficiently remove interference. Extensive research has been carried out to investigate the performance of NOMA in the DL scenario. Some researchers also studied the performance of DL-NOMA using partial CSI at the transmitter side. Fig. 3(b) showcases the classic two-user NOMA network.

Let us consider the DL of a multi-user NOMA network consisting of a set of users denoted by $k = \{1, 2, ...K\}$, total power budget $P$, such that $P = \sum_{i=1}^{K} p_i$ and the BS transmits the signal $x_i$ to the $i^{th}$ user with the power coefficient $p_i$. Then, the signals for multiple users are weighted by the power coefficients $p_i$ and superimposed at the BS is formulated in (1). The received signal at the $i^{th}$ user is formulated in (2).

$$x = \sum_{i=1}^{K} \sqrt{p_i} x_i, \tag{1}$$

$$y_i = h_i \sum_{i=1}^{K} \sqrt{p_i} x_i + \omega_i, \tag{2}$$

where $h_i$ refers to the channel coefficient between the BS and user $i$ and $\omega_i$ associated with the power density $N_i$ denotes the zero mean complex additive Gaussian noise plus the inter-cell interference. Fig. 3(b) shows the two-user case of DL PD-NOMA. User 1 is a near user with a better channel quality (strong user), while user 2 is a far user with a worse channel quality (weak user), such that $|h_1|^2 \geq |h_2|^2$. According to NOMA, the BS will allocate more power to the weak user to provide fairness and facilitate the SIC process $p_1 \leq p_2$. The BS then performs the superposition coding to generate the superimposed signal, which is then sent to all the users simultaneously. In [111], authors derived the closed-form equations for the optimal power allocation for two users. According to the authors in [111], both the upper and lower bounds on the transmit power fraction to achieve the better sum and individual capacities have the form $p_i(x_i) = \left[(1 + Px_i)^{1/2} - 1\right] / (Pxi)$.

According to the NOMA scheme, each user performs SIC at its receiver. As mentioned above, users are arranged according to their channel gains, such as, $|h_1|^2 \geq |h_2|^2$, i.e., user1 and user2 are regarded as strong and weak users. User 1 (strong user), with high channel gain, performs SIC according to the ascending order of channel gains and decodes and subtracts the signal of user 2 before decoding its own signal. On the contrary, user 2 considers user 1's signal as noise as its transmission power is lower than user 2. Subsequently, user 2 can decode its signal directly [112]. The achievable data rates of both users are given below:

$$R_1 = \log_2 \left(1 + \frac{p_1 |h_1|^2}{N_1}\right), \tag{3}$$

$$R_2 = \log_2 \left(1 + \frac{p_2 |h_2|^2}{p_1 |h_2|^2 + N_2}\right), \tag{4}$$

*2) Uplink NOMA:* Multiple users send their individual uplink (UL) signals to the BS in the same RB during UL transmission [113]. With the help of SIC, the BS can identify all user's messages. There are some significant differences between UL-NOMA and DL-NOMA, which are listed as follows:

- **Transmit Power:** In contrast to downlink NOMA, the transmit power of the users in uplink NOMA does not have to be different, and it depends on the channel conditions of each user. If the users' channel conditions are significantly different, their received SINR can be rather different at the BS, regardless of their transmit power [26].
- **SIC Operations:** The SIC operations and interference experienced by the users in the uplink NOMA and downlink NOMA are also rather different. In downlink NOMA, the strong channel users achieve throughput gains by successively decoding and cancelling the messages of weak channel users prior to decoding their desired signals. In the uplink, to enhance the throughput of weak channel users, the BS successively decodes and cancels the messages of strong channel users prior to decoding the signals of weak channel users [114].

- **Implementation Complexity:** Downlink NOMA requires the implementation of sophisticated multi-user detection and interference cancellation schemes at the receiver of each user. This is a cumbersome task, provided the limited processing capability of users. However, in the uplink, it is relatively more convenient to implement multi-user detection and interference cancellation schemes on a centralized entity (i.e., BS) [115].
- **Performance Gain:** The performance gain of NOMA over OMA is different for downlink and uplink. As illustrated in the [116], the capacity region of NOMA is outside OMA, which means that the use of NOMA in downlink has superior performance in terms of throughput. While in uplink systems, NOMA mainly has advantages in terms of fairness, especially compared to OMA with power control. Also, in [111], the authors came up with the concept of fair-NOMA, in which each user is allocated its share of the transmit power such that its capacity is always greater than or equal to the capacity that can be achieved using OMA. This fairness is achieved through power allocation adjustments, where power control is applied to allocate power fractions among users based on their channel gains. Similarly, in [117], the authors investigated the impact of power allocation on the fairness performance of NOMA networks under (i) instantaneous CSI and (ii) average CSI. Furthermore, in the aforementioned work, authors deployed low-complexity bisection-based iterative algorithms to solve the optimization problem that resulted in globally optimal solutions.

Fig. 3(c) depicts the two-user uplink PD-NOMA system. The received signal at the BS from multiple users is formulated as below:

$$y = \sum_{i=1}^{K} h_i \sqrt{p_i} x_i + \omega, \tag{5}$$

where $p_i$ and $x_i$ are the transmitted power and signal transmitted by $i^{th}$ user, respectively, and $\omega$ represents the Gaussian noise that also captures the ICI at the BS. Multiple users transmit their signals simultaneously to the BS over the same frequency band in uplink NOMA. To separate and decode the signals of multiple users, the BS employs sophisticated MUD techniques, such as SIC, to separate the user's signals. As shown in Fig 3(c), in the uplink, user1 is near user and user2 is far user e.g., $|h_1|^2 \geq |h_2|^2$, and BS conducts SIC according to the descending order of channel gains [118]. An important point to note is that, in contract to downlink NOMA, the transmit power of the users in uplink NOMA does not have to be different, it depends on the channel conditions of each user. If the users' channel conditions are significantly different, their received SINR can be rather different at the BS, regardless of their transmit power [114]. For the two-user example, in the first step, the signal of user 1 is decoded, while the signal of user 2 is treated as noise. In the second stage, BS subtracts the decode signal $x_1$ from the received signal and then decodes the signal of user 2.

### B. CD-NOMA

The CD-NOMA method involves the utilization of user-specific spreading sequences to perform signal multiplexing at the transmitter end. This technique offers several benefits, including a low density, low inter-correlation, and the ability to provide grant-free access. The CD-NOMA technique relies on compressive sensing (CS), and sparse codes as its fundamental components. There are five different variations of it, namely: (i) Scrambling-based NOMA, (ii) Interleaving-based NOMA, (iii) Spreading-based NOMA, (iv) Coding-based NOMA, and (v) Lattice and Beam based NOMA. Table IV captures the characteristics of different NOMA variants including PD-NOMA [83], [88], [119]–[121]. Further elaboration on these variants is provided in the subsequent subsections.

*1) Multi-User Shared Access (MUSA):* The MUSA scheme is a NOMA system that operates by employing a concise complex spreading sequence and implementing a receiver based on SIC [122]. Fig. 4(a) depicts the transmitter and receiver structure of MUSA for $k$ users. Following channel encoding and modulation, the data symbols of each user undergo spreading using a complex sequence characterized by elements that reside within the complex field. Subsequently, the spread symbols of each user are transmitted over the allocated radio resources, allowing for shared usage. Upon reception, the multi-user detectors leverage the strategically designed spreading sequences to effectively discriminate and decipher the different data streams of users [123]. Multiple spreading sequences constitute a pool from which each user can randomly pick one of the sequences. Note that for the same user, different spreading sequences may also be used for different symbols, which may further improve the performance via interference averaging [122], [123].

The utilization of short sequence-based spreading serves as a fundamental operation in the MUSA transmitter, playing a significant role in achieving these benefits. The design of spreading sequences in MUSA adheres to the principles of low cross-correlation. Each element of the sequence is chosen from a complex scalar set, specifically consisting of $(\pm 1, 0)$. This selection process ensures that the spreading sequences exhibit minimal correlation with each other, contributing to effective interference management and improved system performance. Complex spreading sequences utilized in MUSA offer lower cross-correlation compared to pseudo-random noise (PN) sequences, even with shorter lengths. The flexibility of selecting arbitrary complex elements enables a large pool of spreading sequences in MUSA, enhancing system performance and interference management within NOMA. [124], [125].

*2) Non-Orthogonal Coded Access (NOCA):* NOCA utilizes the inherent characteristic of low cross-correlation, and this particular property manifests its efficacy in both the time and frequency domains. Before transmitting a specific sequence, the modulated symbols are spread using non-orthogonal sequences in this particular scheme [126]. Its spreading sequences are the LTE-defined reference signals, which are generated by cyclic shifts of a base sequence. Fig. 4(b) presents the transmission structure of NOCA. The serial-modulated symbol sequence is first converted to $P$ parallel subsequences by a $S/P$ converter. $C_i^j$ is defined as the non-orthogonal spreading sequence with length $SF$, where $SF$ denotes the spreading factor. The $j^{th}$ subsequence is then spread on $SF$ subcarriers according to $C^j$. Hence, a total of $PXSF$ subcarriers are required for NOCA [88], [122]. The authors in [127], exhibit the exploitation of Gaussian Approximation-based detection algorithm for NOCA. Furthermore, in the aforementioned work, the efficiency in the sense of complexity and suitable use cases of Gaussian Approximation-based detection were investigated with respect to NOCA and IDMA by link-level simulations. Their numerical results reveal that both IDMA and NOCA are suitable for sporadic traffic. In particular, NOCA exhibits high detection efficiency for earlier iterations, and IDMA exhibits the capability of overloading and performance convergence under very low SNR. In [128], the authors investigated the potential of applying NOCA in uplink GF transmissions with Proactive HARQ retransmission schemes to satisfy the requirements of URLLC.

TABLE IV: CHARACTERISRTICS OF DIFFERENT NOMA VARIANTS INCLUDING PD-NOMA.

| Variants | Name | Organization | Single carrier or Multi carrier | Uplink or Downlink | Receiver used | Technique | Advantages | Limitations |
|---|---|---|---|---|---|---|---|---|
| Short spreading sequence | MUSA | ZTE | Single carrier | Uplink | MMSE-SIC | Short complex sequence | High user capacity, enhanced connectivity | |
| | NOCA | NOKIA | Single carrier | Uplink | MMSE-SIC | Zadoff-Chu sequence | Reduced PAPR, flexible, low complexity | |
| | NCMA | LGE | Single carrier | Uplink | MUD+IC | Grassmanian line packing | Optimal non-orthogonal sequence | High transmitter overhead |
| | SSMA | INTEL | Single carrier | Uplink | MMSE-SIC, MUD+IC | Orthogonal or quasi-orthogonal codes | Low receiver complexity | Actual sequence length affects the performance |
| Long spreading sequence | GOCA | Mediatek | Single carrier | Uplink | SIC | Group-based orthogonal and non-orthogonal sequences | Inter-group orthogonality | |
| | WSMA | Ericsson | Single carrier | Uplink and Downlink | MMSE-SIC | Welch bound coding | low receiver complexity | |
| Coding | PDMA | CATT | Multi carrier | Uplink and Downlink | MPA-IC | Irregular LDS | Provide multidimensional diversity, receiver complexity, low irregular protection | |
| | SCMA | Huawei | Multi carrier | Uplink and Downlink | MPA+IC | Multidimensional modulation | Signal space diversity gain | Optimal codebook design |
| | LDS-SVE | Fujistu | Single carrier | Uplink and Downlink | MPA+IC | LDS and user signature vector extension | Large diversity | Define LDS code and signature vector extension method |
| | LDS-CDMA | Fujistu | Single carrier | Uplink and Downlink | MPA+IC | Sparse spreading CDMA | CSI independent | Redundant coding |
| | LDS-OFDM | Fujistu | Multi carrier | Uplink and Downlink | MPA+IC | Sparse spreading OFDM | CSI independent, provide wide-band signals | Redundant coding |
| Lattice and Beam | LPMA | Nokia | Single carrier | Uplink and Downlink | SIC | Multilevel lattice code and multiplexing in power and code domain | User clustering independent | Specific channel coding |
| | BOMA | | Single carrier | Uplink and Downlink | SIC | Tiled building block | Low receiver complexity | User pairing dependent |
| Interleaving | IGMA | Samsung | Single carrier | Uplink and Downlink | MPA | Bit-level interleaving | Low coding rate, sparse grid mapping | High transmitter overhead |
| | IDMA | Nokia | Single carrier | | Elementary signal estimator (ESE)+IC | Bit-level interleaving | Large diversity gain, low coding rate, less interference | Interleaving design |
| | RDMA | Mediatek | Single carrier | Uplink | SIC | Cyclic shift based time frequency repetition | Easy implementation | |
| | LCRS | Intel | Single carrier | Uplink | ESE-PIC | Bit-level spreading | High code gain | Users separation at receiver receiver depends upon its structure |
| Scrambling | RSMA | Qualcomm | Single carrier | Uplink | Matched filter+IC (MMSE-SIC, ESE-PIC) | Low cross-correlation sequence scrambling | Fit for asynchronized scenario | Not suitable for SE, and requires synchronous multiplexing with OFDM |
| | LSSA | ETRI | Single carrier | Uplink | MMSE-SIC | User specific bit-level interleaving | Low rate FEC code or moderate one with repetition, Large number of signatures | |
| Power domain NOMA | PD-NOMA | NTT-DOCOMO | Single carrier | Uplink and Downlink | SIC | Signals separated through different power levels | High SE and low outage probability | Error propagation |

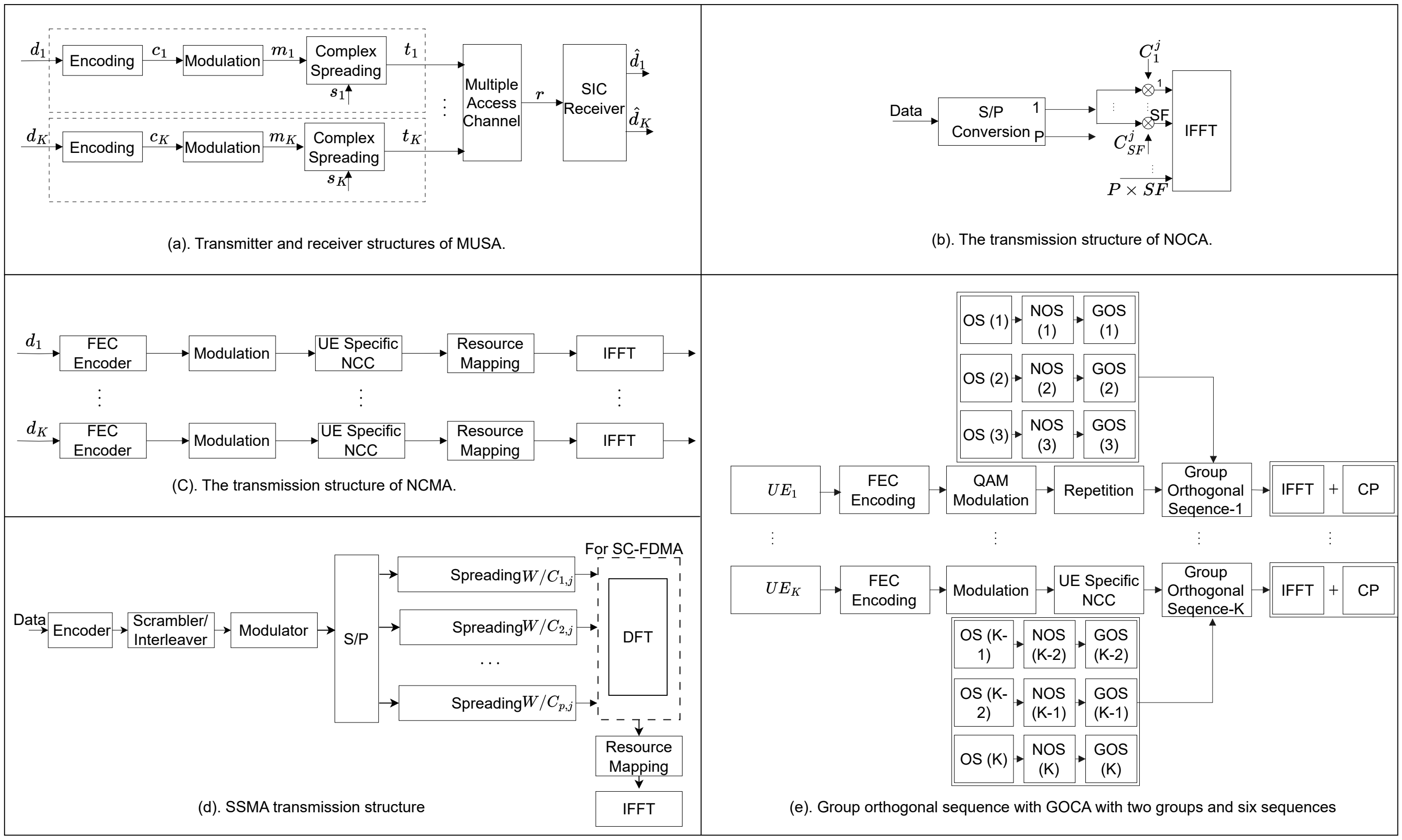

Fig. 4: A comparative representation of different code domain NOMA schemes.

*3) Non-Orthogonal Coded Multiple Access (NCMA):* The NCMA technique adopts dense spreading sequences akin to the MUSA. However, minimizing the multi-user interference (MUI) is the primary objective of NCMA and effectively manages the high overloading regime [123]. Fig. 4(c) presents the transmission structure of NCMA. The acquisition of spreading sequences in NCMA is achieved by solving the Grassmannian line packing problem. The goal is to optimize the chordal distance between spreading codes. The spreading sequences utilized in NCMA are often known as non-orthogonal cover codes (NCC). These codes have been found to be well-suited for predicting the level of interference.NCMA offers two key benefits: i) increased throughput and ii) improved connectivity with reduced BLER.

*4) Group-Orthogonal Coded Access (GOCA):* GOCA adopts a two-stage method to generate grouped orthogonal sequences and then spreads modulated symbols into the same time and frequency resources. Orthogonal sequences and non-orthogonal sequences are used in the first and the second stages, respectively. GOCA's spreading sequences can be divided into different groups for multiple users, and each group has the same orthogonal sequence set and different nonorthogonal sequences. Thus, the users within each group remain orthogonal, while the users in different groups are nonorthogonal. GOCA also takes advantage of the MMSE-SIC algorithm for multi-user decoding [129]. Fig. 4(e) presents the transmission structure of GOCA. According to [130], GOCA's main advantage is its potential to manage inter-user interference more effectively through additional scrambling. This advantage does not significantly translate into better BLER performance under the tested conditions compared to MUSA, NCMA, and NOCA. This suggests that while GOCA may offer theoretical benefits in managing interference, these benefits might not always manifest in practical improvements in performance, especially under non-ideal conditions where blind detection and channel estimation complexities are involved. Besides, the GOCA scheme is also widely acknowledged for its ability to strike a trade-off between the complexity of the transmitter and receiver, thereby delivering superior performance compared to other NOMA variants like IDMA, RAMA, and RDMA [131].

*5) Short-Sequence Based Spreading Multiple Access (SSMA):* SSMA is a variant of CD-NOMA that employs the direct spreading of modulation symbols using multiple codes that are either orthogonal or quasi-orthogonal. Fig. 4(d) presents the transmission structure of SSMA [132]. Non-orthogonal transmission is used to allocate time-frequency resources for transmitting spread symbols. SSMA's transmission structure has a connection to NOCA, as it employs user-specific scrambling to alleviate MUI. The modulation symbols in SSMA are split into several streams, and each stream is spread by utilizing a spreading sequence chosen from a predefined spreading codebook. The codebook consists of a set of sequences with low cross-correlation properties, which can be generated using various methods such as Hadamard matrices, Zadoff-Chu sequences, or Golay sequences. This spreading of the modulation symbols allows for improved separation of signals from users.

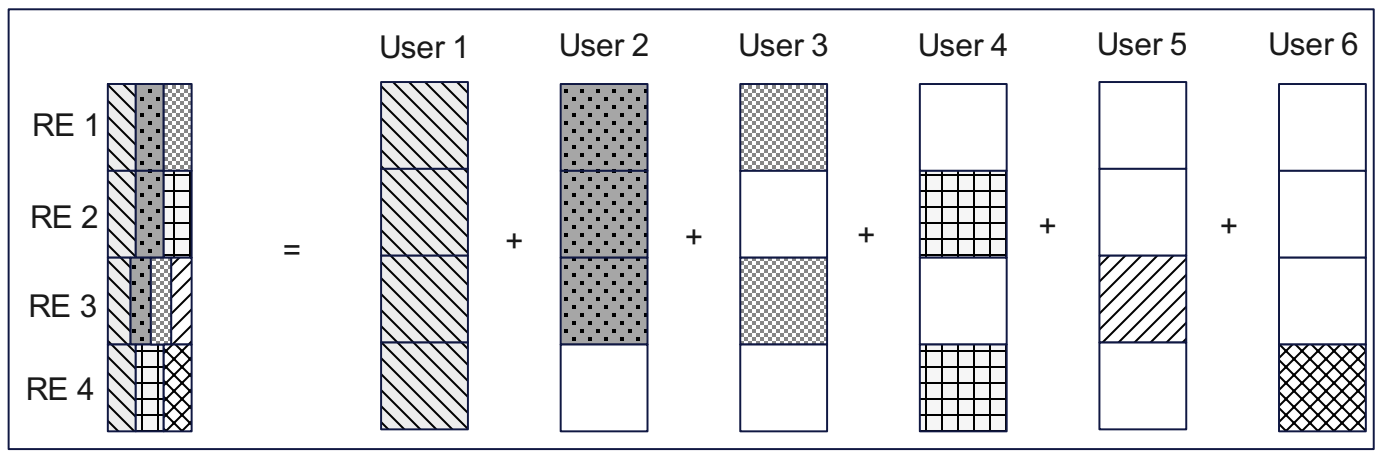

Fig. 5: Resource mapping of PDMA with six users and four subcarriers [55].

Physical resources are assigned to the spread symbols, and in the case of OFDM [133], an operation called Inverse Fast Fourier Transform (IFFT) is performed. Depending on the system's particular requirements, SSMA has the flexibility to employ either time or frequency domain spreading on the modulated symbols. To achieve optimal performance, it is important to determine the length of the spreading sequence in either the time or frequency domain by taking into account the coherence time and bandwidth.

*6) Welch-Bound Equality Spread Multiple Access (WSMA):* The spreading sequence utilized by the WSMA was derived from the Welch bound, resulting in a decrease in the cross-correlation of the spreading sequences. The utilization of MMSE-SIC technique is employed in the decoding of the user's signals at the receiver end. Furthermore, the spreading sequence of WSMA has a lower cross-correlation in comparison to the spreading sequences of MUSA [134].

*7) Pattern-Division Multiple Access (PDMA):* Initial PDMA was introduced to improve 5G radio networks. The PDMA expands on this idea by using a pattern division technique to increase the network's efficiency and capacity even further. The PDMA pattern specifies how transmitted data is assigned to a group of resources, which may consist of time, frequency, spatial resources, or a combination of these. For the purpose of optimizing the transmission of data, the PDMA pattern offers a standardized approach for optimally managing and utilizing these resources. The pattern was meant to aid in distinguishing the signals from different users sharing the same resource block. Resource mapping in PDMA is illustrated in Fig. 5. The pattern was introduced to identify signals from numerous users using the same resource block. Utilizing the pattern benefits from the unique diversity order and sparsity properties of the transmitter and receiver. This design innovation tries to improve system efficiency while maintaining detection complexity at a reasonable level [55].

The order of transmission diversity is determined by the number of mapped resources in a group. Various users' data can be efficiently combined into a shared resource group by employing distinct PDMA patterns. Nonorthogonal transmission is accomplished using this approach. The facilitation of achieving diverse transmission diversity orders among users can be accomplished by assigning the PDMA pattern with varying diversity orders. In short, PDMA utilizes a pattern to construct a sparsely mapped relationship between data and a group of resources. An effective illustration of the PDMA pattern can be obtained by utilizing a binary vector.

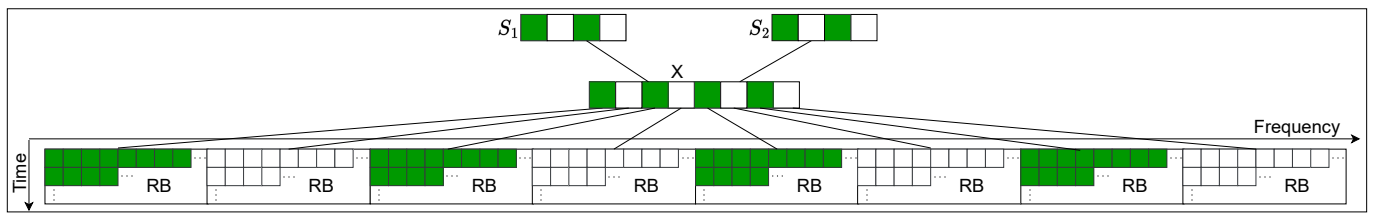

Fig. 6: An illustration of LDS-SVE.

A vector's dimension corresponds to the resources in a group. Each individual element within the vector is associated with a distinct resource within a designated resource group [135].

*8) Low-Density Spreading Signature Vector Extension (LDS-SVE):* LDS-SVE is the LDS-based NOMA scheme [129]. An illustration of LDS-SVE is shown in the Fig. 6. In LDS-CDMA, each modulated symbol is spread separately onto multiple subcarriers. In LDS-SVE, two modulated symbols are transformed and spread together onto twice as many subcarriers. The technique described above entails the multiplication of both the imaginary and real parts of two modulated symbols using a transformation matrix that has been optimized by minimizing the single-user BER. Moreover, the subcarriers that have been mapped can be scattered in order to effectively utilize diversity. Moreover, the mapped subcarriers can be scattered in order to effectively utilize diversity. At the receiver's end, Turbo-equalization and SIC can be jointly employed with Message Passing Algorithm (MPA) to approximate the performance of a single user, that is, without any interference from other users [136].

*9) Low-Density Spreading CDMA (LDS-CDMA):* The LDS CDMA's operational mechanism is a sparse spreading sequence. Interference among users is reduced by using a sparse spreading sequence. It works the same way as a regular CDMA, but here we use LDS instead of conventional sequences. The utilization of the spreading sequence encourages more supported users, thereby enhancing the overall efficiency of the system. The LDS-CDMA system comprises three primary components, namely signal to spread, zero padding, and matrix interleaving. In contrast to the technique of zero-padding matrix interleaving, which serves to improve the sparsity of the signal while retaining processing gain, the method of signal spreading can be used to spread the signal. LDS-CDMA demonstrates enhanced performance in comparison to a conventional CDMA system. LDS spreading employs a randomly generated Hadamard matrix [137].

*10) Sparse Code Multiple Access (SCMA):* Overloaded systems, in which the user numbers exceed signal space dimensions, are crucial for bandwidth-efficient multi-user communications. SCMA was originally proposed in [138], and is an enhanced low density spreading technique (LDS). A typical SCMA transmission scheme is shown in Fig. 7. The SCMA technique is a spreading-based scheme that exhibits low density. This approach has the potential to achieve high overloading while simultaneously maintaining a high level of reliability. The fundamental concept underlying SCMA involves the direct mapping of coded bits onto multi-dimensional modulation symbols, utilizing a pre-established sparse codebook. This approach differs from the conventional method of sequentially performing modulation and low-density spreading. Consequently, the incorporation of both the resource element mapping and the multi-dimensional constellation are fundamental design aspects within the context of SCMA. One notable distinction between LDS and SCMA lies in the fact that SCMA employs a multi-dimensional constellation to produce codebooks, thereby enabling the attainment of a "shaping" gain that is unattainable for LDS. To streamline the multi-dimensional constellation design process, it is possible to create an additional constellation through the minimization of the average alphabet energy while maintaining a minimum Euclidean distance between constellation points. This approach also considers codebook-specific operations, including phase rotation, complex conjugate, and dimensional permutation [139].

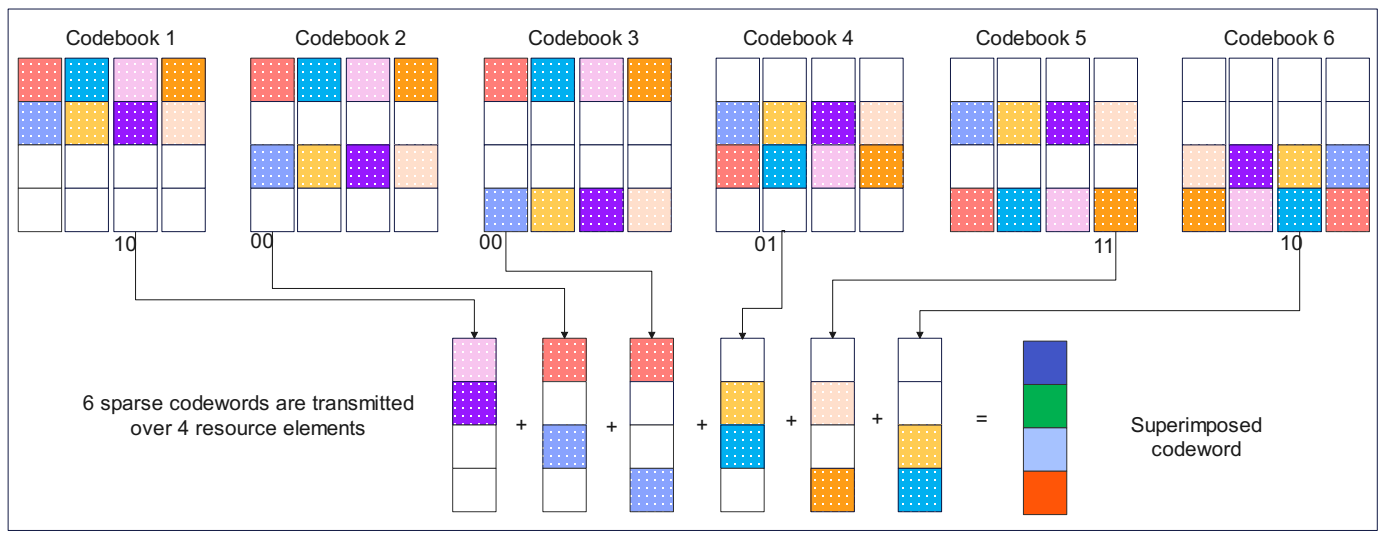

Fig. 7: A typical transmission structure of SCMA with six users and four subcarriers.

*11) Lattice Partition Multiple Access (LPMA):* LPMA enhances the performance of the system by combining the power domain and the code domain [140], [141]. Users in LPMA are each given a unique code that is based on the CSI. Users that have a poor CSI are given codes that have a large minimum distance, which improves detection performance. Users that have a stronger CSI can have codes with a shorter minimum distance without compromising the performance of the network. In addition to this, power domain multiplexing is utilized by LPMA in order to enhance the performance of users with poor CSI. The LPMA system utilizes lattice coding techniques in the transmitter for the purpose of encoding user information and employs SIC at the receiver to decode the encoded information. It increases power domain throughput by superimposing different power streams and code domain security by using linear combinations of lattice codes. The performance gain of LPMA surpasses that of PD-NOMA when the channel quality differences among users increase. Nevertheless, it is worth noting that it demonstrates greater levels of encoding and decoding complexity in comparison to CD-NOMA. The LPMA framework relies on the utilization of multilevel lattice codes, which effectively leverage the design features of lattice codes to effectively handle co-channel interference between multiple users. LPMA multiplexes users by giving them distinct lattice partitions that are isomorphic to the same finite field in cases when users have similar channel conditions. A schematic illustration of LPMA is presented in Fig. 8.

*12) Interleave Grid Multiple Access (IGMA):* IGMA leverages the integration of bit-level interleaving and symbol-level sparse grid mapping as a means to improve the overall efficiency of the entire network. In the IGMA framework, users are differentiated by employing diverse characteristics, including unique bit-level interleavers, different grid mapping patterns, or an integration of both. It uses bit-level interleaving and grid mapping in its transmitter architecture. Forward Error Correction (FEC) with a low coding rate or simple repetition can be used for channel coding. Sparse mapping based on In contrast to the requirement for well-designed codewords or code sequences, the availability of sufficient bit-level interleavers andor grid mapping patterns offers not only the capability to accommodate various connection densities but also the flexibility to achieve a favourable trade-off between channel coding gain and optimizing the use of limited resources. By proper selection, low correlated bit-level interleavers could be achieved. Symbol-level interleaving is a technique that randomizes the order of symbol sequences. This can provide further advantages in mitigating frequency selective fading and inter-cell interference [135]. The transmission structure of IGMA and grid mapping is shown in Fig. 9(a).

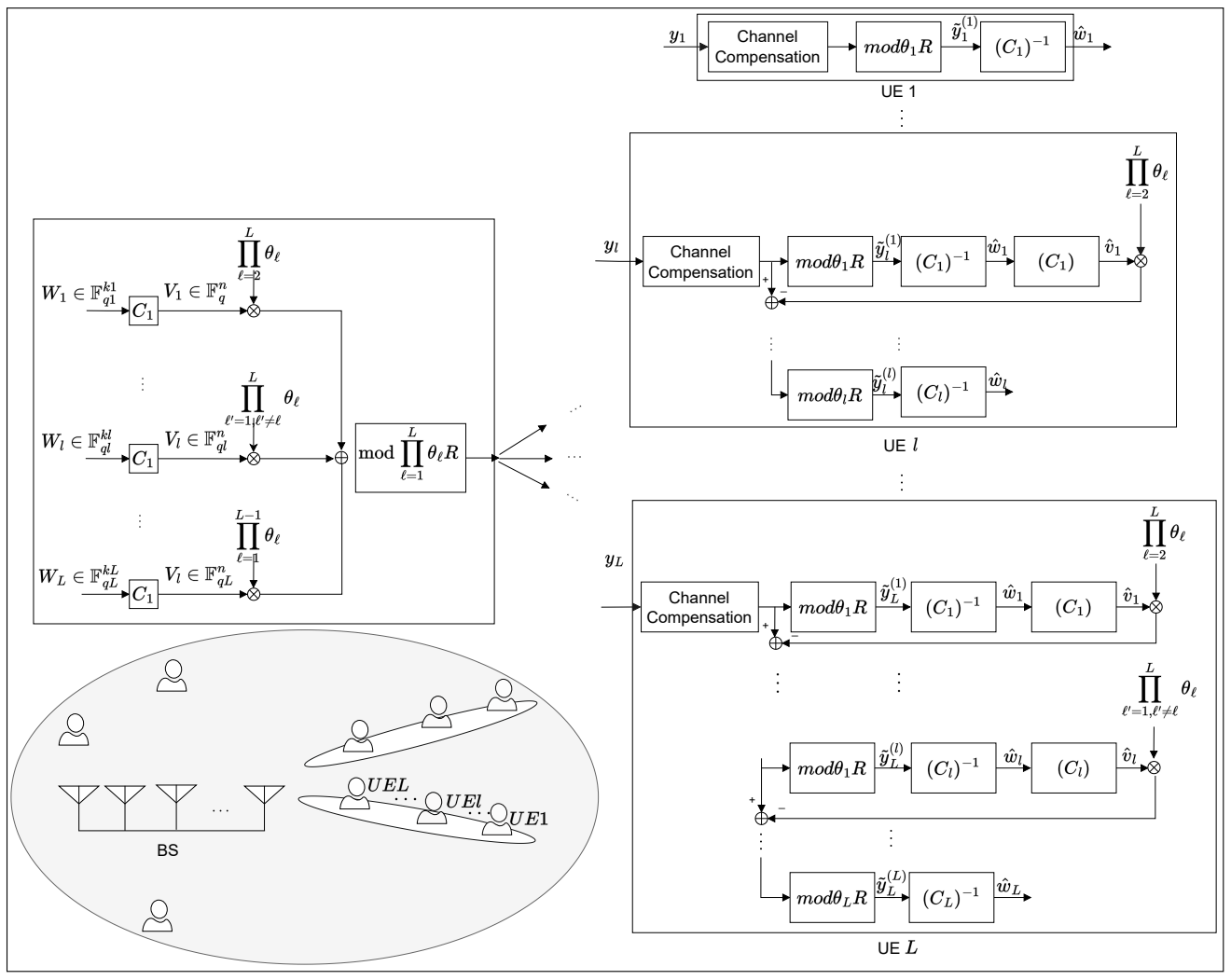

Fig. 8: Schematic illustration of LPMA [140].

At the receiver end, the utilization of interleaving allows for the implementation of a low-complexity multi-user detector known as an elementary signal estimator (ESE) [142]. It should be noted that a decrease in the density of the grid mapping pattern has the potential to further decrease the detection complexity of ESE in the IGMA implementation. Furthermore, the utilization of MAP and MPA detectors in the context of IGMA has been found to significantly enhance detection performance when compared to ESE at the cost of increased complexity. Along with iterative detection and decoding at the receiver, IGMA provides significant user multiplexing capability and reliable transmission. It can also be applied in scenarios such as mMTC and URLLC.

*13) Low Code Rate Spreading (LCRS):* LCRS is a non-orthogonal transmission scheme based on the combination of low code rate transmission and user-specific demodulation reference signals (DM-RS) to differentiate different users at the receiver. Additionally, user-specific scrambling or interleaving can be employed to improve multi-user detection performance further when using more advanced receivers, e.g., turbo equalization-based receivers. A typical LCRS transmitter is depicted in Fig. 9(b) [132]. Each user applies an encoding procedure for its portion of information bits, utilizing repetitions and rate-matching procedures that provide coded bit sequences with low code rates assuming that transmission resources are sufficiently large relative to the transmitted packet sizes. Further, user-specific interleaving or scrambling can be optionally employed for improving multiuser signal separation at the receiver side. This is followed by the QAM modulation procedure. An optional DFT block can be added for the SC-FDMA-based uplink transmissions. After that, resource mapping is applied to associate the transmitted signal with a set of non-orthogonal physical resources assigned to multiple users. The final step is to apply the IFFT operation for the OFDM-based waveform.

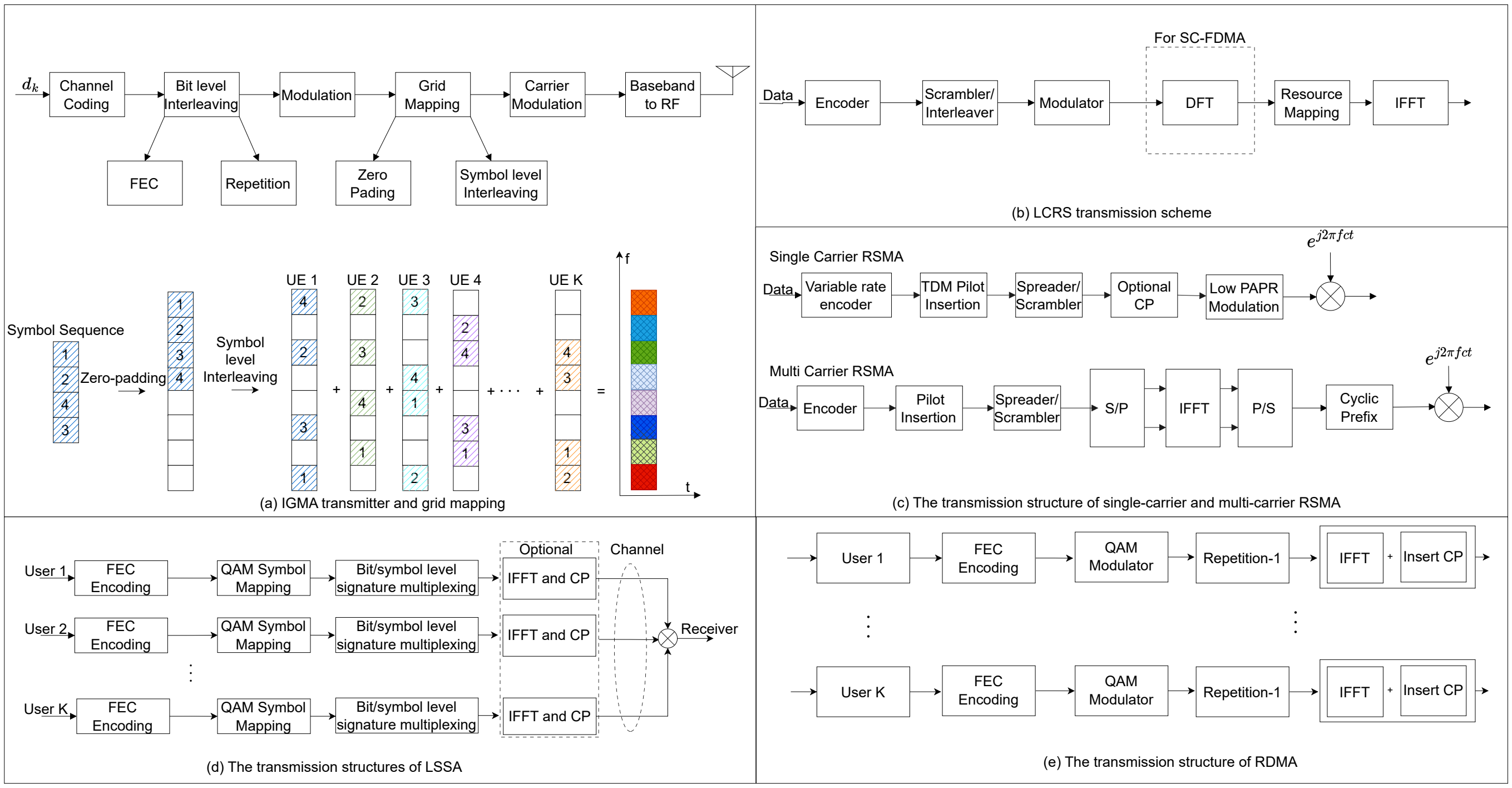

Fig. 9: (a) IGMA transmitter and grid mapping; (b) LCRS transmission scheme; (c) The transmission structure of single-carrier and multi-carrier RSMA. (d) The transmission structure of LSSA. (e) The transmission structure of RDMA.

*14) Resource Spread Multiple Access (RSMA):* The non-orthogonal transmission is achieved through RSMA, which involves spreading each user's signal across the whole frequency or time resource allotted to the group. Fig. 9(c) depicts the transmission structure of single and multiple-carrier RSMA [143]. Different users' signals within the group are not necessarily orthogonal to one another, resulting in the possibility of inter-user interference. The spreading of bits over all available resources facilitates the process of decoding signals at the signal level lower than the background noise. The RSMA technique employs a mix of low-rate channel codes and scrambling codes to effectively separate the signals of different users while maintaining good correlation properties. Depending on the application scenarios, it can include (i) Single carrier RSMA: optimized for battery power consumption and coverage extension for small data transactions by utilization single carrier waveforms, very low PAPR modulations. It supports grant-less transmission and potentially allows asynchronous access. (ii) Multi-carrier RSMA: optimized for low latency access for RRC-connected state users (i.e., timing with eNB already acquired) and supporting grant-less transmissions.

*15) Low Code Rate and Signature-Based Shared Access (LSSA):* LSSA was proposed to support uplink asynchronous massive connectivity [144]. A typical LSSA transmission structure is illustrated in Fig. 9(d). By multiplexing user-specific signature patterns at the bit level, LSSA randomizes the MUI between users. These signature patterns are typically unknown to other users. Each user's data bits in LSSA are encoded using a very low code-rate channel coding technique, which helps to reduce the impact of the MUI. Using a higher-rate FEC code in conjunction with spreading can also be used to replace the low-rate FEC code. Following channel encoding, a user-specific signature bit-level multiplexing scheme would be implemented. The user-specific signature might have some connection to the reference signal, the complex or binary sequence, or the permutation pattern of a short-length vector. The number of simultaneous transmissions can be affected by a number of factors, one of which is the length of orthogonal spreading codes. The receiver in LSSA, fortunately, does not rely on orthogonal multiplexing codes in order to differentiate between the signals transmitted by the target users. Instead, the interference cancellation is taken advantage of in order to provide sufficient support for high user overloading. The signature of LSSA can either be chosen at random by the user or assigned to the user by the gNB/BS. Additionally, LSSA can be optionally adjusted to include a multi-carrier variant in order to take advantage of the frequency diversity provided by larger bandwidth and to accomplish lower latency. This can be done in order to harness the potential for lower latency [122].

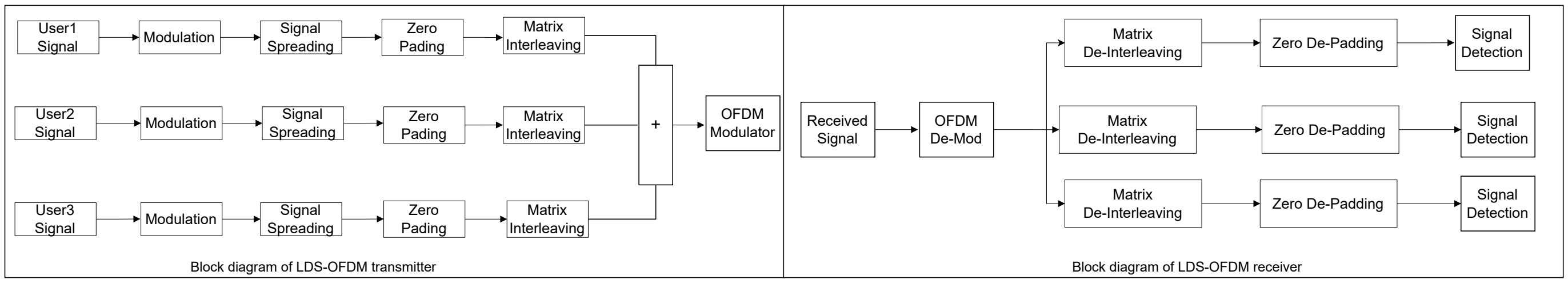

Fig. 10: The block diagram of transmitter and receiver of LDS-OFDM.

*16) Repetition Division Multiple Access (RDMA):* RDMA can be classified as a NOMA scheme utilizing interleave techniques. A typical RDMA transmission structure is illustrated in Fig. 9(e). In contrast to IDMA, RDMA employs symbol-level interleaving using cyclic-shift repetitions. In the context of RDMA, the modulation symbol vector of each user is transmitted multiple times in a repetitive manner, with different cyclic shift indexes assigned to each repetition. Furthermore, users display varying repetition and cyclic-shift patterns, which allow for the complete randomization of Multiple User Interference (MUI) and the achievement of both time and frequency diversities. RDMA has lower system complexity and potentially reduces signalling overhead compared to IDMA and RSMA because it does not require user-specific scrambling and interleaving patterns. In addition, RDMA utilizes a SIC receiver, which effectively manages the trade-off between receiver complexity and detection performance [122].

*17) Low Density Signature OFDM (LDS-OFDM):* LDS-OFDM integrates the concepts of LDS-CDMA with OFDM to efficiently manage situations involving multiple carriers [145]. Fig. 10 captures the block diagram of the LDS-OFDM. Sparse coding sequences are used by the transmitter in this technique, and MPA detection is used by the receiver. There are two processes involved in mapping the data stream in this technique. The data bits are first distributed using methods of low-density spreading. After that, an OFDM modulator is used to transmit each bit of the data stream across a separate subchannel [137]. In [146], authors proposed resource allocation for LDS-OFDM users in underlay cognitive radio networks based on the interference limit with the fairness metric to increase spectrum utilization. The use of the fairness metric was intended to increase the fairness of allocating subcarriers to each user. Furthermore, in [147] a serial schedule was developed for the iterative MUD in LDS-OFDM networks. In the proposed serial schedule, the updated message can be assimilated immediately in the current iteration, which improves the convergence rate. Numeric results show that, by choosing the proper numbers of iterations for serial scheduled MUD and LDPC decoding, it is possible to attain a satisfactory performance with affordable receiver complexity. In [148], a two-stage active user detection scheme, namely the initial active user detection stage and the false alarm correction stage, respectively, is proposed for a grant-free LDS-OFDM system. Then, a centralized and periodic channel estimation mechanism was invoked into the proposed grant-free LDS-OFDM network to obtain sufficiently accurate CSI with marginal overhead.

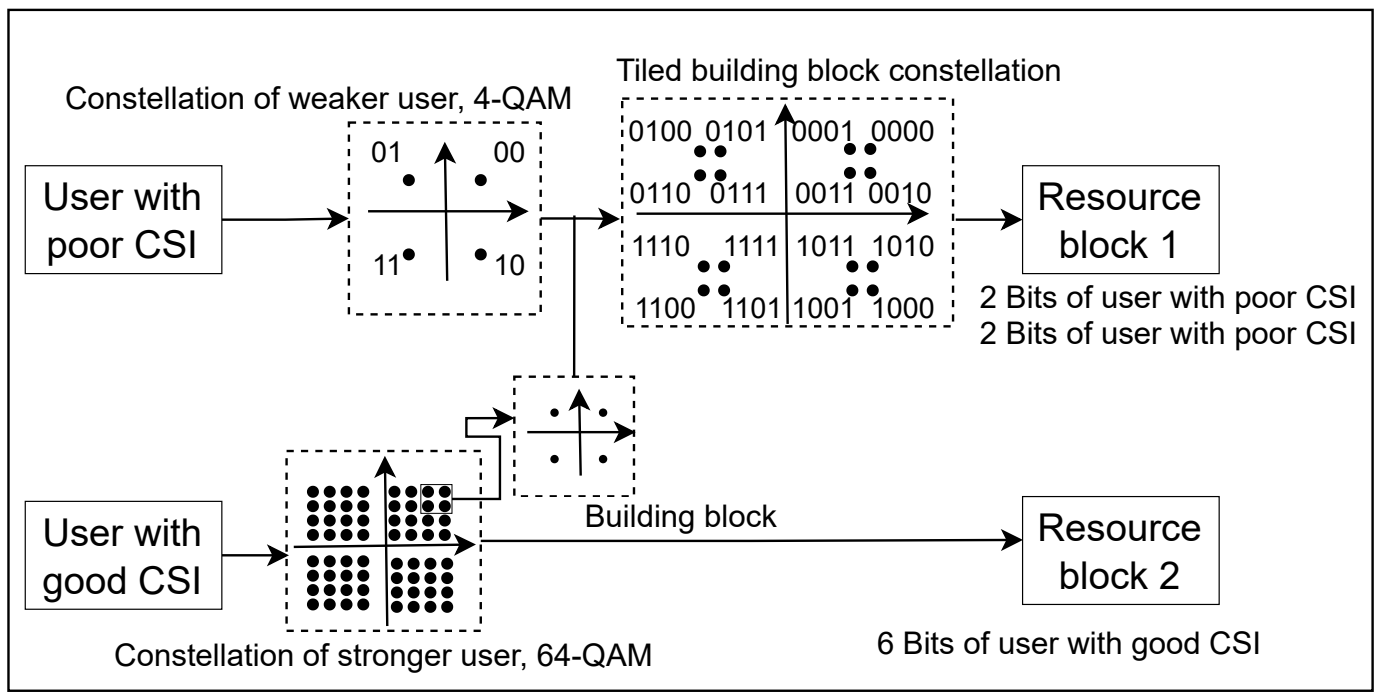

Fig. 11: Building block sparse-constellation based orthogonal multiple access (BOMA).

*18) Building Block Sparse-Constellation Based Orthogonal Multiple Access (BOMA):* BOMA combines the data from a user with high CSI with the symbols of a user with low CSI. As a result, the capacity of a multi-user system grows significantly. Fig. 11 showcases the transmission structure of BOMA. In order for a user with poor CSI to achieve similar BER performance as a user with good CSI, it is recommended that they use a coarse constellation with a larger minimum distance. Small building blocks can, therefore, be used to arrange the data of a user with good CSI in the constellation of a user with weak CSI. In the case of a user with poor CSI, the central point of the building block can be considered as the constellation point, whereas the tiled building block can be viewed as interference. The impact on detection performance is minimal when the building block size is significantly smaller than the minimum distance of the coarse constellation. With a good understanding of CSI, one may quickly locate the points in their own constellation and effectively decode the data. This skill includes detecting all points within the tiled building block constellation [119]. In [149], the BOMA technique was proposed and applied to the downlink of the wireless network. Also, several throughput enhancement methods were proposed. In [150], it is shown that both NOMA and BOMA have the same transmitted signal format that emerges from the building block approach for designing multilevel codes.

*19) Interleave Division Multiple Access (IDMA):* The key characteristic of IDMA is its use of interleaving to differentiate signals from various users. Because of this one-of-a-kind technique, it has been categorized as an interleave-division multiple-access system. The Fig. 12 presents the transmission structure of the IDMA scheme. IDMA shares similar properties with CDMA systems, particularly in its capacity to minimize fading and reduce the interference caused by other users in neighbouring cells. Furthermore, it offers a simple iterative chip-by-chip MUD approach. Regardless of the number of users, the cost per user remains constant in this system. Furthermore, the utilization of interleaving for user separation allows for the complete allocation of bandwidth to coding. This enables an information-theoretic interpretation of the coding-spreading saga of IDMA, leading to great spectral efficiency in the end. However, in IDMA a substantial amount of observations are required to reach optimality. The observation could arise from a multitude of spreading chips or a large frame size [88]. In [151], the authors proposed a novel architecture for the IDMA receiver with low latency while maintaining low complexity. The proposed architecture can perform multi-user detection directly without deinterleaving the received frame in the interference canceller iteration resulting in the decrease of latency by almost half. Additionally, in [152], authors demonstrated that IDMA can mitigate interference among users to a maximum extent and provides high data rates without compromising the required quality of service. Their experimental results show that the BER performance of IDMA was reduced.

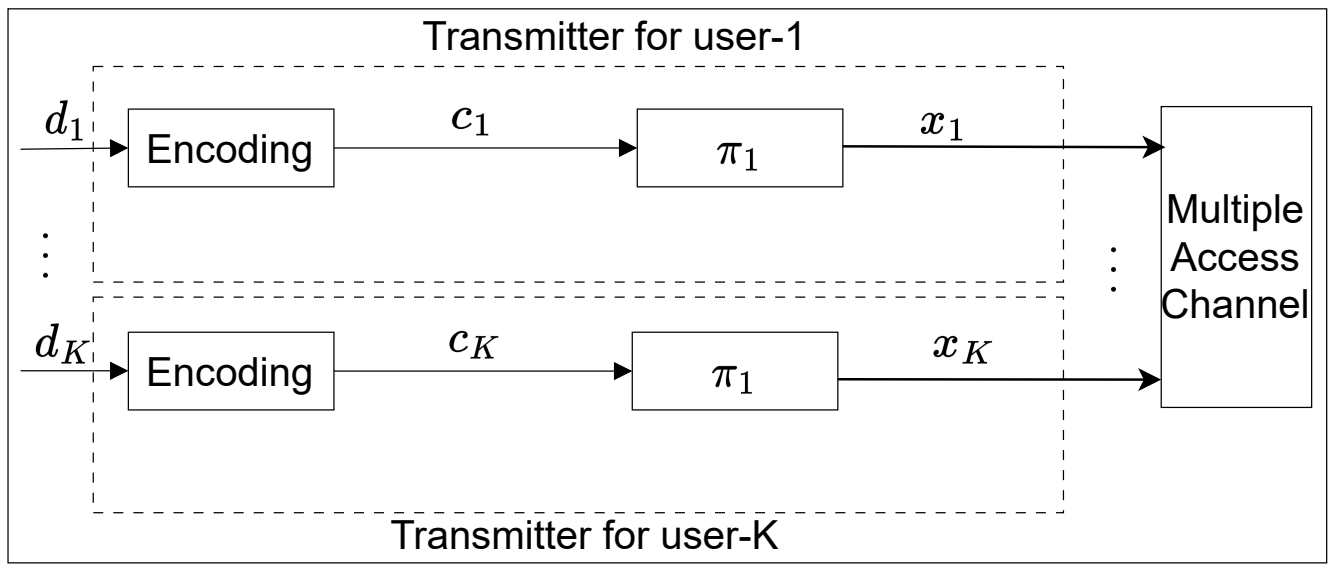

Fig. 12: The transmission structure of IDMA with K multiplexed users.

## III. Key Enabling Technologies

Tomorrow's 6G network is all set to provide a vast array of pioneering technologies, use cases, and applications that will demand uninterrupted connectivity, extremely low latency, and high data rates. NOMA has been widely regarded as a fine option for 6G due to its exceptional spectrum efficiency. This section strives to show off the enabling technologies used in NOMA networks to ensure massive connectivity and support energy-efficient communications. Fig 13 showcases the applications of NOMA in wireless communications technologies, and extensive literature is conducted to explore the performance of NOMA with wireless communications technologies. Subsection III-A presents multi-antenna techniques for NOMA; Subsection III-B explores the artificial intelligence-based optimization techniques in NOMA; Subsection III-C examines the NOMA for ultra-reliable and low-latency communications; Subsection III-D focuses on the utilization of terahertz frequency bands in NOMA framework; Subsection III-E depicts the importance of cooperative communications in NOMA; Subsection III-F illustrates the flexibility of NOMA in backscatter communications; Subsection III-G examines the role of reconfigurable intelligent surfaces to boost the connectivity and coverage in NOMA-based networks; Subsection III-H presents the NOMA-inspired cognitive radio networks; Subsection III-I represents the interplay between MEC and NOMA; Subsection III-J depicts the NOMA for integrated sensing and communications; Subsection III-K depicts the interplay of NOMA and visible light communications followed by Subsection III-L which presents NOMA in non-terrestrial networks.

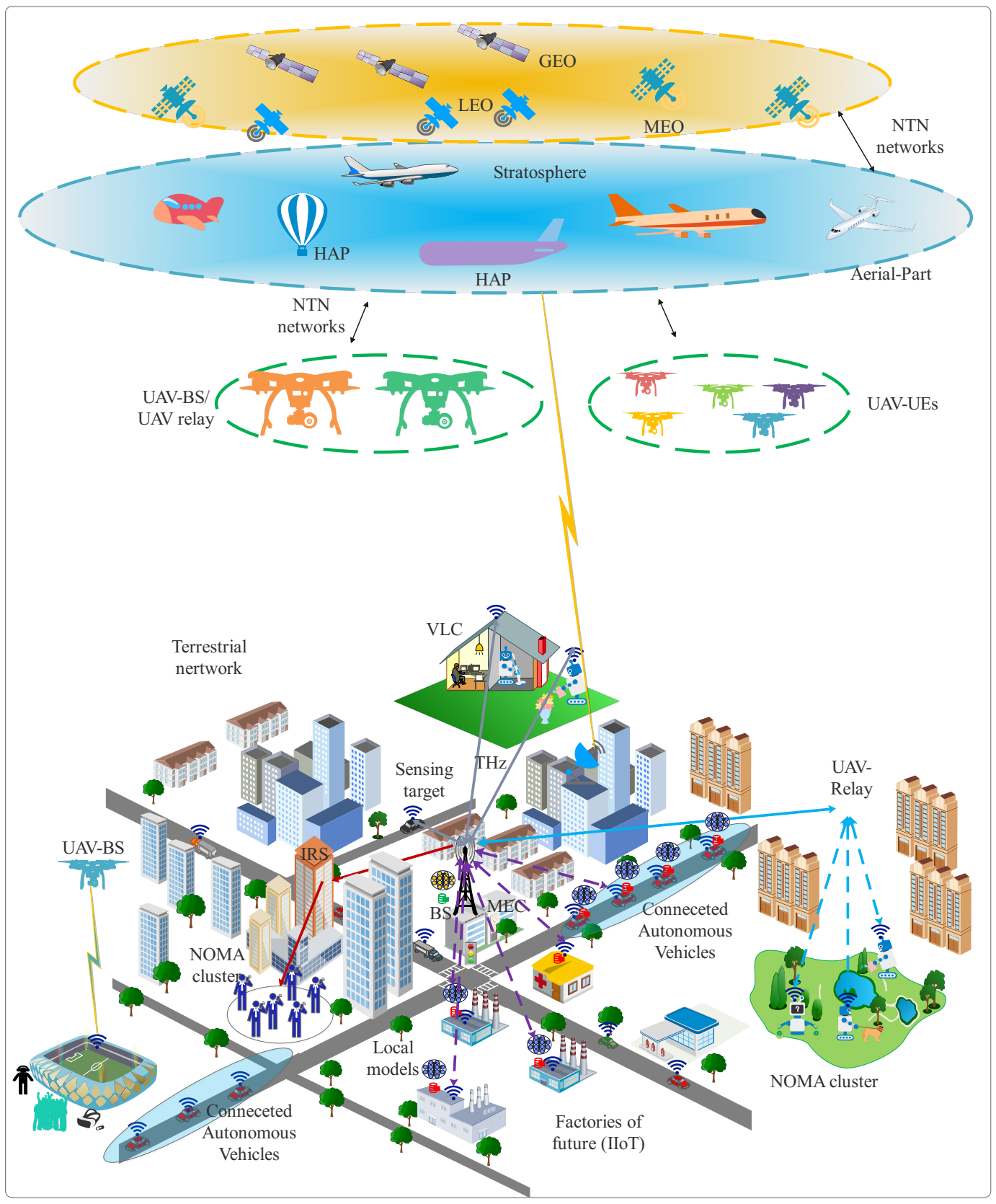

Fig. 13: Interplay of NOMA with other wireless communications technologies.

### A. Multi-Antenna Techniques and Architectures

Future communications systems can eventually be advanced to provide high data rates to wireless sensors and the IoT by implementing NOMA, small cells, and heterogeneous networks (HetNets), as well as Massive multiple-input-multiple-output (mMIMO) [153], [154]. Multi-antenna techniques are envisioned to be an indispensable component of NGMA. Equipping the BS and users with multiple antennas creates additional degrees of freedom (DoFs) in the spatial domain to enhance NOMA performance compared to the case with single-antenna BSs and users [27], [155], [156]. Gains in single-user contexts can be realized through beamforming or spatial multiplexing by employing multiple antennas at the transmitter and receiver. In the context of multiple users, multiple antennas can be deployed to segment the users within the space domain, forming the space division multiple access SDMA [157]. The below sections capture several multi-antenna architectures [27], and Table V shows the analysis of top-notch research works based on the multi-antenna architectures for NOMA.

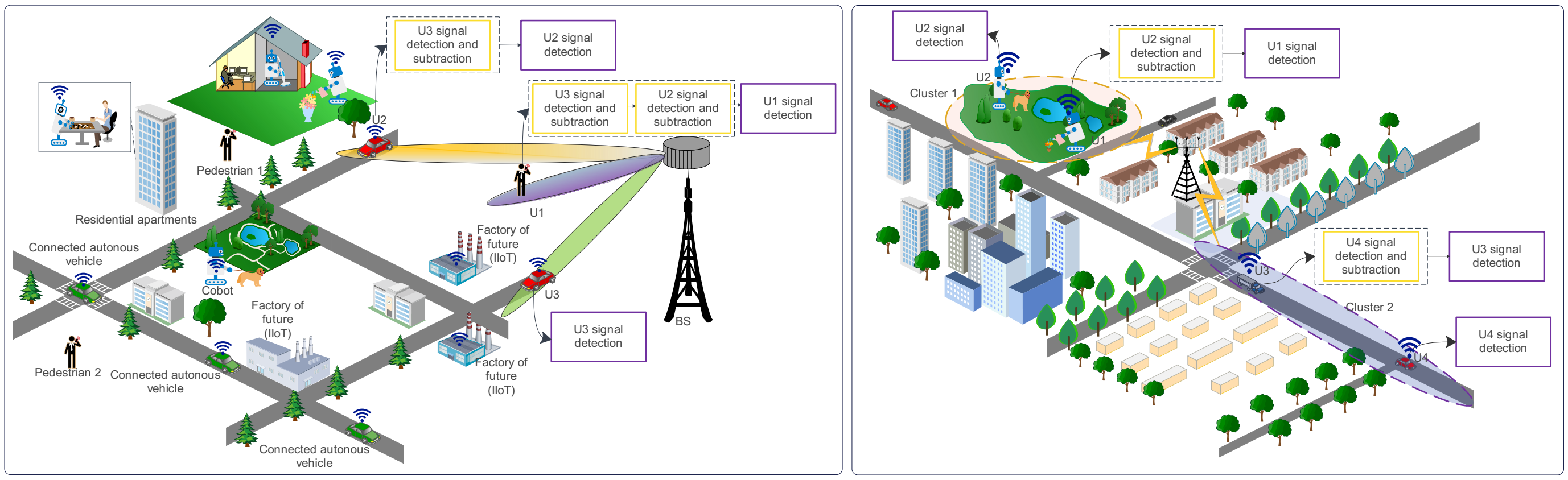

(a) (b)

Fig. 14: (a) Beamforming-based mMIMO-NOMA. (b) Cluster-based MIMO-NOMA.

*1) Beamformer-Based MIMO-NOMA:* Fig. 14 shows the beamforming-based transmission scheme of mMIMO-NOMA. Similar to SDMA, a simple transmit/receive beamforming strategy in MIMO-NOMA is to develop a linear beamformer for each user. However, new constraints need to be added to guarantee the efficacy of SIC. The transmit/receive beamformer at the BS and/or users is a unique ingredient of MIMO-NOMA design compared to single-antenna-based NOMA (i.e., SISO-NOMA), in which the BS and users each have a single antenna [158]. To begin with, the beamformer design, like in the case of MIMO-OMA, notably determines the signal power and the interference power at the various users and thus plays a vital role in the user's SINR [159]. But the SIC performance in MIMO-NOMA is extremely sensitive to the decoding order of the users' data. Therefore, building the beamformer and designing the decoder in conjunction is important for optimal outcomes [160].

We noted that extensive research has been carried out to optimize the performance of beamforming-based MIMO-NOMA networks [161]–[165]. For instance, a three-step framework to solve the multi-dimensional resource allocation problem for beamformer-based MIMO-NOMA was discussed in [161]. Step 1 involves the proposal of a beam-forming method to determine the ideal beam vector for a specific user cluster. In particular, intra-cluster power allocation is carried out using fractional transmitting power control (FTPC). To get a local optimal with less complexity, the limited-memory Broyden–Fletcher–Goldfarb–Shanno (L-BFGS) method was applied. Step 2 takes into account user clustering on the basis of the proposed beam-forming algorithm. In the third step, beamforming and user clustering were used to further optimize power allocation. The simulation results demonstrated the performance of the proposed research. Additionally, in [162], authors developed novel coordinated beamforming techniques to enhance the performance of the MIMO-NOMA networks in the presence of inter-cell interference. The proposed scheme efficiently improved the cell-edge user's throughput and also successfully dealt with the inter-cell interference.

A cluster beamforming strategy to jointly optimize the beamforming vectors and power allocation coefficients for the mobile users in the NOMA-MIMO network to reduce the power consumption was proposed in [163]. In the aforementioned work, the authors opted for the improved coalition game approach to effectively optimize the multi-user MIMO-NOMA clusters. Similarly, in [164], authors evaluated the performance of the network with different linear beamforming techniques. To enhance the network's sum spectral efficiency, a heuristic pairing strategy with minimal complexity was proposed. In [165], the authors investigated the applications of beamforming in the two-user downlink MISO-NOMA network under imperfect SIC. The joint optimization of the power allocation coefficients and the weights of the arrays resulted in a non-convex problem which was solved by using the Sequential Quadratic Programming (SQP) algorithm.

*2) Cluster Based MIMO-NOMA:* In beamformer-based MIMO-NOMA, each user's message is interfered with by all the other users' messages in the network. As a result, optimizing the decoding order and the beamformer simultaneously implies exponential complexity concerning the number of users, which is notably prohibitive in overloaded scenarios. To overcome this limitation, authors in [166]–[169] developed the cluster-based MIMO-NOMA schemes, in which users are divided into various clusters and share the beamformer. The advantages of cluster-based MIMO-NOMA communications architecture are as follows: On the one hand, users with identical spatial features can be clustered together, allowing inter-cluster interference to be considerably reduced or even eliminated by carefully forming beamformers for the clusters. As a result, the limited spatial DoFs can be leveraged more efficiently, which is pivotal for the overloaded scenario. In addition, as SIC only needs to be performed for users within the same cluster, the corresponding decoding order across intra-cluster users can be determined more effectively due to the smaller number of users that need to be considered. In

TABLE V: STATE-OF-THE-ART RESEARCH WORK ON MULTI-ANTENNA TECHNIQUES AND ARCHITECTURES. (All APs, SBS, and MBS are mentioned as BS. IoT devices and UEs are mentioned as users)

| Ref. | System Model | Design Objective | Optimization Techniques | Main Findings |
|---|---|---|---|---|
| *Beamformer-Based MIMO-NOMA* | | | | |
| [161] | One BS with $j$ antennas and $k$ users | Spectrum efficiency | Successive convex optimization | Large clusters can serve more users that results in a gain in spectrum efficiency. |
| [162] | $m$ BS and $k$ users (DL) | Interference | Numerical | Interference alignment based coordinated beamforming methods are proposed. |
| [163] | One BS and $k$ users (DL) | Power allocation | Game theory | The proposed scheme is superior to benchmarks for finding power efficient clusters. |
| [164] | One BS and $k$ users (DL) | Spectral efficiency | Successive optimization approach | NOMA-based network with zero-forcing beamforming gives highest spectral efficiency. |
| [165] | One BS and two users (DL) | Rate fairness | Sequential Quadratic Programming | Proposed algorithm yields improved overall rates as compared to far-field formulation-based modelling and beamforming TDMA solutions. |
| *Cluster Based MIMO-NOMA* | | | | |
| [166] | One BS with $j$ antennas and $k$ users (UL) | Throughput&latency | Numerical | Inter- and intracluster interference plays a substantial role for the performance of cluster-based uplink concurrent transmissions in MIMO-NOMA networks. |
| [167] | One BS with $j$ antennas and $k$ users | Sum secrecy rate | Alternating optimization algorithm | Deploying IRS can bring significant beamforming gains to suppress the eavesdropping. |
| [168] | $m$ BS and $k$ users (UL) | Throughtput | Discrete-Time Markov chains | Authors investigated the system and cluster level performance under diverse traffic conditions. |
| [169] | $m$ BS and $k$ users | Sum secrecy throughput | Approximation algorithm | Proposed algorithm for hybrid-NOMA outperformed the benchmark schemes. |
| *Massive MIMO-NOMA* | | | | |
| [170] | One BS with $j$ antennas and two users | Energy efficiency | Lagrange dual decomposition | Proposed scheme ensured significant gain in energy efficiency. |
| [171] | One BS and $k$ users (DL) | User ordering | Analytical | The proposed scheme is evaluated with perfect user ordering and one-bit feedback. |
| [172] | One BS and $k$ users (DL) | Energy efficiency | low-complexity alternating approach | Proposed scheme optimizes EE with low complexity. |
| [173] | mmWave MIMO-NOMA (DL) | Power allocation | Iterative algorithm | It is shown that only 10 times of iteration are required to achieve convergence. |
| [174] | $m$ BS and $k$ users (Distributed mMIMO) | Sum rate | Graph Theory | Graph-based approach was proposed to optimize the distributed mmWave massive MIMO-NOMA. |
| *Cell-Free mMIMO-NOMA* | | | | |
| [175] | $m$ APs and $k$ users | Spectral efficiency | ML based K means++ algorithms | Proposed scheme ensured improved spectral efficiency. |
| [176] | $m$ APs,$l$ clusters,$k$ users | Power allocation | Iterative algorithm | Large number of users per cluster increases the gain in CF-mMIMO-NOMA. |
| [177] | $m$ APs and $k$ users (SWIPT-CF-mMIMO-NOMA) | Sum rate | Multi-agent deep Q-network | CF-mMIMO and NOMA can enhance the propagation of SWIPT and the proposed ML-based technique can facilitate resource allocation. |
| [178] | $m$ BS and $k$ users (DL) | Power allocation | Deep reinforcement learning | Optimization of user pairing, power allocation, and phase shifts of IRS plays a critical role in enhancing the downlink rate. |
| [179] | One satellite, $m$ APs and $k$ users (DL-SGIN) | Spectral efficiency | SCA & SDP | The conceived SIC order design outperforms the fixed-order design at the same complexity. |

[166], authors investigated the impact of inter and intra-cluster interference on the performance of the uplink concurrent transmissions in the MIMO-NOMA network. To assess the performance, the authors considered the cluster throughput and transmission latency with various network configurations. In [167], an IRS-aided multi-cluster MIMO-NOMA network was examined. The users were divided into clusters according to their channel correlations, and each cluster was provided with a beam through NOMA to enhance the overall secrecy rate. A non-conex problem was successfully transformed into a convex one using the Taylor series expansions and SDP. Subsequently, an alternating algorithm was devised to obtain a feasible solution.

Two random access schemes which enabled intra- and inter-cluster concurrent transmissions for uplink IoT traffic with and without access control were discussed in [168]. To evaluate the performance, the authors created two analytical models using DTMCs. The proposed work thoroughly examined the performance through thorough simulations and considered the system's cluster-level performance in terms of throughput and delay. A sum-secrecy throughput maximization problem based on the mobile user clustering in the downlink HetNet-NOMA network was solved in the [169]. The formulated MINLP problem was tackled using the low-complexity approximation algorithm. The proposed algorithm was compared and evaluated for the HetNets and high-power base station-only networks.

*3) Massive MIMO-NOMA:* Massive MIMO is a vital component for 6G communications because of its unique characteristics, such as the massive number of BS antennas to support a more considerable number of widely dispersed IoT devices and users with massive connectivity and continuous data transfer [180], [181]. Furthermore, NOMA is a possible candidate MA for enhancing future communications systems' spectrum efficiency, energy efficiency, and ultra-low latency. The use of multiple antennas at the transmitter, receiver, or both can considerably increase data rates and has been a crucial component of cellular systems since 4G [182]. mMIMO systems use large-scale antenna arrays at the BS, where the number of transmit antennas dramatically outnumbers the number of users [183].

Extensive research has been carried out to investigate the performance of the mMIMO-NOMA networks as shown in the table V. An energy-efficient resource allocation problem in the mMIMO-NOMA network was solved in [170]. The energy efficiency maximization problem was formulated to effectively determine the optimal resource allocation problem. An advanced method of non-linear fraction programming was utilized to transform the given problem into a convex optimization problem. This problem was then successfully solved using the Lagrange dual decomposition approach. In [171], a low-feedback mMIMO-NOMA scheme was proposed which can decompose the mMIMO-NOMA network into multiple separated SISO-NOMA channels and the analytical results were derived in the perfect user ordering and with one-bit feedback scenarios. Additionally, global energy efficiency (GEE) was optimized in the [172]. The formulated problem of maximizing GEE was a non-convex fractional programming problem. A low-complexity iterative algorithm based on the alternating minorization maximization to solve the non-convex GEE problem. The authors showed that the proposed work has a quicker convergence to a stationary point. The optimization of power allocation and power splitting in the SWIPT-enabled mmWave mIMO-NOMA was proposed in [173] to maximize the achievable sum rate, and the alternating algorithm was developed to solve the non-convex problem. The performance of the developed algorithm was assessed through extensive simulation, which showed that the proposed SWIPT-enabled mmWave mMIMO-NOMA network can achieve higher spectrum and energy efficiency than its OMA counterpart. Sim-ilarly, in [174], graph-theoretic approach was adopted to solve the sum rate maximization problem for the distributed mMIMO-NOMA network, The formulated problem was decoupled into user-access point association sub-problem and pilot resource allocation sub-problem. The first sub-problem of optimal user-AP clustering is solved using a bipartite graph matching approach with a weighting matrix. The second sub-problem of optimal pilot resource allocation is addressed using a graph-theoretic vertex colouring technique. The numerical results indicated that the proposed framework for maximizing the sum rate maintains a considerable portion of optimal performance with feasible computational complexity.

*4) Cell-Free mMIMO-NOMA:* As an alternative to small cells, CF-mMIMO-NOMA system models have distributed APs spread across a region to support the massive number of users [175]. The APs are linked to a central processing unit (CPU) responsible for most of the baseband processing. The integration of NOMA means that distributed antennas are used to establish clusters of users that can be cooperatively served by PD-NOMA in the same orthogonal resource, much like in other MIMO-NOMA scenarios. This allows for more efficient use of the available resources. Therefore, compared to CF-mMIMO-OMA, significant design complexity arises due to the user pairing problem. To tackle this, the appropriate sets of users need to be identified to form a cluster, which leads to the challenging issue of user clustering. On the other hand, the implementation of NOMA results in improvements to a CF-mMIMO system's spectral efficiency in both the DL and UL scenarios [184].

Extensive research has been carried out to investigate and optimize the performance of NOMA-enabled CF-mMIMO. For instance, in [176], a computing model was proposed for CF-mMIMO-NOMA, which utilized two-channel estimation approaches, including individual and linear channel combinations. In addition, a user pairing algorithm was proposed to integrate user power allocation into an optimization problem for maximizing the downlink rate per user. This problem was solved using the inner approximation method. In [177], authors focused on maximizing the sum-information rate maximization design for SWIPT-enabled CF-mMIMO-NOMA networks. The research derived a closed-form SINR expression by utilizing conjugate beamforming. A ML-based solution framework was developed to solve the formulated non-convex and mixed combinatorial problem. In [178], authors proposed a novel approach to optimize the max-min downlink rate of users in an IRS-assisted CF-mMIMO-NOMA network. The optimization problem was solved by using DDPG, and results indicated that incorporating NOMA into IRS and CF-mMIMO yields the best performance in terms of downlink rate per user. Also, in [179], a sum rate maximization problem was formulated for the CF-mMIMO-NOMA-assisted SAGINs. The optimization problem was jointly optimized with the power allocation factors of NOMA DL, uplink power transfer, and both the beamformer of the satellite and of the AP. The formulated non-convex problem was converted into a convex one using SDP and SCA. The simulation results revealed that the proposed system outperforms the FDD and small cell systems in terms of the SE.

### B. Artificial Intelligence

Artificial intelligence (AI) has been acknowledged as the foremost promising technology among the roster of 6G enabling technologies. Many optimization tasks become intractable as mobile networks get more complex and varied, paving the way for cutting-edge machine learning (ML) methods. With the continuous growth of mobile data and advancements in computing hardware, along with development in learning techniques, it is anticipated that intelligent AI-driven approaches will play a crucial role in addressing various challenges in the future 6G networks. These challenges encompass modulation, classification, waveform detection, signal processing, resource allocation, and physical layer design. In recent years, there has been a notable increase in the application of advanced and resilient AI techniques to address a wide range of challenges encountered in wireless networks. For example, the authors in [208] presented state-of-the-art techniques for improving security and privacy in wireless networks. In addition, work in [209] showcased the multi-agent reinforcement learning techniques for mobile edge computing, cell-free massive MIMO, and unmanned aerial vehicles. We noted that researchers are actively designing AI and ML-driven optimization techniques for the NGWN and NGMA. For instance, authors in [97] proposed an innovative cluster-free NOMA framework to achieve the vision of NGMA. Deep learning (DL), reinforcement learning (RL), deep reinforcement learning (DRL), federated learning (FL), and transfer learning (TL) belong to the family of ML. These techniques can be adapted to optimize user pairing [185], resource allocation [195], power allocation [210], and signal detection [211] in a variety of scenarios such as mMIMO-NOMA [186], RIS-NOMA [212], NOMA-IoT [205], UAV-NOMA [213], MEC-NOMA [201] and more. Due to the increasing complexity of next-generation wireless systems, the communication design typically needs to solve challenging and high-complexity problems. To highlight the role of efficient AI-driven techniques for NOMA, we presented a summary of state-of-the-art literature in Table VI, which covers the literature on the significance of applying deep learning [185]–[189], reinforcement learning [190]–[194], deep reinforcement learning [195]–[200] and federated learning [201]–[207] to optimize the performance of NOMA empowered NGWNs.

*1) Deep Learning:* Deep learning (DL), a subfield of machine learning, operates on the principle of acquiring knowledge from data to learn patterns and correlations. DL solutions have been crafted to leverage the insights acquired from the learning process to produce optimum case-specific solutions. When faced with real-world problems, it is common to rely on pre-established heuristics or rules of thumb to find solutions. Such heuristics may not always be optimal or accurate, especially in complex or dynamic environments where the fundamental patterns and relationships are not thoroughly comprehended. On the contrary, DL solutions can acquire knowledge from vast datasets, enabling them to identify intricate patterns and correlations that may elude even the most seasoned human specialists. Subsequently, these solutions may be employed to produce case-specific solutions

TABLE VI: STATE-OF-THE-ART RESEARCH WORK ON ARTIFICIAL INTELLIGENCE-DRIVEN NOMA FOR 6G.
(All APs, SBS, and MBS are mentioned as BS. IoT devices and UEs are mentioned as users.)

| ML Categories | Ref. | System Model | Design Objective | Main Findings | Applications of ML techniques in NGWNs |
|---|---|---|---|---|---|
| Deep Learning | [185] | $m$ BSs and $k$ users (DL) | Energy efficiency | Subchannel allocation and power control in NOMA-HetNets were performed with DNN. | (i). Coverage and capacity, (ii). CSI, (iii). URLLC, (iv). Interference management |
| | [186] | One multi-antenna BS and $k$ users (DL) | Energy efficiency | The proposed scheme achieves 14% improvement in energy efficiency. | |
| | [187] | $m$ BS and $k$ users (DL) | Power allocation | Proposed algorithms are evaluated for sum rate, outage rate, and computational time. | |
| | [188] | One BS and $k$ users (DL) | Min. transmit power | DL based framework is proposed to solve complex power minimization problem. | |
| | [189] | One BS and $k$ users (UL) | Reliability | The proposed scheme ensures gain in the reliability as compared to benchmarks. | |
| Reinforcement Learning | [190] | One BS and $k$ users (DL) | Sum rate | The proposed algorithm achieves better performance over the benchmark scheme. | (i). Resource allocation, (ii). Cooperative relay communications |
| | [191] | $m$ BS and $k$ users (IIoT) | Throughput and SE | The proposed scheme accommodates more users in the system while meeting their QoS. | |
| | [192] | One BS and $k$ users (DL) | Max. sum rate | Q-learning algorithm with user clustering achieves a maximum sum rate than benchmark schemes. | |
| | [193] | Underlay CR-NOMA | Energy efficiency | This work optimizes the power allocation to maximize the energy efficiency. | |
| | [194] | Cache-aided NOMA-MEC | Resource allocation | In RL algorithms, the definition of reward function has a significant influence on performance. | |
| Deep Reinforcement Learning | [195] | $m$ single antennas BSs and $k$ users (UL) | Sum rate | DRL with sparse activations is suitable for heavy network traffic. | (i).Autonomous network management, (ii). end-to-end multiuser channel prediction |
| | [196] | One BS and $k$ users (UL) | Throughput | Proposed scheme maximizes user throughput by optimizing transmit power. | |
| | [197] | $m$ BSs, $k$ users, MIMO-NOMA (DL) | Energy efficiency | Proposed network offered reliable and ultra-high speed connectivity for fast users. | |
| | [198] | UAV-supported NOMA-IoT | Throughput | The proposed algorithm outperforms the benchmarks in throughput and energy saving. | |
| | [199] | Multi-band network (hybrid NOMA) | Resource allocation | The proposed work highlights the benefits of hybrid NOMA in multi-band systems. | |
| | [200] | UAV-BS, $k$ users (NOMA-maritime IoT) | Resource allocation | The proposed algorithm can reduce the whole maritime-IoT network's energy consumption. | |
| Federated Learning | [201] | One BS and $k$ users | Communication efficiency | The proposed FL-NOMA scheme achieved higher FL testing accuracy. | (I). Network-wide learning and optimization, (ii) Privacy, (iii). Distributed learning on edge |
| | [202] | One BS and $k$ users (UL) | Min. system cost | The proposed NOMA assisted FL scheme outperformed FDMA assisted FL by 78%. | |
| | [203] | One BS and $k$ users (UL) | Computational fairness | This paper provided tradeoff between convergence round and fairness. | |
| | [204] | $m$ UAV-BS and $k$ users | Resource allocation | AI techniques are developed for UAV deployment and joint resource allocation. | |
| | [205] | One BS and $k$ users (UL) | Min. energy consumption | This near-optimal scheme uses 6, 4, and 2 times lower energy than benchmarks. | |
| | [206] | One BS,$k$ users,$O$ STAR-RIS elements (UL) | Interference mitigation | The proposed algorithms outperformed benchmarks in training loss and test accuracy. | |
| | [207] | $m$ BS and $k$ IIoT users | Resource allocation | The goal of the proposed scheme is to realize the efficient and green HFL for IIoT. | |

for the particular task [214].

The realm of wireless communications has witnessed remarkable success through the implementation of DL techniques, mainly performed by convolutional neural networks (CNNs), artificial neural networks (ANNs), and deep neural networks (DNNs). These methods have yielded exceptional outcomes, thereby solidifying their efficacy [215]. The implementation of DL has demonstrated its efficacy as a valuable tool in strengthening the performance of wireless networks, streamlining resource allocation, channel estimation, signal detection, and security and privacy.

The advent of 6G-based applications, use cases, and fascinating technologies are on the horizon, and among the contenders for providing reliable and low latency communications, as well as enhanced spectrum and energy-efficient performance, is NOMA. This technology is particularly promising for facilitating massive connectivity. At the core of NOMA's concept lies the ability to manage multiple users sharing the same resource block. The prospect of employing DL-based techniques in NOMA-based wireless communications networks is exciting, as the use of DL makes it possible to realize the full potential of NOMA. For example, in [185], the authors introduced a deep learning-based algorithm to address the issue of energy-efficient user association, subchannel allocation, and power allocation in the context of NOMA networks. Furthermore, the authors in [185] have chosen to employ the Lagrange dual decomposition method as the basis for their user association technique. In addition, they have utilized DNN and semisupervised learning to tackle the challenging tasks of subchannel allocation and power control in the NOMA networks. The authors in [186] introduced an innovative approach that incorporated deep learning with the MIMO-NOMA system. DNN was used to optimize power allocation in the proposed method. To solve the power allocation problem, a communications deep neural network (CDNN) was introduced with separate activation functions at each layer. Innovative learning methods were employed to extract spatial features of MIMO-NOMA and train the proposed CDNN framework for offline and online learning. The advanced capabilities of deep learning in representation and mapping provided precise CSI and enhanced SIC performance for users in complicated MIMO systems. In addition, the CDNN's approximation ability helped optimize power allocation.

DNN-based technique for the joint user selection and power allocation in the downlink of the MISO-NOMA network was proposed in [187]. Using unsupervised learning and a unique training model based on SINR information, the proposed technique outperformed the benchmark schemes while reducing the computational complexity. Authors in [188] investigated the use of deep learning techniques in the context of the MC-NOMA network. Their objective was to minimize the overall transmit power of the network while simultaneously guaranteeing the QoS for the users. Furthermore, a sophisticated deep learning model, specifically a deep belief network (DBN), was utilized. The detailed procedure of the proposed approach can be divided into three distinct components: data preparation, training, and running. The simulation results show that the proposed DBN-based approach can achieve superior performance in power consumption than the exhaustive search method. In [189], an end-to-end neural network model based on the deep variational auto-encoder (VAE) was developed for the grant-free NOMA network. The proposed network consisted of a NOMA encoding network, which models the random user activation and symbol spreading, and a NOMA decoding network, which jointly estimates the user activity and the transmitted symbols. Their simulation results demonstrated that the proposed scheme outperforms conventional grant-free NOMA schemes in terms of reliability.

*2) Deep Reinforcement Learning:* The integration of DL and reinforcement learning (RL) has given rise to a subfield of ML known as deep reinforcement learning (DRL). Through interaction with an environment, an agent in RL learns to choose actions that result in greater rewards. The agent's ultimate objective is to optimize its cumulative reward over time. In DRL, DNN is used to approximate the optimal value function or policy, allowing the agent to learn more complex and abstract representations of the state and action spaces, which can help solve complex decision-making problems.

DRL-based NOMA has gained a lot of attention [192], [193], [216]. Authors in [190] considered the application of DRL based on a shallow neural network structure to solve the power allocation problem in the downlink of energy harvesting NOMA network. The authors demonstrated that using the optimal power allocation policy in the multiuser broadcast downlink channel increases the harvested energy, channel gains of all users, and remaining battery. In [191], authors put forward a hybrid model that combines OMA and NOMA. They utilized Q-learning and matching theory to effectively optimize the resources. In their work, the authors focused on minimizing inter-device interference by optimizing various factors such as the number of devices per sub-band, the power of devices in each sub-band, and the width of each sub-band. This approach ensured that each device could achieve its target data rate efficiently. The simulation results confirmed the effectiveness of their proposed scheme compared to other benchmark schemes. Additionally, their scheme can handle a larger number of devices in the system without compromising the quality of service.

A reinforcement learning-based power allocation algorithm for the NOMA network was developed in [192]. The proposed Q-learning algorithm supported the efficient power allocation to maximize the sum data rate. In [194], authors proposed a cache-aided NOMA-MEC framework. A joint long-term optimization problem, subject to caching and computational resource allocation constraints in the AP, is proposed to minimize the overall energy usage in cache-aided NOMA MEC networks. The defined MINL problem optimizes the three parameters. A SAQ-learning based resource allocation algorithm technique was proposed to make decisions about user task offloading and resource allocation for the AP. A Bayesian learning automata (BAL)-multi-agent Q-learning (MAQ)-learning algorithm was developed for the task offloading problem, wherein each agent adopts an action selection strategy based on BLAs to acquire the best possible action in each state. Simulation results demonstrated the performance of the proposed framework and algorithm.

DRL was also applied to investigate the resource allocation in the uplink NOMA. For instance, in [216], the authors proposed three models that utilize DRL technologies for resource allocation in uplink NOMA networks. These models include continuous DRL-enabled resource allocation (CDRA), discrete DRL-enabled resource allocation (DDRA), and combined DRL resource allocation (CDRLA). The three models were jointly trained to determine the optimal power and subchannel allocation policy. When compared to other benchmarks, these models offered improved resource allocation performance with less computation time. In [217], authors proposed a cooperative multi-agent DRL approach to acquire decentralized offloading policies. These policies were utilized to allocate task offloading and local execution powers to each user to minimize the long-term average network computation cost regarding power consumption and buffering delay. In addition, utilizing a centralized training and distributed execution approach, the proposed approach not only acquired proficient decentralized policies but also alleviated the user's computational load and exhibited satisfactory performance in managing the interference of NOMA-based networks. Besides aforementioned works, DRL is also being used in the NOMA-IoT [218], RIS-NOMA [219], NOMA-MIMO [220], and more.

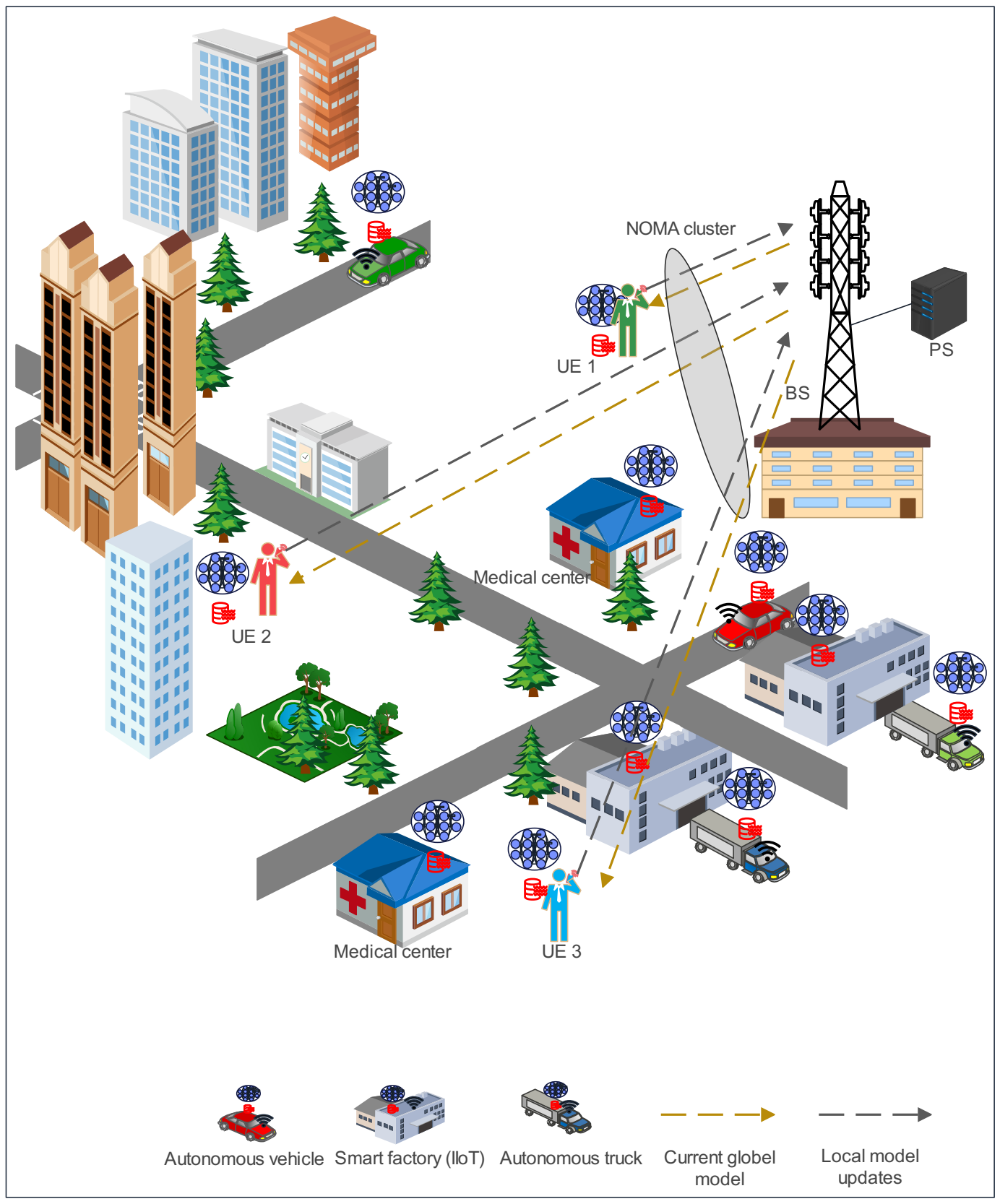

Fig. 15: NOMA-assisted FL in smart city.

*3) Federated Learning:* Conventional ML approaches typically demand centralized learning in a data centre or a cloud [221]. Due to privacy concerns and limited communication resources for data transmission, it is problematic for all users to transmit all of their collected data to a data center or the cloud. This, in turn, favours the emergence of distributed learning frameworks that enable devices to train a learning model locally utilizing individually collected data. One of the most potent distributed learning frameworks is federated learning (FL) [201]. FL envisages training of statistical models on remote devices or siloed data centers, such as mobile phones or hospitals, while keeping localized training data. FL enables training in diverse and ultimately massive networks, encouraging a radical break from conventional approaches for large-scale machine learning, distributed optimization, and privacy-preserving data analysis [202]. Besides, FL allows phenomenal flexibility in data collection and model training. For instance, a crowd of smart devices can actively sense and collect data throughout the day. They jointly provide feedback and update the global model at night to increase efficiency and accuracy for the next day [222]. In the context of a smart city, FL presents a promising solution for preserving the security and privacy of user's data. This is achieved by enabling devices to conduct local data training and then updating models to a BS that is equipped with a parameter server for model aggregation. Fig. 15 depicts the NOMA-assisted FL in the smart city.

TABLE VII: STATE-OF-THE-ART RESEARCH WORK ON ULTRA-RELIABLE AND LOW-LATENCY COMMUNICATIONs.
(All APs, SBS, and MBS are mentioned as BS. IoT devices and UEs are mentioned as users)

| Ref. | System Model | Design Objective | Optimization Techniques | Main Findings |
|---|---|---|---|---|
| [223] | One BS and $k$ users (UL) | Average latency | DQN | Proposed scheme improves mURLLC's low latency and high reliability. |
| [224] | One BS and two users (DL) | User fairness | Sub-optimal heuristics | Sub-optimal algorithm is proposed which converges to near-optimal solution. |
| [225] | One BS and two users | Min. power | BCD algorithm | The numerical results show that NOMA leads to lower power consumption. |
| [226] | One BS and $k$ users (UL) | Reliability | SARSA | This paper addresses user clustering, optimal RA, and instantaneous feedback system. |
| [227] | One BS and $k$ users (UL) | Reliability | Heuristics | This paper analyzes antenna deployment for reliability and latency. |
| [228] | One BS and $k$ users | Power allocation | Stochastic geometry | The simulation results indicate the effectiveness of the proposed model. |
| [229] | One BS and $k$ users | Resource allocation | Heuristics | Power control and channel blocklength allocation are optimized globally. |
| [230] | One BS, one relay and user | Energy efficiency | Iterative algorithm | Proposed a packet re-management framework for direct and relay links. |
| [231] | One source, relay and destination node | Min. power consumption | Sub-optimal heuristics | Authors prove that the algorithm converges to a Karush-Kuhn-Tucker point of the problem. |

Researchers from academia and industry are also focusing on NOMA-assisted FL to boost the performance of wireless communications systems [203]–[205]. The authors of [206] proposed a unified architecture that uses concurrent communication to integrate NOMA-STAR-RIS with over-the-air federated learning (AirFL). The authors developed a MINLP problem by jointly designing the transmit power at users and configuration mode at the STAR-RIS to minimize the optimality gap, from which they derived the closed-form expression. Their simulation results demonstrated that the learning performance in terms of training loss and test accuracy may be effectively enhanced with the help of the STAR-RIS. An alternating optimization algorithm was developed to obtain the sub-optimal solution. In addition to the research works mentioned earlier, the authors in [207] proposed hierarchical federated learning with MEC servers for the NOMA-enabled IIoT to lower latency and network burden. To reduce latency, energy consumption, and model accuracy while taking into account the computational capacity and transmission power of IIoT devices, the authors developed a multiobjective optimization problem in their research. Extensive simulations were carried out to validate the efficacy of the proposed research. FL significantly improves the user's security and privacy because users train their models locally and send them to the BS for aggregation without revealing the user's private data to BS. Better training accuracy can be achieved by utilizing a large dataset and extensive communications rounds. However, the vast amount of data that is trained and transmitted within the system results in significant dimensionality, complexity, computation, and communications costs. An important direction for future research is evident in designing approaches that can skillfully balance the trade-off between computation and communication.

### *C. Ultra-Reliable and Low-Latency Communications (URLLC)*

URLLC is a key concept in 6G since it allows for machine-type communications, the process by which machines can communicate wirelessly with one another without the intervention of a person. In addition to its traditional uses in areas like manufacturing, public service, and autonomous driving, it is increasingly being employed in cutting-edge applications made possible by the widespread deployment of robots, UAVs, and advanced human-machine interfaces (HMIs)[11]. URLLC was considered one of the new application scenarios in the 5G mobile communications systems. On the other hand, URLLC has the goal of achieving an end-to-end latency on 1 ms time scale with a guarantee of ultra-reliability (e.g., with a 99.999% packet success probability). Such a performance guarantee is to meet the requirements of mission-critical applications [232], [233]. Nevertheless, 5G URLLC does not fulfill all the KPIs of diverse mission-critical applications like industrial automation, intelligent transportation, telemedicine, Tactile Internet, Virtual and Augmented Reality (VR/AR), and Meta-Universe. To provide satisfactory services to these applications, the 6G communications systems must meet additional requirements on some of the following KPIs in combination with URLLC: high spectrum efficiency, reliability, throughput, energy efficiency, network availability, security, AoI, and latency. These new requirements pose unprecedented challenges regarding design methodologies and enabling technologies in 6G.

IoT-based applications place a premium on low-latency communications [234], making URLLC an essential part of the ever-evolving communications environment to IoE. Such applications prioritize the least latency and reliable communications and employ coding block lengths that are both finite and short. The finite number of channel observations reduces coding gain and increases the Shannon's limit gap in the finite block length regime [40]. Short packet limit communications is required for URLLC and mMTC to reduce latency when resources are limited [235]. a number of works considered NOMA-inspired short packet transmission schemes to minimize network latency and maximize network performance [236], [237].

Retransmission techniques were also considered together with NOMA in the finite block length regime [238]. Particularly, mMTC benefits from retransmissions as time diversity improves reliability and, consequently, coverage. This is primarily due to the fact that many mMTC applications need to handle sporadic traffic with small payloads from each device; consequently, a packet may be retransmitted multiple times before it is effectively delivered to the BS [239]. Automatic repeat request (ARQ) and its variants, including hybrid ARQ (HARQ), are typically utilized by wireless networks for retransmissions when necessary [240]. The receiver sends a feedback message to the transmitter to indicate whether retransmission is required or not. In Chase combining (CC) HARQ [241], the entire codeword is transmitted during each retransmission, and repeated packets are combined using the maximum ratio combining (MRC) technique to improve the effective SNR. In incremental redundancy (IR) HARQ [242], the original codeword is split into multiple sub-codewords that are transmitted in sequential retransmissions to boost the coding gain. In the finite block length regime, both subclasses are being intensively studied. The authors of [243] proposed a dynamic retransmission strategy for the uplink NOMA with two users, HARQ, and multiple authorized retransmissions. Moreover, in [243], whenever both users simultaneously receive negative acknowledgement (NACK) from the BS, the high-power user is prompted to retransmit its previous packet, while the low-power user is prompted to send a new packet the next time slot.

[11] *6G URLLC: Unlocking the Factories of the Future and Self-Driving Cars.*

We noted that researchers also focused on designing efficient short-packet communications using grant-free NOMA-inspired communications systems [244], [245]. Grant-free NOMA (GF-NOMA) is an exciting multiple-access candidate for the forthcoming NGWNs to enable the URLLC to ensure low access latency and high connectivity density. GF-NOMA allows multiple users to use the same time-frequency resource block and communicate with the BS without waiting for the BS to assign them access on-demand [246], [247]. In [248], authors used GF-NOMA-based uplink transmission in developing massive-URLLC, which prioritizes high rate-reliability and fairness for massive users with strict latency constraints. To develop a spectral-efficient and reliable resource allocation policy, authors in [248] formulated resource optimization problems for the GF-NOMA scheme to maximize reliability and minimize latency under joint power and channel allocation.

Table VII shows the top-notch research on NOMA-inspired URLLC. NOMA is considered the best candidate multiple access technique because of its high spectrum efficiency and throughput to fill the gap between 5G URLLC and the diverse KPI requirements of the 6G URLLC. Unlike OMA, users can get more access to the power domain using NOMA, which is particularly helpful for high-frequency communications across low-rank channels. Therefore, NOMA provides two solutions to meet the new, stringent requirements of the URLLC. NOMA improves the spectral efficiency of URLLC. Moreover, by serving multiple users in the same resource block, NOMA can significantly minimize the latency for URLLC, especially in the case with the massive number of users. In [249], the authors proposed a downlink MIMO-NOMA architecture for the URLLC system. The design initially included only two users, but it may easily be scaled to accommodate more users and complex scenarios.

URLLC has the objective of facilitating a diverse array of multimedia services and applications that are delay-sensitive and aim to meet the strict demands of users in terms of the delay-bounded QoS. The theory of the AoI is utilized to describe the timeliness of information. It measures the time difference between the current time and the time stamp of the most recent observation. This theory has been suggested as a means to examine the latency of information in URLLC. In [250], authors investigated the timeliness performance of DL-NOMA and proposed an adaptive transmission policy under hybrid automatic repeat request with chase combining (HARQ-CC) aided NOMA systems. Furthermore, in [251], authors minimized the AoI in NOMA-assisted vehicular networks using the Markov decision process. A novel learning framework for multiple active configured agents for GF-NOMA to address the challenges of efficient, delay-bounded, and reliable communications for a massive number of UEs was discussed in [223]. In [224], the authors focused on corporative NOMA in the context of short-packet communications. The performance of decode and forward relaying, as well as the strategies of selection combining and maximum ratio combining at the receiver, were analyzed.

The downlink RAN slicing of the eMBB and URLLC traffic employing the NOMA technique was analyzed in [225]. Authors proposed a novel heuristic algorithm to address the non-convexity of the formulated optimization problem. Uplink resource allocation in NOMA-aided URLLC to reduce the mean-decoding error probability in the time-varying environment using the DRL-based algorithm was proposed in [226]. A novel resource allocation algorithm for the NOMA-assisted URLLC in the distributed antenna systems was proposed in [227]. The authors analyzed the rates and reliabilities of distributed antenna systems with OMA and NOMA, considering the impact of latency. In [228], authors derived the closed-form uppar bound for the delay target violation probability in the downlink MIMO-NOMA aided URLLC by applying the stochastic network calculus to the Mellin transforms of service processes. Also, authors proved that the infinite-length Mellin transforms resulting from the non-negligible interferences of NOMA are Cauchy convergent and can be asymptotically approached by a finite truncated binomial series in the closed form.

NOMA has great potential to increase spectral efficiency and reduce latency. However, reliability problems limit its applicability in the URLLC. To address this challenge, authors in [229], investigated the optimal resource allocation schemes for both the UL and DL of NOMA-inspired URLLC. The authors optimized power control and channel blocklength allocation to optimize transmission reliability and reduce error probability. In [230], a framework for packet re-management is proposed for C-NOMA-aided URLLC-based NGWNs. By optimizing block length, power, and re-managed packet size, an energy-efficient transmission framework is proposed to reduce power consumption while maintaining the stringent URLLC requirements and maximum power constraints. Another packet re-management-based C-NOMA transmission scheme was proposed for URLLC in [231]. In their work, authors jointly optimized the block length, power, re-managed packet size, and block error rate to minimize the power consumption to meet the strict URLLC requirements.

TABLE VIII: STATE-OF-THE-ART RESEARCH WORK ON TERAHERTZ.
(All APs, SBS, and MBS are mentioned as BS. IoT devices and UEs are mentioned as users)

| Ref. | System Model | Design Objective | Main Findings | Main Findings |
|---|---|---|---|---|
| [197] | $m$ BSs, $k$ users, MIMO-NOMA (DL) | Energy efficiency | Deep reinforcement learning | Proposed network offered reliable and ultra-high speed connectivity for fast users. |
| [252] | One BS,$k$ users (DL) | Energy efficiency | ADMM | The proposed scheme outperformed THz-OFDMA and NOMA without THz in EE. |
| [253] | $m$ BSs, $k$ users, MIMO-NOMA (DL) | Energy efficiency | Machine learning | Power optimization and low power consumption improve THz enabled network EE. |
| [254] | One BS, $k$ users, MISO-CNOMA (DL) | Energy efficiency | Hungarian algorithm | Proposed schemes outperformed benchmark schemes in EE and sum rate. |
| [255] | One BS, 2 users, THz-NOMA (DL) | Outage probability | Heuristics | This paper presents SC and MC THz-NOMA network outages. |
| [256] | One BS and two users | Beam alignment | Analytical | Proposed THz-NOMA scheme outperformed conventional THz-OMA. |
| [257] | One BS,$k$ users,$M$ radar targets(NOMA-ISAC) | Max. throughput | Successive convex approximation | The NOMA-ISAC achieves considerable gain over conventional ISAC. |
| [258] | One BS $k$ users (DL) | EE resource allocation | Dinkelbach algorithm | The system's total sum rates decrease with the increase of AWGN. |
| [259] | One BS and two users RHS-NOMA (DL) | Outage probability | Analytical | Closed form expressions of outage probability are provided. |
| [260] | One BS and $k$ users (DL) | Max. throughput | Iterative algorithm | The findings indicate the optimal range for the sub-band assignment coefficient lies between 0.47 and 0.7. |
| [261] | RIS assisted THz-NOMA | Max. sum rate | Sub-optimal heuristics | The sum rate increases with the number of RIS elements. |
| [262] | One SBS, one MBS, $k$ users (NOMA-HetNets) | Max. energy efficiency | Whale optimization algorithm | The more the number of the SBSs, the higher the energy efficiency. |
| [263] | Cooperative NOMA-THz for V2X (DL) | Max. data rate | Analytical | Authors derived the closed-form expressions for optimal power allocation. |

### *D. Terahertz*

The detailed summary of recent research work on THz-NOMA is presented in Table VIII. Terahertz communications is a technology that holds promise for next-generation mobile communication networks. The demand for rapid data transfer necessitates ever-increasing bandwidths. Future networks may look to use higher frequencies as a possible solution. Millimeter-wave bands might be sufficient for now, but higher frequencies will be required for the next generation of wireless communications networks. Researchers are considering utilizing the Terahertz (THz) band for future generations of wireless communications. The range from 0.1 THz to 10 THz is usually used to define the THz band, and its available bandwidth is more than one order of magnitude than mmWave [264]. Additionally, the THz band shares the features of a narrow beam and large communications capacity. In addition, the THz communications technique is considered a core technology in the 6G mobile communications network. Although the THz band has plenty of available spectrum, 6G networks will need to accommodate many users and a growing number of use cases, such as those in the mobile, industrial, and healthcare sectors. However, the THz band's primary limitation is its smaller coverage area. Because of this, there will be a significant increase in energy consumption and transmission costs [255].

We noted that researchers have made significant contributions in the THz-band-based NOMA network, focusing on optimizing the key performance metrics such as sum rate [261], throughput [257], [260], energy efficiency [197], [252]–[254], [258], [262], outage probability [255], [259], and beam alignment [256]. By allowing multiple UEs to use the same sub-channel, NOMA increases the efficiency of THz networks by employing their available bandwidth to support massive data services. A user's power allocation in NOMA is proportional to its channel gain. The spectral efficiency of a NOMA system can be enhanced by using SIC at the receivers to decode and demodulate the superposed signals of multiple UEs precisely. Furthermore, THz-band-based NOMA systems may certainly improve massive connectivity, but it is hard to acquire flexible sub-channel allocation and power optimization strategies with high user intensity. Therefore, it is essential to investigate THz-NOMA network resource optimization in order to ensure energy-efficient systems.

Compared with mmWave communications, the difference is not only a frequency shift or transmission bandwidth but also the channel propagation conditions and transmission beamwidths. Moreover, different from mmWave, the THz communications are susceptible to unique challenges such as molecular absorption noise, molecular absorption at different frequencies leading to serious path loss peaks, and also the mathematical structure of the channel propagation model based on Beer's-Lambert Law adds to the challenges [255]. The authors in [265] characterized the large-scale best-direction and omni-directional path losses in indoor scenarios. Moreover, free-space path loss was computed by invoking the Friis' law, given by

$$\mathrm{FSPL}\left(d_0\right)=-20\log_{10}\left(\frac{r}{4\pi f d_0}\right), \tag{6}$$

where $d_o$ represents the reference distance, $r$ demotes the speed of light and $f$ represents the carrier frequency. Furthermore, in [266] investigated the use of NOMA in THz networks to support secondary users with pre-configured spatial beams. The goal was to enhance network capacity while minimizing any disruption to the current legacy network. The beam and power allocation problem was formulated as a non-convex optimization problem. It was solved using different methods, including branch-and-bound and Successive Convex Approximation. Their research revealed a significant problem associated with strong inter-beam interference. Thus, the need for developing sophisticated solutions that can successfully handle this problem is imperative.

### *E. Cooperative Communication*

Cooperative relaying is a promising approach to alleviate channel imperfections like fading, path loss, and shadowing and, therefore, significantly improve the communications coverage [280]. The simplest cooperative relaying network consists of three nodes: the source, the relay, and the destination. With the help of the relay, the source communicates indirectly with the receiver. Two well-known relay protocols exist amplify-and-forward (AF) and decode-and-forward (DF) protocols. The AF relay simply forwards a scaled version of its observation to the destination, which means that the noise at the relay is also forwarded to the destination. In contrast, the DF relay first attempts to decode the received signal and then re-transmits the decoded information to the destination, where decoding errors could be passed to the destination [281]. Cooperative transmission schemes have opportunities due to their capacity to increase the network's coverage area and the users' signal-to-interference-plus-noise ratio (SINR).

TABLE IX: STATE-OF-THE-ART RESEARCH WORK ON COOPERATIVE COMMUNICATION.
(All APs, SBS, and MBS are mentioned as BS. IoT devices, and UEs are mentioned as users)

| Ref. | System Model | Design Objective | Main Findings | Main Findings |
|---|---|---|---|---|
| [267] | One BS, two users (Relay assisted) | System sum rate | Lagrange dual method | The proposed scheme improves the system sum rate. |
| [268] | One BS and two users(User assisted) | Sum throughput | Markov chain based stochastic model | The two proposed schemes outperform the cooperative FDR and HDR NOMA. |
| [269] | One BS, $k$ users | System sum rate | SCA, BCA Hungarian algorithm | The proposed scheme outperforms benchmark techniques in system sum rate. |
| [270] | One BS, two users (Relay assisted) | Ergodic capacity | Analytical method | NOMA outperformed OMA both for AF and DF relaying. |
| [271] | One BS and $k$ users (DL)(D2D) | Sum rate | Iterative algorithm | Additional power constraints are introduced to D2D pairs for NOMA decoding order. |
| [272] | One BS, $k$ users (DL)(relay assisted) | Outage probability | Analytical | Closed-form expressions for both exact and asymptotic outage probabilities are derived. |
| [269] | One BS, $k$ users | System sum rate | SCA, BCA Hungarian algorithm | The proposed scheme outperforms benchmark techniques in system sum rate. |
| [273] | One BS, two users | Throughput | Polak Ribiere conjugate gradient | The proposed schemes can greatly improve network outage and performance. |
| [274] | One BS, two groups of users | Energy efficiency | Bat algorithm | The proposed scheme performed better than the benchmarks in both SE and EE. |
| [275] | One BS, $k$ users (relays assisted) | Achievable rate | Bisection based power allocation | The proposed scheme outperformed the benchmark schemes. |
| [276] | One BS and two users (relay assisted) | Sum rate | Analytical | The hybrid relay's average achievable rates are derived for both AF and DF modes. |
| [277] | One BS and two users (DL)(relay assisted) | Block error rate | Analytical | Authors propose new FD NOMA protocol for short-packet communication. |
| [278] | One BS and $k$ users (DL)(relay assisted) | Sum rate | Analytical | The proposed technique improves user fairness and reduces inter-user interference. |
| [279] | One BS and two users (user-assisted) | Power allocation | Analytical | The authors derive the closed-form expression for the achievable rate of the system. |

Recently, NOMA has drawn significant interest from academia and industry [267]. Table IX presents the state-of-the-art research work on the cooperative NOMA. Fig. 16 illustrates UAV-relay-aided cooperative NOMA transmission. The combination of cooperative relaying and NOMA is important because the near users are aware of the information for the far users after carrying out SIC. Thus, system spectral efficiency and the number of supported users increase [268]. In the literature, several types of cooperation schemes have been investigated by researchers, such as user-cooperative, dedicated relay-cooperative, and D2D transmissions. In the user-cooperative NOMA, the near user (strong user) acts like a relay to assist the far user (weak user) in receiving the message from BS. In the first phase, the source transmits a superimposed signal to the users. The strong user uses AF or DF modes to forward the signal to the weak user, which employs SIC to decode its message [269]. In relay cooperative NOMA, the dedicated relay assists the data transmission between BS and users. In this type of cooperation, BS sends the superimposed signal to a relay, and then the relay forwards the received signal to users through AF or DF modes. The users then perform SIC to decode their signals [270]. Researchers also have considered utilizing D2D cooperative communications in NOMA networks to boost spectral efficiency and reduce latency [271].

In the past, most research on NOMA-enabled cooperative communications has been done using half-duplex (HD) relays due to their simplicity and lower self-interference. In HD mode, a device can either transmit or receive data at a given time, but not both simultaneously. This mode is simple to implement, and it inherently avoids the problem of self-interference, which arises when a device simultaneously transmits and receives data [272]. However, recent advancements in self-interference cancellation technology have made it possible to use FD communications mode in NOMA-enabled cooperative communications systems [273]. In the full-duplex (FD) mode, a device can simultaneously transmit and receive data, improving network performance and spectral efficiency. However, implementing FD mode requires advanced techniques to cancel self-interference, which can increase system complexity and cost. A hybrid half/full-duplex cooperative system has been proposed to balance the advantages and disadvantages of each mode [272]. This system can dynamically switch between half-duplex and full-duplex modes to achieve optimal performance. The decision to switch between the two modes depends on several factors, such as network conditions, channel quality, and available resources. The system can exploit the tradeoff between the efficiency loss of the half-duplex mode and the inherent self-interference of the full-duplex mode. While using a dedicated node for relaying in FD relaying can increase system cost, user-assisted communications with energy harvesting technology can be used to avoid draining the user's batteries and reduce system cost. This means that the users can help in relaying the data, and the energy harvested from the environment can be used to power the user's transmission.

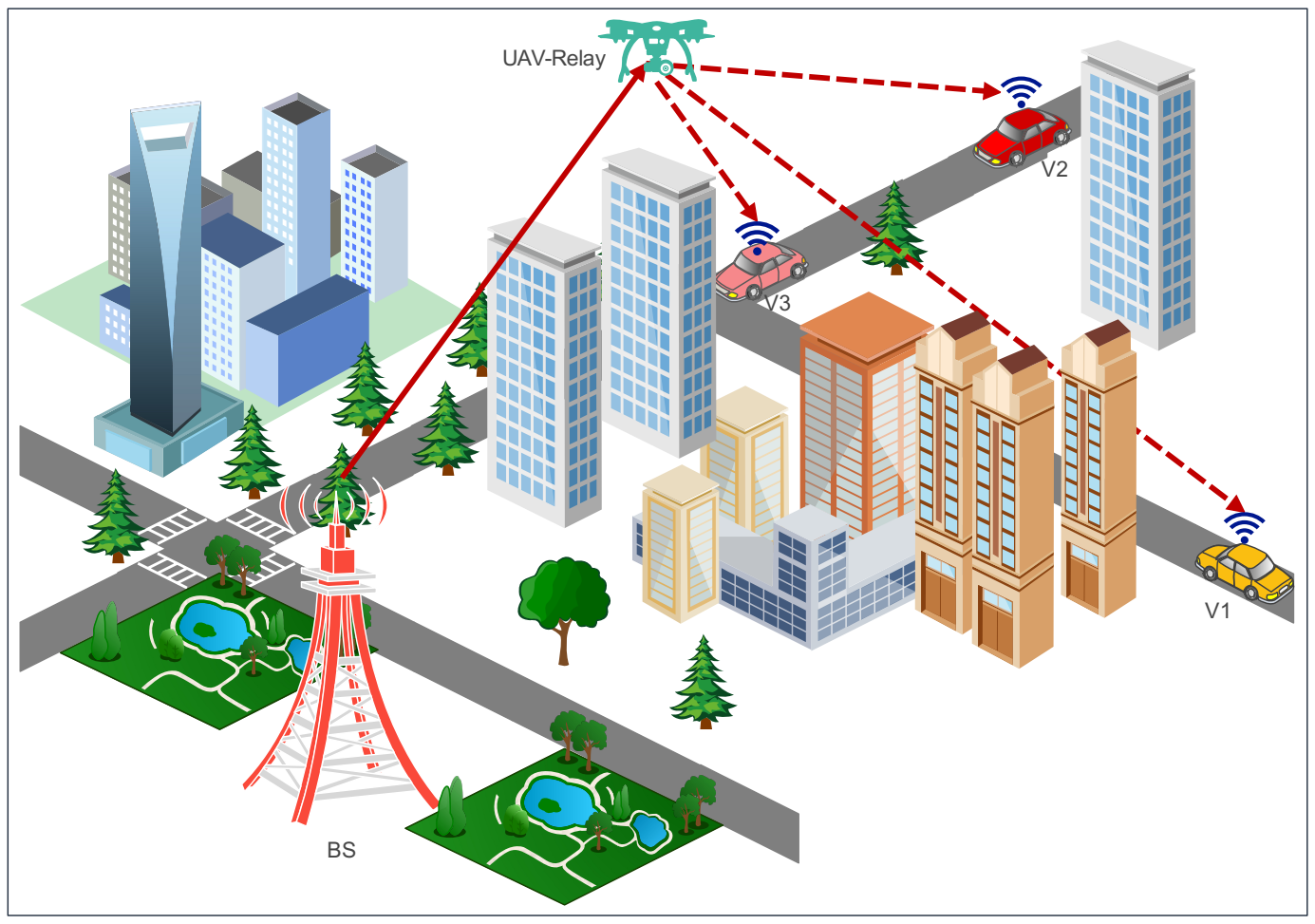

Fig. 16: UAV-relay aided cooperative NOMA transmission.

A cooperative NOMA scheme with wireless power transfer with the objective of optimizing the EE was considered in [274]. A sub-optimal solution was obtained by decoupling the original problem into two sub-problems; an iterative algorithm was used for user grouping, and the Bat algorithm was used for the power allocation. In [275], authors proposed two relay-assisted NOMA communications schemes for V2X communications, i.e., HD relay-assisted NOMA and FD-relay-assisted NOMA and examined the optimal power allocation problem. The formulated problem was transformed into a sequence of convex feasibility problems and the solution was by a bisection-based power allocation algorithm. In [276], A NOMA scheme was proposed for the cooperative relay system, where a source node communicates with a pair of energy-harvesting UEs through multiple antenna relay nodes. A hybrid protocol was implemented at the relay station, where successful signal decoding led to the adoption of the DF protocol for signal forwarding to the users. Otherwise, the AF protocol was implemented.

TABLE X: STATE-OF-THE-ART RESEARCH WORK ON BACKSCATTER COMMUNICATIONS.
(All APs, SBS, and MBS are mentioned as BS. IoT devices, and UEs are mentioned as users)

| Ref. | System Model | Design Objective | Main Findings | Main Findings |
|---|---|---|---|---|
| [282] | One BS, one BD and one user (DL) | Spectrum efficiency | Heuristics | Closed-form and asymptotic expressions for the outage probability are derived. |
| [283] | One BS, $q$ BDs and one BR (DL) | Throughput | Successive convex optimization | Larger throughput-gain is achieved for higher BS transmit power. |
| [284] | $m$ BSs, two users, $p$ BDs (multicell DL-NOMA) | Max. secrecy rate | Sub-gradient method | Proposed NOMA scheme significantly outperforms the traditional TDMA scheme. |
| [285] | One BS, two users, and one RIS | Outage probability | Analytical | Authors validated the analytical outage probabilities via Monte Carlo simulations. |
| [286] | One BS, $p$ BDs and $k$ users (DL) | Max. sum rate | Iterative algorithm | Increase in the number of users results in an improved sum rate at high transmit power. |
| [287] | One BS,two users,one BD,and one eavesdropper | Physical-layer authentication | Analytical | Authors derived the analytical expressions for the OP of AmBC-based NOMA networks. |
| [288] | One BS, $p$ BDs and one RIS | Power reflection coefficients | Successive convex approximation | This work jointly optimizes RIS phase shifts and power reflection coefficients at BDs. |
| [289] | One BS, two users and one BD (DL) | Energy efficiency | Dinkelbach's method | Higher min. SINR increases transmission rates and power consumption, declining EE. |
| [290] | One BS and two BDs (DL) | Bit error rate | Analytical | Analytical BER expressions for imperfect SIC in NOMA-backcom system are presented. |

A cooperative NOMA scheme for the short-packet communications over the Rayleigh channels was proposed in [277]. In this work, the near user acts as the relay operates in FD mode and employs the DF protocol. Two types of residual self-interference (RSI) models, one assuming no fading and the other assuming Rayleigh fading, are considered. Researchers have also considered using cooperative NOMA in the CRNs. For instance, in [278], the authors proposed time-sharing-based HD/FD CRNOMA schemes to efficiently use the spectrum of unpaired users. In [279], the authors explored a wireless-powered FD cooperative NOMA system, where self-energy recycling is employed at the near user. A new power allocation scheme was proposed that prioritizes the nearby user, utilizing partial CSI. Later on, the authors derived closed-form expressions for the outage probability and the diversity order.

### *F. Backscatter Communications*

Backscatter communications (BackCom) is a wireless communications technology that enables devices to communicate with each other by reflecting the radio waves already there. Instead of sending out its own signal, a BackCom device modulates and reflects RF signals back to the sender. This is done in place of sending out its own signal. This technique can be used in various applications, including low-power IoT devices, RFID tags, wireless sensors, and many more [291]. Communication via backscatter offers several advantages over conventional wireless communication, including lower power consumption, lower cost, and longer communication range. It is especially helpful in scenarios where conventional wireless communications methods cannot be implemented, such as in remote regions or in underground structures [292].

Table X depicts state-of-the-art research work on integrating NOMA and backscatter communications. The NOMA-inspired backscatter communications can achieve reliable and effective communication [282]. This strategy involves having numerous backscatter devices sharing the same frequency band to reflect the ambient signal with various modulation patterns. At the receiver, NOMA is used to decode the signals that have been reflected. This interplay has several advantages over conventional OMA-based backscatter communications. First, it can support multiple users, which is important for applications such as IoT networks. Second, it can achieve higher data rates by allowing multiple devices to transmit simultaneously. Third, it can improve communication reliability by using advanced signal processing techniques to mitigate interference and noise. The authors of [283] maximized the throughput of a NOMA-enhanced bistatic backscatter communications network. In this network, multiple backscatter devices transmit data to a backscatter receiver using a protocol that integrates NOMA with dynamic time-division-multiple-access.

An optimization framework for ensuring the link security in the BackCom-aided NOMA network was investigated in [284]. In [285], the authors considered an IRS-assisted BackCom-NOMA network with a direct link, which consists of a BS, two users, and one IRS. In this work, the authors conducted Monte Carlo simulations to validate the performance of the proposed framework in terms of power allocation coefficients, the number of IRS reflecting elements, and the backscatter signal. In [286], the authors explored a transmission scheme that combines FD-NOMA and BackCom. In this scheme, the FD source serves a multiuser downlink NOMA network while BDs transmit their information to the source simultaneously. In this work, the authors jointly optimized the power allocation coefficient and reflection coefficient to maximize the downlink sum rate.

A unified ambient BackCom-assisted NOMA network to validate the physical layer authentication was developed in [287]. In the work mentioned above, three different authentication tags were created: one for physical layer authentication using shared authentication tags, another for physical layer authentication using space division multiplexing authentication tags, and a third for physical layer authentication using time division multiplexing authentication tags. The authors of [287] validated the proposed scheme through simulations, and it was determined that the physical layer authentication scheme with the shared authentication tag demonstrated superior robustness when compared to the other schemes. Additionally, in [288], the authors proposed a RIS-aided BackCom-enabled NOMA network and jointly maximized the system's total rate by optimizing the power reflection coefficients at the BDs and the phase shifts at the RIS, all while making sure that each BD met the minimal QoS criterion. To solve the formulated problem, the authors proposed a joint optimization algorithm based on alternative optimization, SCA, and manifold optimization approaches.

An energy efficiency-based resource allocation problem with QoS guarantee in a BackCom-NOMA network was investigated in [289]. In the mentioned study, the authors focused on enhancing the energy efficiency of users while considering factors such as the minimum SINR requirement, maximum transmit power constraint, and successful decoding order constraint. In [290], the authors conducted an analysis of the BER performance of a NOMA-enhanced BackCom system with imperfect SIC. The mentioned research work focuses on deriving exact closed-form analytical BER expressions for BPSK in two BDs. The derived BER expressions are validated through Monte Carlo simulations for a range of reflection coefficients.

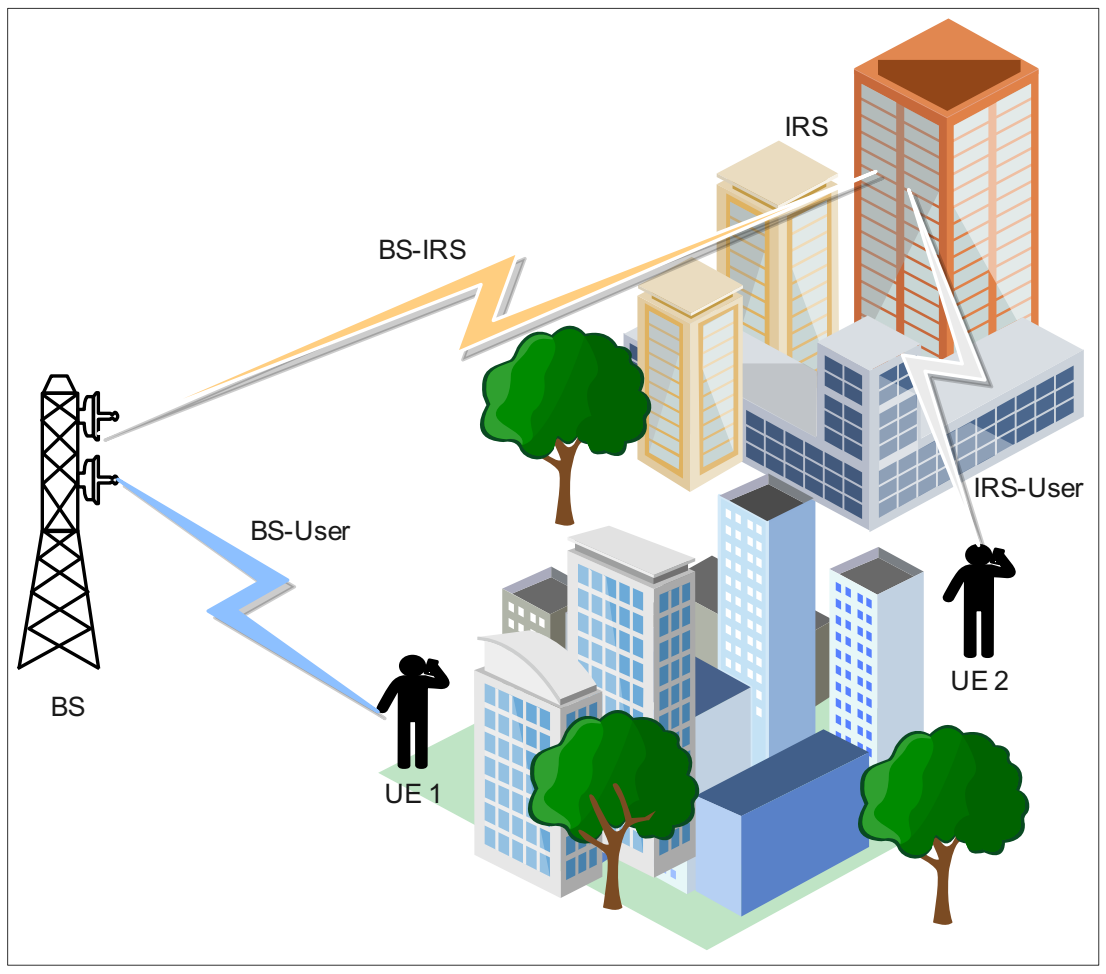

Fig. 17: RIS assisted NOMA networks.

### *G. Reconfigurable Intelligent Surface*

To meet the future demands of 6G, a novel concept called intelligent radio environment proposes that random and time-varying wireless channels can be dynamically controlled and reconfigured to improve wireless communications performance, which leads to the digitally controlled low-cost RIS or its equivalents, such as reconfigurable intelligent surface (RIS). RIS allows wireless communications to create effective line-of-sight (LoS) links, improve the channel rank, reshape channel realizations or statistical distributions, and more by dynamically tuning signal reflection amplitude/phase.

The remarkable capabilities of RIS and NOMA offer a persuasive approach to tackling the challenges encountered by next-generation networks. Therefore, it can be inferred that the incorporation of RIS with the NOMA network could present a remarkably effective approach to fulfilling the data rate demands of 6G networks. The combination of NOMA with RIS introduces an innovative framework that could provide a potential multi-access solution to meet the rigorous requirements for latency, extensive connectivity, and data rates anticipated in the upcoming 6G networks. Furthermore, incorporating RIS and NOMA reduces power consumption, extends network coverage and improves overall network performance in dense or high-demand network scenarios. The latest research indicates that the integration of RIS and NOMA may significantly impact the performance of wireless networks, promoting scalable and energy-efficient communication.

In [293], the authors explained the simple design of the RIS-assisted NOMA transmission network. Here, spatial division multiple access allows for serving more customers in each orthogonal spatial direction. Additionally, the performance of real-world transmission and the effect of hardware impairments on RIS-assisted NOMA design were analyzed. In [294], the correlation between individual users' transmit power and phase shift variables was analyzed to tackle the joint power control and RIS phase shift optimization problem. The phase shift determination problem was subsequently tackled using the sequential rotation algorithm. In a traditional communication network, attaining quasi-degradation is challenging, as the propagation environments control the channels and are not subject to reconfiguration. To address this challenge, a system utilizing RIS in a MISO-NOMA configuration was examined in [295], where the wireless channels can be effectively changed, and to reduce transmission power, the optimization of beamforming vectors and the RIS phase shift matrix were optimized. In [296], a network utilizing RIS-assisted NOMA for IoT applications was examined. In this context, a resource allocation scheme was introduced to enhance the overall throughput by optimizing the time allocation factor and phase shift matrices of the RIS. In [297], a system utilizing RIS assistance for uplink NOMA was introduced to optimize the total data rate for all users while adhering to individual power limitations. This issue required a collaborative approach to power regulation among users and the design of beamforming at the intelligent reflecting surface, presenting a non-convex challenge. SDR was employed to address the issue, yielding a solution that approached optimality. A network assisted by the RIS was designed in [298] to improve communication coverage and optimize energy efficiency. This study introduces an algorithm designed to enhance the overall sum rate while simultaneously reducing total power consumption in an energy-efficient manner. To enhance the energy efficiency of the system, a joint optimization of the transmit beamforming at the base station and the reflecting beamforming at the intelligent reflecting surface was conducted. This involved an iterative approach where the transmit beamforming and the phases of the cost-effective passive elements on the intelligent reflecting surface were optimized alternately to achieve communication convergence.

The summary of the other state-of-the-art literature on the utilization of RIS in NOMA transmission is presented in Table XI. Integrating RIS with NOMA networks has been identified as a promising technique for meeting the high data rate demands of 6G networks [299]–[301]. By incorporating RIS into NOMA networks, a new paradigm is proposed to offer a potential solution for addressing the stringent requirements on latency, massive connectivity, and data rates that are expected of future 6G networks [302]. In a conventional NOMA network without the RIS, the performance gain of NOMA heavily depends on the channel conditions of the two users. Traditionally, these channel conditions are assumed to be fixed and determined solely by the users' propagation environment. However, using RIS presents opportunities for intelligently reconfiguring the users' propagation environment to facilitate the implementation of NOMA, which can result in significant performance gains [303], [304]. In particular, the combination of RIS with NOMA provides greater degrees of freedom for system design when the channel gains of both users are the same. SIC is a crucial component of NOMA systems and is key to mitigating multiple access interference. The selection of the SIC decoding order can be based on either the user's CSI or their QoS requirements. By incorporating RIS, the design flexibility of NOMA networks is increased, enabling a transition from channel condition-based NOMA to QoS-based NOMA. Without RIS, the performance gain of NOMA over OMA is critically dependent on the two user channels. By intelligently tuning the phase shifting matrix, RIS-NOMA can introduce more DoFs for system design. This can be exploited to not only generate a significant difference in the users' channel gains but also to customize the users' effective channel gains according to the users' QoS requirements [105], [305]. Consider a scenario where there are two users in RIS-NOMA: one strong user and one weak user. If the weak user requires a higher data rate and should be prioritized as the stronger user, this can be achieved by adjusting the phase shifting matrix of RIS [305] Combining NOMA and RIS emphasizes the importance of actively adjusting the channels to meet QoS requirements, ultimately improving network flexibility, efficiency, and user satisfaction [306]. Moreover, using RIS can relax constraints on the number of antennas at the transmitters and receivers in conventional MIMO-NOMA networks. Certain constraints on the number of antennas may need to be satisfied in such networks. However, these constraints can be relaxed with the additional passive array gains the RIS provides. Given these potential benefits, researchers have begun investigating the use of RIS-NOMA and addressing the challenges to optimize its performance including energy efficiency [298], [299], [301], outage probability [300], [307], transmit powers [302], [308], [309], throughput [106], [303], [310], data rate [303], coverage probability [311] and secrecy rate [312].

TABLE XI: STATE-OF-THE-ART RESEARCH WORK ON RECONFIGURABLE INTELLIGENT SURFACE. (All APs, SBS, and MBS are mentioned as BS. IoT devices and UEs are mentioned as users)

| Ref. | System Model | Design Objective | Main Findings | Main Findings |
|---|---|---|---|---|
| [299] | One BS, $k$ users, $O$ RIS reflecting elements | Energy efficiency | DC programming | Proposed algorithm gives sub-optimal EE with minimal computational complexity. |
| [300] | One BS, 2 users, $O$ RIS reflecting elements | Outage probability | Laguerre series | Non-ideal RIS with four or more bits for quantization performs same as ideal RIS. |
| [301] | One BS, 2 users, $O$ RIS reflecting elements | Energy efficiency | Heuristics | The EE of the proposed scheme increases with the number of reflecting elements. |
| [302] | One BS, $k$ users, $O$ RIS reflecting elements | Min. transmit power | SDR and SCA | The RIS-assisted SWIPT NOMA network can decrease BS transmit power by 51.13%. |
| [303] | One BS, $k$ users, $O$ RIS reflecting elements | Sum throughput | ECA and MSGD | The proposed scheme outperformed the benchmark schemes. |
| [304] | One BS, $k$ users, $O$ RIS reflecting elements | Max. data rate | DRL | Proposed scheme shows that RIS can improve the grand free user's network sum rates. |
| [212] | One BS, $k$ users, $O$ RIS reflecting elements | Max. sum rate | DQN and K-GMM | Three algorithms are proposed for user clustering, mobility, and phase shift matrix. |
| [310] | One BS,$k$ users, $O$ RIS reflecting elements (DL) | Max. system throughput | Many-to-one matching algorithm | The proposed RIS-NOMA scheme surpassed benchmark schemes in throughput. |
| [308] | One BS, $k$ users and one eavesdropper | Min. transmit power | SROCR based alternating algorithm | Signal power highly depends on QoS constraint of legitimate users. |
| [313] | One BS,$k$ users, $O$ RIS reflecting elements (DL) | Max. sum rate | Successive convex approximation | 3-bit phase shifters nearly match ideal RIS performance in RIS-NOMA system. |
| [307] | One BS,2 users, $O$ RIS reflecting elements | Outage probability | Analytical | The closed-form expressions for the outage probability and ergodic rate are derived |
| [312] | One BS,$k$ users, $O$ RIS reflecting elements (DL) | Secrecy rate | Successive convex approximation | Equal allocation of reflecting elements enhance secrecy in distributed RIS. |
| [298] | One BS,$k$ users, $O$ RIS reflecting elements (DL) | Max. Energy efficiency | Alternating optimization | The proposed scheme outperformed benchmark schemes. |
| [314] | One BS,$k$ users, $O$ RIS reflecting elements | Max. sum rate | Iterative algorithms | Achievable sum rate increase with an increase in transit powers. |
| [309] | One BS,$k$ users, $O$ RIS reflecting elements (DL) | Min. transmit power | DC programming | the proposed scheme is able to induce exact rank-one optimal solutions. |
| [311] | $m$ BSs,$k$ users, $O$ RIS reflecting elements (DL) | Coverage probability | Analytical | Closed-form expressions for coverage probabilities of the NOMA users are derived. |
| [106] | One BS,$k$ users (STAR-RIS-NOMA) (DL) | Max. average throughput | Federated learning | The proposed algorithms effectively reduce or exempt the training overhead. |

To better understand the concept of RIS-NOMA networks, Fig. 17 showcases a classic example of RIS-assisted NOMA transmission. A downlink transmission scenario of the RIS-NOMA network is considered. BS simultaneously transmits the superimposed signals to the multiple UEs via RIS, assuming that the direct link between BS and $UE_2$ is highly attenuated. This scenario exists in urban environments where communications between BS and users is blocked due to sky-rise infrastructure. By utilizing RIS, $UE_2$ can receive the signals from BS via RIS, improving coverage, spectral efficiency, and fairness. Several design guidelines should be considered while designing the RIS-NOMA network. Efficient control over the order of users in the RIS-NOMA networks can be achieved through appropriate configuration of the RIS's meta-atoms. To facilitate this capability, it is imperative that the RIS be dynamically optimized in accordance with the instantaneous channel realizations. The implementation of RIS-NOMA presents considerable challenges. One challenge is the potential for optimization to become exceedingly intricate as the quantity of reflecting elements grows.

In contrast to the conventional optimizing techniques [313], [314], ML-based techniques are gaining popularity due to their ability to adapt to instantaneous channel realization based on the radio environment [212], [219], [304]. Moreover, ML-empowered optimization techniques can be designed to further improve RIS-NOMA performance. For instance, the conventional optimization techniques are very complex as the number of reflecting elements of the RIS is large. To overcome this problem, researchers proposed ML-based optimization techniques for the RIS-NOMA networks. For instance, authors in [304] proposed a semi-grant-free transmission scheme for the RIS-NOMA networks. The authors investigated ideal and non-ideal RIS cases in their proposed network and modelled a joint-optimization problem to maximize the long-term data rate. Moreover, authors in [304] proposed three multi-agent deep reinforcement learning-based frameworks to solve the problem under different RIS cases.

In [212], authors proposed a novel protocol to achieve the maximum sum rate for all users in the RIS-aided MISO-NOMA network by jointly optimizing the passive beamforming vector at the RIS, decoding order, power allocation coefficient vector, and the number of clusters. To solve the optimization problem, the authors invoked LSTM and K-Gaussian mixture model (GMM) for user position estimation and clustering and then combined them with the DQN algorithm to assist in the phase optimization of RIS. In [315], authors designed a RIS-NOMA network for energy transfer and data transmission. Moreover, authors in [315] formulated a joint resource allocation problem to minimize the total energy consumption for the RIS-aided federated learning-based wireless power transfer for the NOMA network. Furthermore, authors in [316] proposed using UAV with RIS-NOMA for IoT and deployed the reinforcement learning-based algorithm to optimize the transmit power and sub-carriers, trajectory, and phase shift matrix to minimize the average AoI. In addition to the works mentioned above, it is possible to explore other learning-based techniques to develop efficient and low-complexity methods to improve the performance of the RIS, making the research field more promising.

TABLE XII: STATE-OF-THE-ART RESEARCH WORK ON COGNITIVE RADIO NETWORKS.
(All APs, SBS, and MBS are mentioned as BS. IoT devices and UEs are mentioned as users)

| Ref. | System Model | Design Objective | Main Findings | Main Findings |
|---|---|---|---|---|
| [317] | $m$ BS,one primary user and $k$ cognitive users | Security-reliability tradeoff | Analytical | Proposed cooperative scheme outperformed the benchmark schemes. |
| [318] | Secondary BS, two users and one relay (DL) | Energy efficiency | Analytical | Each NOMA destination's exact closed-form outage probability is derived. |
| [319] | One BS, $k$ clusters(primary and secondary users) | Secrecy rate | Analytical | The proposed scheme boosts system secrecy. |
| [320] | One CBS, one PBS, $k$ PUs and SUs (DL) | Energy harvesting | Search algorithm | NOMA provides superior performance compared to the conventional OMA. |
| [321] | One PN, one SN, $k$ users and $O$ RIS devices | Max. sum rate | Iterative algorithm | Simulation results proved the efficacy of the proposed algorithm. |
| [322] | One PBS, one SBS, $k$ users (PU&SU) | Power allocation | Genetic algorithm | Proposed scheme is evaluated based on throughput, power level and access probability. |
| [323] | $m$ BS (SBS&MBS),one macro and $p$ small cells | Throughput | DC programming based algorithm | Proposed scheme maximizes throughput within NOMA cluster using KKT conditions. |
| [324] | One BS and $k$ users (DL) | Power allocation | Heuristics | The computational complexity of proposed scheme is linear. |
| [325] | One BS, one PU and $k$ SUs (DL) | Throughput | Dichotomy method | Simulations proves the performance in overlay and underlay modes. |
| [326] | NOMA-CRN based on $k$ users and eavesdropper | Power allocation | Successive convex approximation | Greedy algorithm used for secondary user scheduling and SCA for power allocation. |

### H. Cognitive Radio Networks

Cognitive radio network (CRN) has garnered considerable interest owing to its capacity to optimize the effective utilization of the radio frequency spectrum. This technology enables unlicensed (secondary) users to operate by using the same spectrum as licensed (primary) users without causing interference to the primary network. The performance of NOMA systems and CR networks has been investigated extensively in the literature [317]–[321]. Table XII presents the state-of-the-art research on NOMA-inspired CRNs. The incorporation of NOMA into CRN is anticipated as a potential solution to tackle challenges, such as spectrum efficiency and the ability to support massive connectivity. This integration aims to ensure reliable transmission for both primary and secondary users. The integration of CRN and NOMA networks presents significant advantages for both networks [322]. Despite the benefits, this integration also causes some challenges [327]. In CRN-based NOMA networks, perfect SIC implementation at the receiver requires perfect detection of weaker users' signals by stronger users. CRN-based NOMA networks cannot guarantee this, resulting in inadequate SIC implementation and residual inter-user interference at the near user. Moreover, when CR is integrated into NOMA networks, the QoS requirements for SUs should be considered. Thus, ensuring a set of predefined QoS requirements in CR-based NOMA networks for all users should be further investigated.

A resource allocation problem for a two-tier cognitive heterogeneous network in interweave spectrum sharing mode was studied in [323]. Furthermore, in [323] formulated a mixed integer non-linear programming problem with the objective of maximizing the sum throughput of a second-tier small cells network. In the mentioned work, the authors have successfully developed a novel clustering algorithm with the primary objective of maximizing the throughput of the clusters. Subsequently, they have proposed a power allocation mechanism for NOMA clusters, employing the well-known Karush-Kuhn-Tucker optimality conditions. In [324], the authors proposed a novel power allocation algorithm for CRN-NOMA networks, and the computation complexity of the proposed algorithm was only linear. A CRN-NOMA scheme was proposed in [325], and the proposed scheme was developed for overlay and underlay modes. The authors formulated a problem to optimize the system throughput and subsequently allotted the time slot and transmit power in an optimal manner to maximize the system throughput. Then, an optimal algorithm based on dichotomy was developed to address the optimization problem. In [326], the authors addressed a downlink resource allocation problem in the CRN-NOMA network that takes into account a delay constraint and security considerations. Then, the authors examined the probability upper limit of the maximum the maximum packet delay by utilizing a queuing model. An algorithm for scheduling secondary users was developed using the greedy approach, and a power allocation algorithm was proposed using the successive convex approximation method.

### I. Mobile Edge Computing (MEC)

6G communications networks rely heavily on NOMA and MEC to keep up with the escalating need for data transfer rates [339]. The state-of-the-art research on integrating NOMA in MEC is summarized in Table XIII and Figure 13 shows the NOMA-assisted MEC system. With the increasing use of mobile devices for computing-intensive tasks such as augmented reality (AR) and virtual reality (VR), there is a need for more computing resources than mobile devices can offer. Offloading tasks to a remote cloud for processing is a feasible solution. However, it can lead to long delays and degrade the user's QoE due to the long transmission distance. Moreover, it can also put a heavy burden on the core networks. MEC was proposed to address these challenges. MEC is a framework that places edge servers with computing resources near mobile users network access points e.g., a base station equipped with a MEC server, to provide high-speed computing units. By bringing computing resources closer to the users, MEC can improve the QoE of the users and relieve the burden of transmission tasks on the core network. MEC can provide several benefits, including reducing latency, improving QoE, and reducing the load on the core network. By computing tasks at the edge, MEC can reduce the distance between the user and the computing resources, reducing the latency and improving the QoE. MEC can also reduce the load on the core network by processing tasks at the edge, where network congestion can be avoided, and the network's overall performance can be improved.

Integrating and implementing NOMA and MEC in NG-WNs can yield significant benefits [328]–[330]. While MEC can effectively reduce task processing latency and overhead, NOMA can accommodate a larger number of users and enable higher task offloading transmission rates. This can ultimately enhance the overall performance of computation offloading in NGWNs. From the perspective of emerging technologies and applications such as CAVs, ITS, and eHealth, it is crucial to consider the cost of task processing and obtaining satisfactory task offloading services as well. To this end, authors in [331] proposed a new model for vehicular edge computing (VEC) that utilizes NOMA and aims to minimize the overall cost of the system. This includes the expenses associated with both MEC and cloud processing. The approach takes into account the decision to offload jointly. The primary objective of [331] was to guarantee the maximum task processing latency. Moreover, the proposed work analyzes and optimizes various aspects of VUEs, including VUE clustering, wireless communications, computation resource allocation, and transmission power control.

TABLE XIII: STATE-OF-THE-ART RESEARCH WORK ON MOBILE EDGE COMPUTING.
(All APs, SBS, and MBS are mentioned as BS. IoT devices and UEs are mentioned as users)

| Ref. | System Model | Design Objective | Main Findings | Main Findings |
|---|---|---|---|---|
| [328] | $m$ BSs and $k$ users | Energy efficiency | Particle swarm optimization | The proposed scheme outperformed benchmarks in energy efficiency. |
| [329] | $m$ BSs and $k$ users | Energy efficiency | Deep reinforcement learning | The simulation results depicts the superiority of proposed scheme over FDMA. |
| [330] | $m$ BSs and $k$ IoVT users (UL) | Minimize delay | Game theory | MEC server with more computation resources will serve more IoVT devices. |
| [331] | One macro BS, $w$ APs with MEC servers, and $k$ users (UL) | System cost | Heuristic | Proposed algo. decreased the task processing cost under the premise of ensuring the maximum task processing delay requirement. |
| [332] | One BS and $k$ users (DL) | Min. transmit energy | Greedy algorithm | Proposed scheme consumes 3-10 dB less energy in a MEC network of 256 IoT devices. |
| [333] | One MBS, one SBS, and $k$ users | Min. total cost | Hybrid-genetic hill-climbing algo | Offloading certain tasks to the MEC server for processing reduces total cost. |
| [334] | One BS and $k$ single antenna users | Energy efficiency | Successive optimization algorithm | The simulation results verify the Pareto-optimality of the obtained solution. |
| [335] | One MBS, $m$ SBSs and $k$ users | Energy efficiency | Deep Q learning | The proposed algorithm outperformed benchmark schemes. |
| [336] | One BS and $k$ single antenna users | Energy efficiency | Slicing-aware clustering and RA algo | Network slicing technique reduces the unnecessary allocation of RBs for NOMA. |
| [337] | One BS and $k$ users | Energy efficiency | Matching theory | The proposed algorithm achieves near-optimal performance with low complexity. |
| [338] | One BS and $k$ single antenna users | Energy efficiency | Bisection search iterative algorithm | The proposed iterative algorithm matches the derived optimal solution. |

Multicarrier-NOMA based MEC framework for the downlink IoT networks was proposed in [332]. Next, a comprehensive framework was developed to minimize energy consumption in downlink transmission while considering latency and power constraints. This involved optimizing various factors such as SIC decoding order, power control, frequency blocks, computation resource allocation, and user cluster scheduling. Next, the authors provided an efficient greedy algorithm for allocating frequency blocks and a min-max issue solution for computational resources. They used a potential game to explain the MC-NOMA MEC network and found a suboptimal IoT device cluster scheduling solution. An alternating optimization algorithm was used to design power control, resource allocation, and cluster scheduling together. In [333], the authors explored the task offloading of MEC-NOMA networks in small cells. To decrease the costs associated with task offloading, a problem was devised to reduce the energy and delay, taking into account their respective weights. A hybrid genetic hill-climbing algorithm was eventually developed to solve the formulated problem.

A multi-objective optimization problem was formulated in [334] to minimize the energy consumption for MEC offloading in the hybrid-NOMA-MEC networks. Furthermore, a low-complexity resource allocation solution was derived and shown to be Pareto-optimal. In [335], the authors developed a NOMA-MEC model for the HetNets to optimize the offloading decisions, base station selection, and channel resource allocation. DQN was deployed to efficiently assign the resources in the developed framework. In [336], a network slicing technique for NOMA-MEC network was proposed, which can reduce both service latency and unnecessary allocation of resources. In [337], the authors focused on resource allocation, i.e., the joint power, time allocation, and user grouping, in a hybrid NOMA-MEC network to reduce the weighted sum of energy consumption and delay.

The task delay minimization problem for the NOMA-MEC network was investigated in [338]. The optimization problem was formulated to minimize the task delay, which is a nonconvex problem. Later, the formulated optimization problem was transformed into a quasi-convex form. The bisection searching-based algorithm was proposed to find the globally optimal solution. Most of the existing works focused on conventional optimization techniques, as discussed above and also presented in Table XIII. Researchers can further explore intelligent algorithms, such as RL and DL, to design more efficient algorithms for adapting to the heterogeneous environment and can guarantee reduced latency and cost.

### J. Integrated Sensing and Communications

Future wireless networks need to prioritize sustainability, high resource efficiency, and intelligent environment awareness. In the past few decades, research on environmental sensing and wireless communication has adopted different spectral and signalling resources in two separate ways. Integrating sensing capability with communications is possible now owing to developments in communications and signal processing techniques, which allow for more reliable communications and sensing hardware through novel waveform design, transmission tactics, and resource allocation. ISAC, in which radar sensing and wireless communications are combined to share the same spectrum and equipment, has become increasingly popular in both industries and academia [348], [349].

Wireless sensing in communications networks is challenging. Coexistence, scheduling, and interference concerns will arise when wireless sensing applications, and consequently signals, will increase network traffic [350]. Despite the fact that ISAC holds a lot of potential for 6G, it is difficult to design due to the need to strike a suitable performance tradeoff between the two functions. Because of the inherent nature of the hardware platform and radio resource sharing, ISAC may experience severe inter-functionality interference. This necessitates research into effective methods for reducing interference and optimizing the use of the available resources [351].

NOMA is a better solution than OMA as SIC fits the ISAC scenarios. There are existing works on the fusion of PD-NOMA into ISAC to improve spectral and energy efficiency to ensure reliable and greener aspects of communications [340]–[342]. Fig. 18 shows both the uplink and downlink of the NOMA-assisted ISAC communications framework. Implementing NOMA into ISAC networks results in several obvious benefits. First, NOMA helps to increase spectrum efficiency by dividing each resource block to serve several users. Because the SIC scheme has been studied for some time, we also have a solid foundation for conducting the fundamental analysis necessary to effectively utilize NOMA in ISAC networks. Additionally, for traditional ISAC networks, the BS may require expected radar echoes to boost computation accuracy, increasing the variance in the powers of the communications and radar signals. In the case of power allocation, we can design the transmit powers of the users (for communications signals) and the BS (for radar signals) to improve the accuracy of radar prediction and provide ISAC networks with a new degree of freedom. ISAC is a new technology, and its integration with NOMA also brings some new challenges. Low-complexity and efficient resource allocation schemes should be designed to ensure swift and reliable communications and sensing while maintaining the user's data integrity and privacy.

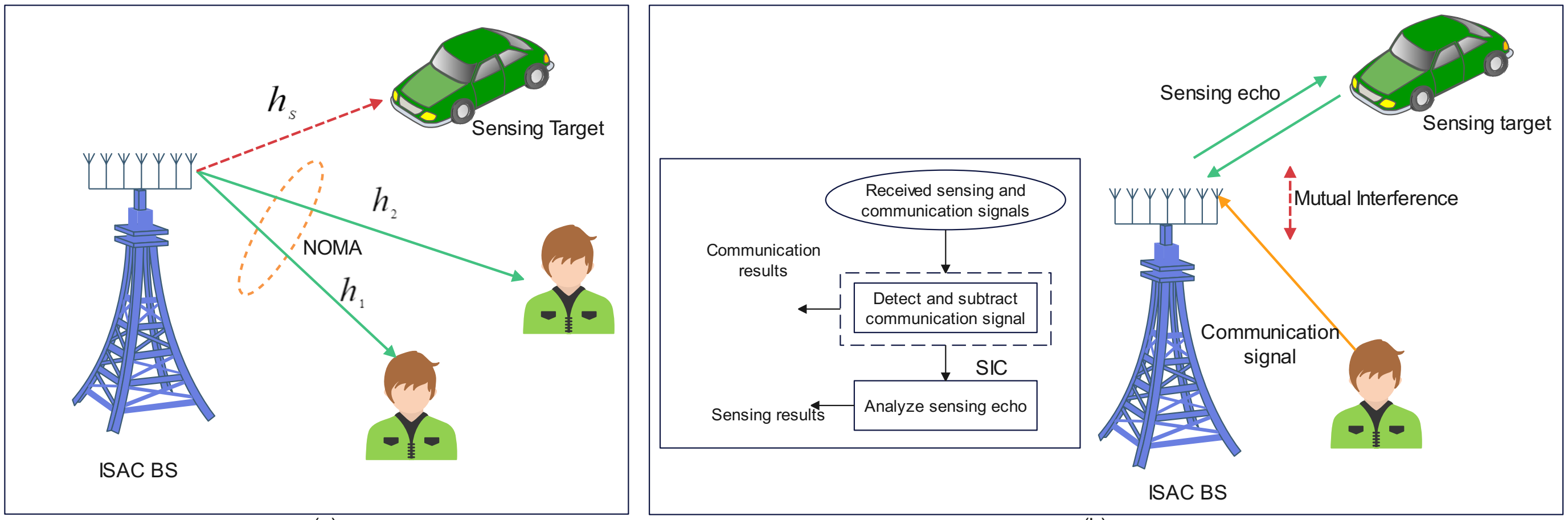

(a) (b)

Fig. 18: (a) NOMA-inspired integrated sensing and communications. (b) NOMA-inspired uplink integrated sensing and communications.

TABLE XIV: STATE-OF-THE-ART RESEARCH WORK ON INTEGRATED SENSING AND COMMUNICATIONS. (All APs, SBS, and MBS are mentioned as BS. IoT devices and UEs are mentioned as users)

| Ref. | System Model | Design Objective | Main Findings | Main Findings |
|---|---|---|---|---|
| [340] | Semi-ISAC-NOMA | Outage probability | Analytical | Authors derived the analytical expressions for ergodic radar estimation information rate. |
| [341] | One BS and $k$ users | Beam pattern | Successive convex approximation | The numerical results show superiority of ISAC-NOMA over conventional ISAC. |
| [342] | One BS and one user | Interference | Iterative algorithm | The proposed multi-domain NOMA-ISAC scheme has significant advantages in efficiency. |
| [343] | One BS and $k$ users | Max. weighted sum | Successive convex approximation | Proposed scheme outperforms the conventional ISAC in the underloaded and overloaded regimes with highly correlated channels. |
| [344] | Multi-antenna NOMA-ISAC | Sum secrecy rate | Successive convex approximation | The proposed secure NOMA-ISAC scheme ensures secure transmission. |
| [345] | RIS assisted NOMA-ISAC | Beam pattern | Successive convex approximation | A joint optimization problem over active beamforming, power allocation coefficients, and passive beamforming was formulated. |
| [346] | One BS, $k$ users (FD-NOMA ISAC)(UL&DL) | Max. sensing SINR | Semidefinite relaxation approach | The proposed framework demonstrates the tradeoff between sensing and communication. |
| [347] | One BS, $k$ users, one radar target | Max. sum secrecy rate | Iterative algorithm | The proposed scheme outperformed the benchmark schemes. |

Table XIV shows the most recent work on ISAC-NOMA. In [343], the authors developed the NOMA-empowered ISAC framework. An optimization problem was formulated to maximize the weighted sum of the communications throughput and the effective sensing power. Later, an efficient algorithm based on SCA was utilized to solve the optimization problem. In [344], a joint optimization problem was formulated for the NOMA-ISAC network using artificial jamming. The objective was to maximize the sum secrecy rate while satisfying various constraints, which include the demodulation order, SINR constraint for NOMA, rate thresholds for NOMA users, sum transmit power constraint, and echo signal SNR requirement for sensing. In [345], the authors formulated the problem of joint optimization of active beamforming, power allocation coefficients, and passive beamforming in the RIS-NOMA-ISAC network. The objective was to maximize the minimum beampattern gain. To solve the formulated nonconvex problem, an alternating optimization-based algorithm was proposed.

In [346], a beamforming optimization problem was formulated for the FD-NOMA-ISCA network. The objective was to maximize the sensing signal-to-interference-plus-noise ratio while ensuring that the downlink SINR requirement for each downlink user was met. The formulated non-convex problem was solved using SDP, and the numerical results were presented to showcase the performance improvement achieved in the FD-NOMA-ISAC network. In [347], the authors formulated a joint precoding optimization problem for the NOMA-ISAC network with the aim of maximizing the sum secrecy rate for multiple users via artificial jamming. The formulated non-convex problem was transformed into a convex one using the SCA, and then an iterative algorithm was proposed to solve the original optimization problem.

TABLE XV: STATE-OF-THE-ART RESEARCH WORK ON VISIBLE LIGHT COMMUNICATIONS.
(All APs, SBS, and MBS are mentioned as BS. IoT devices and UEs are mentioned as users)

| Ref. | System Model | Design Objective | Main Findings | Main Findings |
|---|---|---|---|---|
| [352] | $m$ BS, $k$ users (DL) | Acheiveable throughput | Analytical | No.of handovers can be significantly decreased with the tunable field of views strategy. |
| [353] | One BS and two users (DL) | Max. sum rate | Harris hawks optimization | Proposed algorithm outperformed alternative schemes and metaheuristic algorithms. |
| [354] | One BS and two users (DL) | Bit error rate | Analytical | BER gap among users decreases with the increase of the modulation order. |
| [355] | $m$ BS, $k$ users and one eavesdropper (DL) | Sum secrecy rate | Iterative algorithms | This paper studied convergence and complexity analysis of the proposed algorithms. |
| [356] | One BS, $k$ users and one optical-RIS (DL) | Achievable sum rate | Iterative algorithm | Optical intelligent reflecting surface improves the sum rate of NOMA-VLC system. |
| [357] | One BS and $k$ users(DL) | Sum rate | Geometric programming | This paper investigate the QoS-guaranteed power allocation within each cell. |
| [358] | One VLC AP and $k$ users (DL) | Power allocation | Deep Q learning | Algorithm performed better than other schemes by achieving faster convergence times. |

## K. Visible Light Communications

Visible light communications using light-emitting diodes (LEDs) has become an exciting technology in optical wireless communications and has recently received much attention from both industry and academia because of the multitude of offered benefits, such as simple installation, high bandwidth, low power consumption, and improved security [359]. It has been anticipated that VLC will show itself to be a successful technique for enhancing spectral efficiency while also providing support for massive connectivity [360]. The VLC spectrum spans from 430 to 790 terahertz and has an excessive amount of spectrum resources that are currently underutilized [361]. Many works have utilized different techniques to improve spectrum efficiency and reduce interference among LEDs to boost the performance of VLC.

Table XV captures the latest ongoing research summary on the interplay of NOMA and VLC. A typical indoor NOMA-assisted VLC for IoRT is shown in Fig. 19. NOMA is gaining ground as a potential future multiple-access technique for future wireless networks. It has the potential to enhance both spectrum efficiency and connectivity significantly. VLC and NOMA studies have recently garnered considerable interest. The first integration of NOMA with downlink VLC was examined in [352] to increase the achievable throughput. In [353], a joint problem of power allocation and UAV's placement with the aim of maximizing the sum rate of all users was formulated for UAV-assisted NOMA-VLC network. The formulated problem was solved by using the swarm algorithm. The proposed algorithm funded the solution for UAV's placement and power allocation simultaneously. In [354], the authors proposed an analytical framework for the NOMA-VLC network, which was based on the bitwise-decision axis, and signal space allows for obtaining closed-form BER expressions. To tackle the increased analysis complexity challenge, the authors proposed a novel analysis method that successfully decreased the complexity. Also, the closed-form expressions for the different modulation schemes were derived.

The secrecy performance of NOMA-inspired VLC network in the presence of eavesdropper was investigated in [355]. In [356], the authors investigated the optical RIS-VLC network employing the NOMA, and the achievable sum rate was maximized by optimizing the optical RIS reflection matrix. In [357], a user grouping mechanism based on the user location was proposed to reduce the interference from SIC for multi-cell NOMA-VLC. In [358], the authors formulated an optimization problem that considers both power allocation and LED transmission angle tuning. The objective is to maximize the average sum rate and the average energy efficiency. According to the aforementioned research works, it is evident that VLC systems that use NOMA can provide greater spectral efficiency than their OMA counterparts.

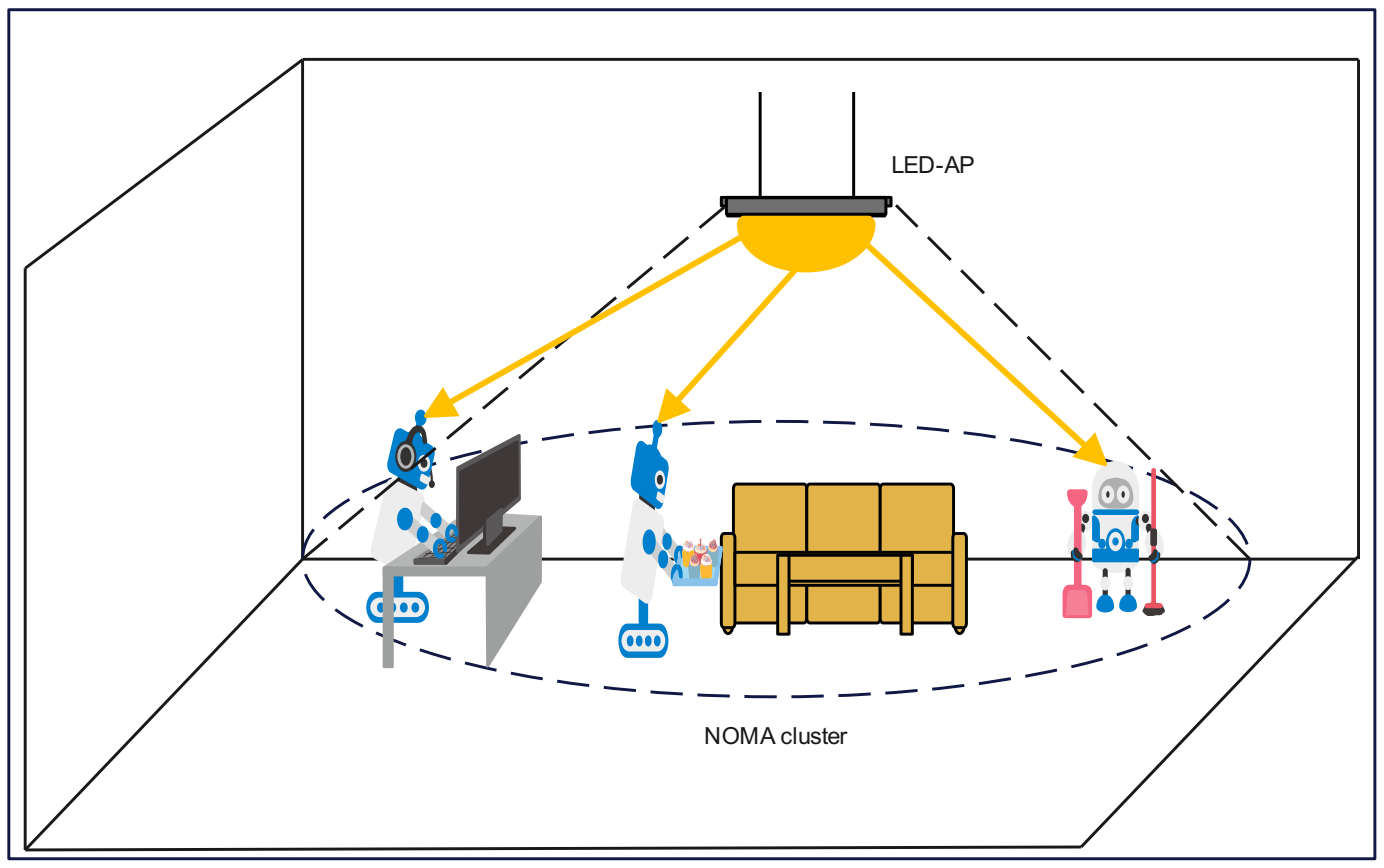

Fig. 19: NOMA assisted VLC.

## L. Non-Terrestrial Networks

Non-Terrestrial networks (NTNs), which include UAVs, high altitude platforms (HAPS), and satellite networks, are traditionally used for certain applications such as disaster management, navigation, television broadcasting, and remote sensing [362], [363]. Furthermore, the transition from 5G to 6G will result in an explosion of IoT devices that provide pervasive and constant connection across all spheres of human activities. In this regard, NTN will play a crucial role in supporting and supplementing terrestrial systems in order to cope with such a vast number of IoT devices and to fulfil the massive capacity needs of the most sophisticated of them [364]. NTN's contribution to future communications infrastructure is founded on a number of major benefits [365]. First, space and aerial platforms provide wider coverage than ground-based platforms. Second, NTNs can be deployed faster than ground-based infrastructure and provide faster responses to important applications such as emergency communications. Third, NTNs can easily manage huge data streams and deliver high bandwidths, making it excellent for 4K video streaming, virtual reality experiences, and other data-intensive workloads.

On the other side, NOMA enables users to utilize resources on a subcarrier concurrently in a non-orthogonal manner by allocating codes or sending data concurrently on the same frequency at different power levels. NOMA has already proven its efficiency in terrestrial systems [258], [286], [299]. Now, researchers are considering applying NOMA in the NTNs [257], [366]–[368]. Table XVI captures the latest ongoing research on the efficacy of NOMA in NTNs. Fig. 20 depicts the UAV-enabled BS communicating with multiple ground users in the NOMA way, where users have heterogeneous QoS requirements.

TABLE XVI: STATE-OF-THE-ART RESEARCH WORK ON NON-TERRESTRIAL NETWORKS.
(All APs, SBS, and MBS are mentioned as BS. IoT devices and UEs are mentioned as users)

| Ref. | System Model | Design Objective | Main Findings | Main Findings |
|---|---|---|---|---|
| [257] | One UAV-BS and $k$ users (DL) | Energy efficiency | Iterative algorithm | Proposed scheme achieved three times higher energy efficiency than benchmarks. |
| [366] | Satellite BS, $k$ users, $r$ relays (DL) | Outage Probability | Analytical | Power allocation factor plays a key role in improved outage probability. |
| [367] | Satellite BS, $k$ users (DL) | Ergodic sum rate | Analytical | The system performance can be impacted by factors such as channel estimation error and residual interference. |
| [368] | $m$ UAV-BS and $k$ users (UL) | NOMA user pairing | Block coordinate descent | Proposed scheme provides the tradeoff between UAV flight time and total energy. |
| [369] | $m$ UAV-BS and $k$ users | User clustering | K-means clustering and SCA | Proposed schemes are sub-optimal with less computational complexity. |
| [370] | One UAV-BS and $k$ users (UL) | Weighted sum rate | Successive convex optimization | The algorithm outperformed benchmark systems in throughput. |
| [371] | One UAV-BS and $k$ users (UL) | Max. sum rate | Reinforcement learning | Proposed algorithm provides an online solution to resource allocation. |
| [372] | One UAV-BS and $k$ gateway clusters | Energy efficiency | Geometric programming | Proposed an algorithm to group the gateways into multiple gateway clusters. |
| [373] | One UAV-BS and 2 users (DL) | Power allocation | Successive geometric programming approximation | The optimal hovering height and power allocation are provided with numerical results. |
| [374] | One UAV-BS and $k$ users (DL) | Sum rate | Iterative algorithm | The secrecy rate depend on the transmit powers of users. |
| [375] | One BS and $k$ users | Max-min rate | Inner convex optimizations | Jointly optimizing all parameters yields higher rate gain than optimizing subsets. |

Applying NOMA to UAVs can offer numerous advantages over terrestrial systems [369]. First, AUs have the advantage of their high altitude, which often results in different propagation conditions than terrestrial users (TUs). This can lead to higher probabilities of experiencing stronger communication links. These unique channel characteristics offer an exciting opportunity to enhance the overall system performance by utilizing NOMA to achieve a larger rate region. Furthermore, NOMA can enhance system performance with AUs moving in 3D space. Secondly, the downlink can efficiently cater to the higher data rate needs of TUs for multimedia applications while ensuring reliable connectivity for AUs with lower data rate requirements for command and control links by exploiting their asymmetricity. Third, in the uplink, AUs have the potential to achieve a larger macro-diversity than TUs, which can lead to increased degrees of freedom for the system [376], [377]. This feature has great potential to enhance overall performance by enabling AUs to utilize multiple base stations for cooperative SIC or apply precoding to improve throughput or fairness [378].

An adaptive RL-based framework was developed for the NOMA-UAV network to maximize the sum rate in [370]. In [371], the authors formulated a joint optimization problem of 3D trajectory design and power allocation to maximize the throughput. A K-means-based clustering algorithm was developed to partition the user periodically, and then a mutual DQN algorithm was developed to solve the formulated problem. An iterative optimization algorithm based on the geometric programming-based algorithm was developed to optimize the UAV-BS energy efficiency in the NOMA-UAV network in [372]. A signomial programming problem was formulated to optimize the hovering height and power allocation of UAV in NOMA-UAV network in [373], and to solve the optimization problem a successive geometric programming approximation algorithm was developed. In [374], the authors proposed two secure transmission schemes for UAV-NOMA networks, including a solution for a single user and another for multiple users, leveraging UAV positioning, power allocation, and beamforming techniques. In [375], a path-following algorithm was proposed to solve the max-min rate optimization problem under total power, total bandwidth, UAV altitude and antenna beamwidth constraints in the NOMA-UAV network.

Although researchers have shown interest in NOMA-UAV-based communications networks as shown in the aforementioned literature however, despite its benefits, NOMA in a 3D aerial environment faces several challenges as well, including selecting the best SIC order, devising the most efficient power allocation scheme, swift and reliable user pairing, minimizing ICI, and incorporating MIMO-NOMA to boost overall performance [365], [379], [380]. These challenges stem from the aerial channel's peculiarities, such as its dynamic and non-stationary channel statistics, severe ICI, and LOS/NLOS links, which influence small-scale and large-scale fading [381]. In addition, NOMA's performance is constrained by imperfect SIC and partial CSI, calling for novel approaches and solutions adapted to the aerial settings. Therefore, overcoming these challenges is critical to experiencing the full potential of NOMA in UAVs-assisted NGWNs.

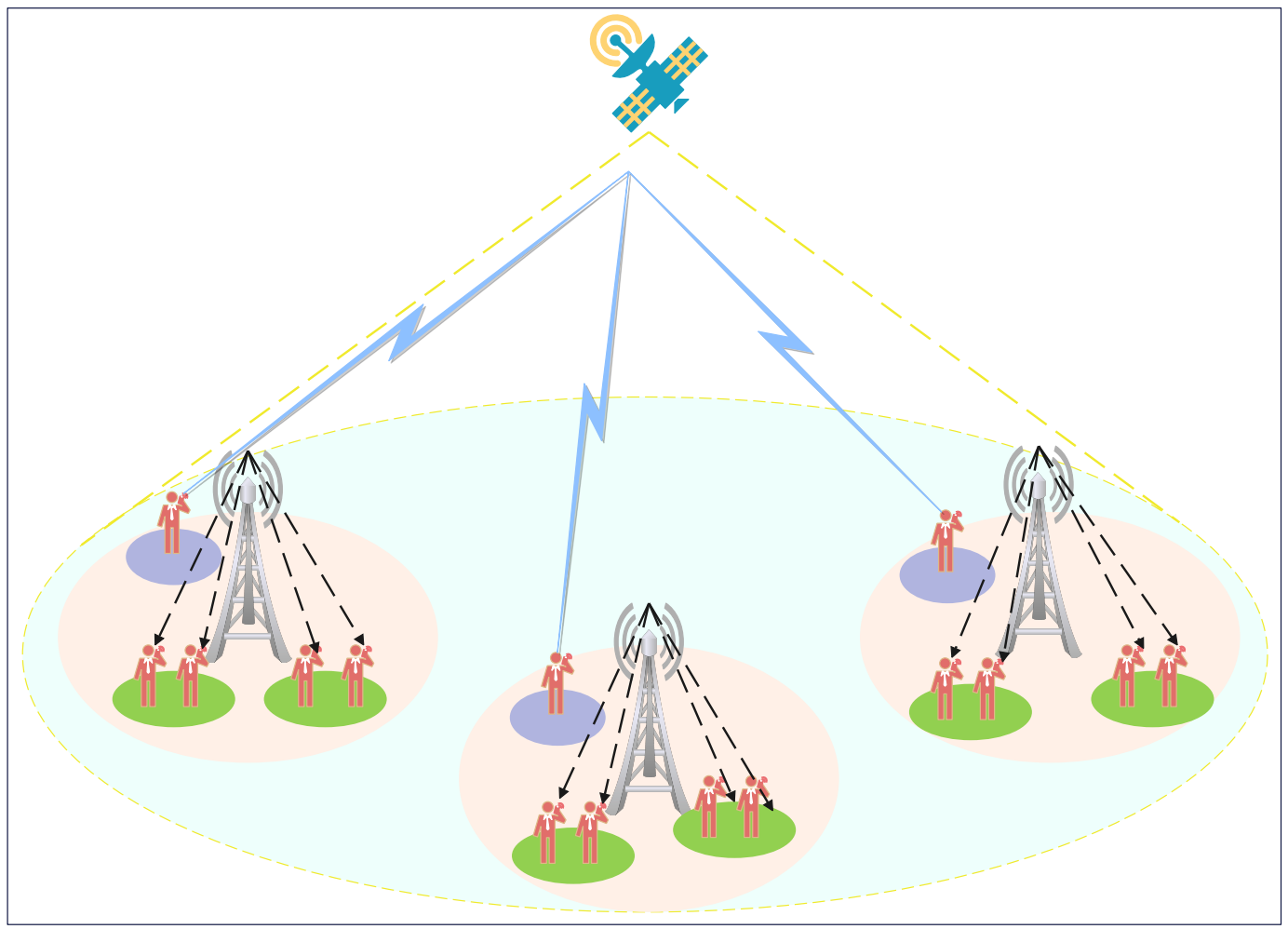

Fig. 20: NOMA-based NTN.

## IV. Trends

The IoE is a pivotal force in propelling sustainability initiatives forward. By facilitating the implementation of cutting-edge and economical technological solutions in a range of sectors and industries, including manufacturing, transportation, and supply chains, the IoE is helping to pave the way toward a more sustainable world. Achieving a sustainable and interconnected ecosystem on a global scale is contingent upon advancing energy-efficient and reliable multiple-access techniques. These techniques play crucial roles in fulfilling the limitless connectivity potential of the 6G network. This section is designed to showcase the recent trends of NOMA for the forthcoming 6G covering the Grant-Free NOMA in Subsection IV-A; satellite communications in Subsection IV-B; 6G space-air-ground-integrated networks in Subsection IV-C; simultaneous transmitting and reflecting surfaces in Subsection IV-D; semantic communications in Subsection IV-E; synergy of NOMA and green communications in the 6G era in Subsection IV-F; orthogonal time-frequency space modulation in Subsection IV-G; digital twin networks in Subsection IV-H; Tactile Internet in Subsection IV-I; and zero-touch networks in Subsection IV-J.

### A. Grant-Free NOMA (GF-NOMA)

NGWNs are anticipated to fulfill the rigorous QoS demands of the three main network usage scenarios, namely eMBB, mMTC, and URLLC [382]. eMBB is primarily driven by human-type communications (HTC). The proliferation of IoT devices necessitates cellular networks to support massive connectivity, hence imposing reliability and latency requirements beyond human expectations. In existing cellular networks, radio resources are allocated to users/devices using contention-based mechanisms. This allocation is done orthogonally through the physical random access channel (PRACH), where a four-way handshake is performed by the base station [383]. The four-way handshake fulfils multiple purposes, including the initial access of newly admitted users, uplink synchronization of users, transmitting data, acknowledgement response, and handover management, among others [384]. While the signalling overhead associated with the grant-based (GB) method may have minimal impact on eMBB, it is acknowledged as a significant cause of excessive delay and a primary performance limitation in mMTC. This is particularly evident in scenarios where a large number of IoT devices intermittently transmit short packets [385].

NOMA, unlike OMA, is the robust and spectrally efficient multiple access technique to enable numerous users or IoT devices on the same RB, boosting spectral efficiency, huge connection, and limitless access over a limited number of RBs. As previously stated, NOMA has garnered significant interest from both industry and academics [386]. In the context of PD-NOMA, it is necessary for users to have the CSI of other users who share the same RB. It is worth noting that PD-NOMA introduces a considerable amount of signalling overhead, particularly in the UL direction. The non-convex and combinatorial nature of power control and clustering makes PD-NOMA difficult to implement in a real network, even if the CSI is known. Due to this particular rationale, there has been a recent surge of interest in grant-free-NOMA (GF-NOMA). This approach aims to render NOMA more feasible by using simplistic or no power control mechanisms, hence enabling devices to transmit data as needed without the need for scheduling requests.

The rationale for GF-NOMA for the massive connectivity regime is as follows: (1) The omission of the "asking-for-grant" approach for packet transmission aims at boosting effective throughput and minimizing transmission latency. Grant-free access should be deemed necessary in situations when a network is tasked with accommodating a substantial volume of users, as is the prevailing circumstance in the context of the IoT. In these circumstances, a grant-based transmission would result in reduced spectral efficiency due to excessive signalling overhead and transmission delays. (2) Achieving higher spectral efficiencies than OMA thanks to the superimposition of several packet transmissions over the same bandwidth and collision resolution through powerful MUD techniques at the demodulator. Authors in [185] applied DRL in the decision-making for GF-NOMA systems to mitigate collisions and improve the system throughput in an unknown network environment. In order to mitigate collisions in the frequency domain and minimize the computational complexity of DRL, subchannel and device clustering was developed. This approach involves grouping devices together into clusters, where a cluster of devices compete for a cluster of subchannels following GF-NOMA. In addition, the authors in [387] introduced a hybrid grant (HG) NOMA scheme in order to mitigate the challenges associated with signalling overhead and power collision. The primary goal of this approach was to attain a fair trade-off between signalling overhead, access failure probability, and access time. Along with that, the authors in [388] devised an innovative approach that treats collisions as interference with the remaining received signals. In addition, the authors in [388] employed Poisson point processes and ordered statistics to derive simple expressions for approximating the outage probability and throughput of the system in the context of successive joint decoding (SJD) and SIC. In general, GF-NOMA is an important factor in enabling massive access in the forthcoming NGWNs and serving the massive number of tiny IoT devices.

### B. Satellite Communications

Satellite communications make it possible for ground users, passengers sitting in the aeroplane, and astronauts in space to connect. Satellite communication can provide coverage in remote or relatively unpopulated regions in which terrestrial networks are costly and not doable. This is especially important for achieving global connectivity, such as for IoT devices and smart cities, where stable and everywhere connectivity is key. Satellite communications can rapidly connect mobile and remote users with the least latency. This is important for many 6G applications, like high-definition video streaming, VR/AR, CAVs, telemedicine, and more. Third, satellite communications can help in emergency services and disaster response, where rapid and reliable communication is needed to save lives. Lastly, satellite communication can be an important alternative for terrestrial networks in case of natural disasters or other network breakdowns. This keeps communication services running even when terrestrial networks are down.

Recently, researchers have carried out various works in utilizing NOMA in satellite communications [56], [389]. Satellite communications systems utilizing NOMA can offer low-latency, spectral-efficient, and energy-efficient communications, which is critical for the aforementioned technologies. NOMA can optimize the performance of satellite communications systems by enabling multiple users to share the same RB. Furthermore, NOMA can potentially enhance energy efficiency in satellite communications systems through power-efficient transmission. In general, NOMA has the potential to serve as a key MA technique for enhancing the performance and energy efficiency of satellite communications systems.

### *C. 6G Space-Air-Ground Integrated Networks*

6G will lead to new IoT and AI services that will be harnessed on social and environmental concerns producing a large-scale sustainable impact and game-changing technological improvements in society. Aiming to give an unparalleled user experience via seamless hyper-connectivity, the convergence of satellite and terrestrial communication will focus on making communication unlimited, ultra-fast, and everywhere.

The space-air-ground integrated network, which combines mega-constellation satellites, aerial platforms, and terrestrial networks, has gained a lot of interest as a game-changer technology of 6G communications because of its capability of accommodating ubiquitous access and flexible deployment. Broadband communications between satellites and ground users is becoming increasingly important as low-earth-orbit (LEO) constellations continue to evolve fast in the quest for greater bandwidth and improved QoS. UAV-based satellite terminals have been viewed as a viable solution for satellite access for ground users, owing to the benefits of broader coverage and fast deployment brought by their natural mobility, particularly in scenarios such as rural areas, emergency deployment, disaster relief, and so on. In November 2020, the first 6G experimental satellite developed by the University of Electronic Science and Technology of China was successfully launched [12]. Its task is to test communications from space using a high-frequency terahertz spectrum.

UAV communications improve coverage and link quality, but massive connectivity communications efficiency remains an issue. Thus, B5G/6G radio access architecture relies on NOMA to improve spectrum efficiency and provide access for an enormous number of users. The combination of NOMA and space-air-ground integrated networks is expected to increase system throughput, especially in multi-user scenarios. NOMA becomes more advantageous when the channels between the transmitter and multiple receivers are distinct, as is typically the case in space-air-ground integrated networks. The authors in [390] proposed an energy-efficient UAV-enabled space-air-ground integrated network, whereby a phased-array antenna is mounted on the UAV to enhance the satellite's signal, and NOMA is applied to UAV-to-ground communications. Because of NOMA's capability to support multiple users in the same resource blocks, implementing its functionality in a space-air-ground integrated network would open up new avenues of opportunity, particularly in HetNets, E-health, and autonomous vehicles.

[12]https://www.jagranjosh.com/current-affairs/china-launches-worlds-first-6g-experimental-satellite-1605166418-1

### *D. Simultaneous Transmitting and Reflecting-Reconfigurable Intelligent Surfaces (STAR-RIS)*

STAR-RIS is an emerging technology that uses reconfigurable intelligent surfaces to simultaneously transmit and reflect signals from multiple sources, enabling more efficient and flexible use of spectrum resources. NOMA and STAR-RIS, together, can further improve the performance of STAR-RIS by enabling multiple users to transmit and receive signals simultaneously while optimizing the resource allocation and power management for each user [206]. NOMA-inspired STAR-RIS has the potential to enhance spectral efficiency, expand coverage, and increase network capacity, making it useful in a variety of scenarios. More efficient and affordable network connectivity is possible because of NOMA's ability to provide seamless communications between STAR-RIS and other wireless technologies, including CAVs, UAVs, and the IoT. We noted that most trending research on the STAR-RIS-assisted NOMA includes the integration of STAR-RIS and GF-NOMA [391], joint beamforming optimization for STAR-RIS aided NOMA ISAC systems [392], and the adaptation of STAR-RIS-aided NOMA in the URLLC systems [393]. Also, researchers are actively deploying ML-based solutions to optimize the performance of STAR-RIS-aided NOMA networks [394], [395].

### *E. Semantic Communications*

With the rapid breakthrough of emerging semantic communications and the steady development of the currently employed bit-based communications, future networks are expected to simultaneously support both the semantic and bit transmissions to provide ubiquitous, customized, and intelligent connectivity among different types of devices (e.g., human, machine, and their interactions). However, since the available radio resources are limited, one of the most fundamental problems is designing an efficient MA scheme to facilitate the heterogeneous semantic and bit transmission for multi-user communications. In contrast to conventional bit-based communications systems, the main idea of semantic communications is to transmit the semantic meaning contained in the source data. By doing so, the source data can be dramatically compressed, and the required communication resources can be significantly reduced [396].

Given the significant breakthrough of artificial intelligence in semantic information extraction, many efficient approaches have been developed for facilitating semantic communications, such as text/speech/image/video transmission, image classification, machine translation, and visual question answering. Semantic communications-enhanced NOMA can provide sustainability, reliability, and intelligence in next-generation wireless networks. Researchers from both academia and industry are working on the semantic-empowered NOMA. For example, in [397] a novel channel-transferable semantic communications (CT-SemCom) framework is proposed for OFDM-NOMA systems, which adapts the codecs learned on one type of channel to other types of channels. In [398], authors proposed an innovative two-user semantic-empowered NOMA scheme for resource efficiency management.

### F. Synergy of NOMA and Green Communication in the 6G Era

6G wireless networks will integrate space, aerial, terrestrial, and submarine networks for three-dimensional communication. The goal is to provide ubiquitous and unlimited connectivity to a massive number of IoT and machine-type devices with diverse quality of service requirements, support substantial and heterogeneous traffic demands, and reduce energy consumption with highly energy-efficient communications protocols, transceivers, and computing technologies. Worries about the extraordinarily high energy usage, which worsens the greenhouse effect, are prompted by the massive IoT devices. This calls for "green communications", whereby the next generation of massive IoT networks needs to find a win−win solution between low energy consumption and high data throughput [399]–[404]. Green communications refers to the practice of employing computers and other information technology tools in ways that minimize their impact on the environment.

NOMA has been commonly regarded as an important candidate for ensuring the massive connectivity needs in IoT networks, as it enables multiple devices to connect to the network using the same time-frequency resource. This reduces the need for multiple transmissions, which can save energy. Additionally, NOMA can prioritize transmissions from devices with weaker signals, which can further save energy by reducing the need for retransmissions. Extensive academic research has been dedicated to exploring the viability of NOMA as a multiple-access technique that promotes energy efficiency in NGWN [246], [405]–[407]. Given its potential to increase spectrum efficiency, NOMA has also been incorporated into the 3GPP long-term evolution advanced standard. Also, the NOMA scheme can be combined with many other technologies to meet the needs of future 6G networks in terms of high spectral efficiency, very low latency, massive device connectivity, ultra−high reliability, excellent user fairness, high throughput, supporting different QoS, low energy consumption, and low operating costs [35], [408]. Green NOMA or energy efficient NOMA is a potential candidate for the next-generation 6G mobile network that can support the mission of building a cheap, self-sufficient, environmentally friendly wireless network [13].

### G. Orthogonal Time Frequency Space Modulation

The forthcoming intelligent digital society is anticipated to rely heavily on reliable connectivity in high-mobility scenarios, particularly at high carrier frequencies. The anticipated outcome of 6G wireless networks is to offer global coverage comprising terrestrial and non-terrestrial networks. Orthogonal time-frequency space (OTFS) is a newly developed modulation technique that has been proposed as a potential solution for high-mobility communications [409].

The OTFS modulation technique utilizes the delay-Doppler (DD) domain for information modulation instead of the conventional time-frequency (TF) domain as in OFDM systems. This feature offers robust delay and Doppler resilience while leveraging the benefits of full diversity, which is essential for ensuring reliable communications [410], [411]. The OTFS technique involves the mapping of data symbols into the DD domain. Each symbol has the capability to be expanded completely in the TF domain. This enables the channel to have a similar impact on every data symbol, effectively minimizing fading and interference. To clarify, this particular technique transforms channels that exhibit fading in the time-frequency domain into a relatively constant channel in the DD domain. In highly dynamic situations, OTFS minimizes interference with information symbols to increase reliability and accuracy.

We noted that most existing research on NOMA considered low mobility or static UEs [412], [413]. Recently, OTFS modulation has been considered in cases featuring double-dispersed channels. Compared to OFDM, the time-invariant channel gains of OTFS in the DD plane facilitate channel estimation and signal detection under high mobility scenarios. In [414], the authors proposed an OTFS-NOMA scheme in which a single high-mobility user employing OTFS in the delay-Doppler domain paired with a group of low-mobility OFDM users for non-orthogonal channel sharing.

### H. Digital Twin Networks

Digital twin networks (DTNs) are a remarkable innovation that allows the creation of virtual models of physical objects or systems. This approach involves the creation of an emulated software replica of the mobile network, which facilitates ongoing development, testing, and optimization processes. The DTNs for NOMA-driven NGWNs can significantly benefit the development and operational stages of upcoming 6G applications and technologies due to the scale, complexity, and high impact of their downtime and failures [415]. Mobile network providers can utilize their assistance in predicting and evaluating network events, conducting testing procedures, providing network upgrades, and doing other related tasks. DTNs have the potential to enhance the adoption of advanced functionalities and services, such as the integration of artificial AI with NOMA-assisted NGWNs. DTNs are a valuable tool for the development of AI algorithms in circumstances when real networks are unable to provide sufficient real data or when it is deemed hazardous to apply AI algorithms directly to real networks. Furthermore, considering the continuous advancements in 5G and the anticipated emergence of 6G, future NGWNs are expected to be denser and necessitate a swift response. It is crucial to note that even little alterations within the network might potentially trigger a domino effect within a compressed timeframe. In [415], authors integrated DTN and MEC in the NOMA-based industrial IoT networks to minimize the total task completion delay of industrial IoT devices by jointly optimizing the industrial gateway's subchannel assignment as well as the computation capacity allocation, edge association, and transmit power allocation of industrial IoT device. In [416], authors proposed the joint optimization for the NOMA and multi-tier hybrid cloud-edge computing empowered industrial IoT that results in improved utilization of communications and computing resources. The successful deployment of DTN technology has the potential to significantly enhance the performance of NOMA-assisted NGWNs in the forthcoming 6G era and beyond. This advancement offers mobile network operators the ability to simulate diverse situations, analyze potential solutions, allow comprehensive network analysis, and optimize mobile network operations.

[13] *Green G: The path towards sustainable 6G-The Green G working group's true objective is to have North America take the reins as the sustainable environmental frontrunner in next-generation wireless networks.*

### *I. Tactile Internet*

The term "Tactile Internet" refers to the forthcoming generation of internet technology that will facilitate haptic communications and control among humans, machines, and devices. The objective is to offer connectivity with exceptionally low latency and high reliability, enabling immersive and interactive experiences like telepresence, remote surgery, and augmented reality. The Tactile Internet enables a surgeon to remotely operate a robot with haptic feedback, thereby facilitating a tactile experience during surgical procedures.

NOMA has the potential to contribute to the Tactile Internet's advancement. It can facilitate the optimal utilization of available resources, enhance mMTC, and enhance the reliability and security of communications channels. By leveraging NOMA, it is possible to unlock the full potential of the Tactile Internet, thereby enabling novel applications and use cases that can revolutionize industries and societies. NOMA and Tactile Internet-enabled framework can assist eMBB, URLLC, and mMTC to achieve low latency, high bandwidth, service availability, spectral efficiency, and end-to-end security. In [417], authors highlighted several NOMA and Tactile Internet-based solutions for 5G and beyond. The authors of [418] presented a queuing model for the Tactile Internet in the context of CRAN architecture. This was achieved by utilizing PD-NOMA to serve multiple pairs of tactile users. The findings indicate that implementing dynamic adjustments of access and fronthaul delays can result in significant power savings compared to OMA-based approaches.

### *J. Zero-Touch Networks*

Zero-touch networks aim to automate network infrastructure management and operation without any human intervention. Zero-touch networks use various technologies, including SDN, NFV, and AI. These technologies enhance the network infrastructure's flexibility, agility, and adaptability to accommodate evolving network conditions and requirements. The future of automated networks promises to revolutionize connectivity by delivering more efficient, intelligent, and self-managing systems, ensuring minimal human intervention while enhancing performance. NOMA, the key contestant for NGMA, can play a vital role in zero-touch networks. NOMA can be used for resource allocation in zero-touch networks. NOMA can also facilitate spectral efficient, reliable, and low-latency communications in zero-touch networks, which is critical for 6G applications such as CAVs, IoT, and more. By allowing multiple users to access the network simultaneously, NOMA can reduce latency and enhance performance. Moreover, NOMA can improve energy efficiency in zero-touch networks by enabling power-efficient transmission. By allocating power more efficiently to users, NOMA can reduce the overall energy consumption of the network. We noted that researchers are actively working on developing effective NOMA-based industrial automation systems [419], NOMA-based intelligent transportation systems [3], [372], NOMA-based smart healthcare [420], [421] to push the NOMA's applicability towards the new era of NGWNs.

## V. Research Challenges and Open Problems

Identifying and addressing the key challenges in the initial stages of NGMA development is of the utmost importance for strengthening the aim of accomplishing massive access. This section highlights the challenges we identified through a thorough literature review, as well as current trends in NOMA and its integration into NGMA-assisted NGWNs. The challenges were selected based on their potential to impede progress toward massive access, their importance in the ongoing evolution of NOMA, and their significant impact on the performance of NGWNs. The increasing demand for massive connectivity, minimal latency, and ubiquitous coverage has led us to concentrate on the challenges that significantly affect the scalability and efficiency of NOMA in this scenario. Specifically, subsection V-A presents CSI: Estimation and Acquistion in the NGMA era, user grouping and power allocation in subsection V-B, successive interference cancellation in subsection V-C, high-mobility users in subsection V-D, and signal superposition: security and privacy risks in subsection V-E. Addressing these challenges is essential for optimizing the performance of NGMA, thereby paving the way for a robust framework capable of supporting the future of NGWNs.

### *A. CSI: Estimation and Acquistion in the NGMA era*

In NOMA, one of the most important parameters is CSI. CSI is the knowledge of the channel characteristics between the transmitter and receiver, such as channel gains and fading. In NOMA, the transmitter assigns different power levels to different users based on their CSI to improve the system's capacity and reliability. NOMA uses sophisticated signal processing algorithms, which depend on precise CSI estimates, to differentiate the signals of different users and segregate them at the receiver. Estimating the CSI in NOMA systems can be accomplished using a variety of methodologies, such as pilot-based methods or blind estimation. In pilot-based methods, the transmitter is responsible for sending known pilot symbols to the receiver. The receiver is responsible for estimating the CSI based on the received pilots. In blind estimation, pilots are not used, and the receiver estimates the CSI based on the statistical aspects of the received signal.

Multiantenna techniques, such as massive MIMO and CF-mMIMO, present a promising avenue for enhancing forthcoming NGWN's data rate and capacity. Combining these techniques with NOMA gives rise to new opportunities and design challenges. In massive MIMO-NOMA, a large number of spatial degrees of freedom offered by massive MIMO can be utilized to cater to a greater number of users through power-domain multiplexing. The acquisition of precise CSI is of utmost importance for the optimal functioning of MIMO-NOMA. In the case of massive MIMO, the process of CSI acquisition necessitates significant channel training and feedback overhead, particularly for frequency-division duplex (FDD) systems. Most of the existing literature focused on the underloaded regime in the case of massive MIMO. However, forthcoming 6G-based NGWNs will require connecting the massive number of users, which means that spatial DoFs provided by massive MIMO may not be sufficient [27]. NOMA provides a spectrally efficient solution by serving multiple users on the same RB and enhancing the performance of massive MIMO networks in the overloaded regime.

NOMA-assisted multiantenna networks also bring challenges such as resource allocation, signalling issues, and security and privacy issues. In massive MIMO-NOMA, clustering algorithms are essential solutions for the issues of user pairing and clustering. ML-based k-means clustering algorithms have successfully paired users into clusters based on their channel conditions and correlations. Low complexity-based ML and matching theory-based techniques can be utilized to achieve effective user pairing. PA issues can be further researched through stable, convergent, and self-organizing solutions using game theory in overloaded massive MIMO systems. In scenarios where the acquisition of perfect CSI at the transmitter is critical, promising ML techniques can be adopted. Efficient CSI estimation techniques for supporting a large number of antennas at the BS can be achieved through the application of unsupervised or reinforcement learning ML models, which can utilize the vast CSI datasets available in the cell for training. The use of big data analytics can also aid in applying ML in NOMA-enabled scenarios.

CF-mMIMO, which is a distributed and scalable version of mMIMO networks, has been proposed to mitigate the large propagation losses, provide better coverage to cell edge users, and improve QoE. However, in CF-mMIMO networks, the effect of network interference is acute. To this end, in [422], authors proposed the k-means++ and improved k-means++ for efficient user clustering in CF-mMIMO-NOMA networks. These algorithms overcome the problems with the k-means algorithm, which are caused by the randomness of the starting centroids, and ensure that each cluster has the most UEs possible. The works in [423]–[428] investigated the DL performance of power domain NOMA-based CF mMIMO systems. In particular, the authors in [423]–[426] analyzed the system performance under CB precoding. Aiming at attaining high DL rates, the work in [427] applied more sophisticated DL precoding techniques, namely, FZF and modified RZF. Furthermore, while the aforementioned works considered a Rayleigh fading channel between users and APs, the work in [428] investigated the system performance under correlated Rician fading channels. More specifically, the aforementioned works compared the DL performance of NOMA-based CF mMIMO and the OMA-based counterpart, considering the joint detrimental effects of intra-cluster pilot contamination, inter-cluster interference, and imperfect SIC effects.

CSI in NOMA is especially required for optimal SIC decoding order, user pairing and power allocation. Continuous availability of CSI for these signal processing results in increased overheads and computational complexity, which leads to delayed feedback [79], [376]. To overcome these issues, several solutions such as statistical CSI, imperfect CSI, and limited bits used for feedback channels using different techniques have been used in literature [429]–[431]. Accurate CSI for each user poses a significant problem for NOMA networks, necessitating channel estimation and feedback. When the number of users is substantially more than the number of antennas at the base station, estimating CSI and getting feedback could be challenging in a high-traffic scenario [25], [27], [432]. Recently machine learning techniques have been utilized to predict the CSI based on statistical models of the channels [433], [434]. This method can reduce the amount of required feedback, but it requires a larger size of training data and can be computationally expensive.

In sum, it is essential to have accurate and swift CSI in order to fulfill the stringent criteria of the next phase of NGMA and NGWNs. Efficient user grouping and power allocation serve as essentials for optimal performance of NOMA. However, obtaining CSI in real-world scenarios like CAVs, SAGINs, IoRT, IIoT, UAVs, etc., poses a major challenge. Hence, it becomes essential for researchers to prioritize the design of effective approaches that can effectively tackle the challenges of CSI acquisition and comply with the criteria of NGMA.

### *B. User Grouping and Power Allocation*

Optimal resource allocation is the backbone of NOMA. Users are grouped together based on their respective CSI, each acquiring a different amount of power. As a result of appropriate resource sharing, interference can be decreased, and system capacity is increased when users are properly grouped, and optimal power allocation techniques are implemented. User clustering/grouping is a critical aspect of NOMA, where users of different channel gains are multiplexed in the power domain on a single RB. NOMA allocates higher power to weaker users and lower power to stronger users, resulting in fair user treatment. The weaker user decodes the superimposed signal by treating the stronger user's signal as interference or noise. In comparison, the stronger user decodes the intended signal by applying SIC in DL. The greater the difference between the channel gains of paired users, the better the SIC implementation, which leads to an overall increase in sum capacity. However, in a given cell, if cell-center users (high gain) are paired with cell-edge users (low gain), mid-gain users will either be paired together or left unpaired. Pairing mid-gain users together may lead to capacity reduction due to SIC performance degradation, whereas leaving them unpaired means they cannot benefit from the capacity gains achieved by NOMA. Therefore, user pairing needs to be carefully designed to ensure optimal performance. Another challenge in user pairing is the number of users that can be multiplexed together. Ideally, a given RB should be capable of pairing a large number of users to achieve maximum benefit from the SE of NOMA. However, as the number of users increases beyond two, the complexity also increases. Therefore, achieving a tradeoff between the number of users and complexity is crucial in PD-NOMA-based systems, where the matching theory has been shown as a low-complexity and effective user clustering method.

Besides user pairing, power allocation is another challenge. Inappropriate power allocation techniques can result in interference, system outage, and energy inefficiency. Therefore, appropriate PA is necessary to enhance the overall system performance. Various factors, such as channel conditions, availability of CSI, QoS requirements, and total available system power, characterize power allocation in NOMA. Hence, in order to enhance the NOMA performance in the upcoming era of limitless connectivity, researchers should design more robust, efficient, and intelligent resource allocation techniques to improve the performance and to meet the strict requirements of the applications and use cases of 6G.

### C. Successive Interference Cancellation

With the significant increase in the number of connected devices and data-centric ultra-low latency applications, there is a growing demand for spectrum resources. NOMA is a spectrally efficient multiple-access technique that accommodates multiple users in a single RB. To bridge the gap between theory and practical implementation of NOMA solutions, certain signal processing issues have been identified. In [435], [436], the authors first time came up with the idea of NOMA with SIC for cellular radio access. They proposed to use the power domain for multiplexing different users in one RB and suggested the need for a complex receiver such as SIC for decoding the received superimposed message sent from BS to UE. Implementing SIC has hardware complexities with high power consumption but this theoretical idea has only been made possible due to advancements and evolution of signal processing capabilities following Moore's law. One of the impacts on NOMA performance gain is due to imperfect SIC, which is because of receiver complexity and sensitivity, decoding order, analogue to digital quantization error, imperfect channel estimation and error propagation due to interference residue. In SIC previously decoded signals are subtracted from the superimposed signal, SIC imperfection can also be due to an error in bit decisions made in the detection phase of the previously decoded users. Furthermore, the authors in [437] discussed the issue of SIC receivers in decoding information for different users. The case of imperfect SIC implementation when the decoding order does not match the sequence of cancellation is discussed and simulated. The mismatch causes incorrect message detection, and a request for retransmission has to be made to the BS, which increases the number of iterations for decoding, thereby prolonging the computation time. The prolonged computation time due to erroneous messages might also be due to potential external network attacks.

Besides, the authors in [438], [439] emphasized the SIC as the crucial component of NOMA. In particular, selecting the SIC decoding order based on the users' CSI and the users' QoS, respectively. For instance, in the uplink PD-NOMA serves the two users simultaneously and applies SIC at the BS for signal separation. A natural SIC strategy is to first decode the signal from the strong user and then decode the weak user signal. In this way, strong user may suffer from severe interference because its signal is decoded in the presence of weak user, hence it deteriorates the strong user's QoS. This QoS issue becomes much more severe in the presence of more than two users. Another issue of PD-NOMA is that user's channel conditions need to be sufficiently different in order to yield a reasonable performance gain. if the users' channel qualities are identical, the performance of NOMA is the same as that of OMA, but the system complexity of NOMA is higher than that of OMA. These disadvantages can be avoided by using QoS-based SIC. The benefit of QoS-based SIC is that QoS-based SIC can be easily extended to a general case with more than two users while guaranteeing the users' QoS requirements.

The SIC algorithm is susceptible to the propagation of errors at every detection stage, as incorrect decisions made in earlier layers result in less-than-ideal SIC decoding. As a result, when a user's signal is decoded incorrectly, the system fails to accurately recognize the signals from other users. This issue significantly diminishes the effectiveness of PD-NOMA users. To address the challenges associated with SCI in NOMA and enhance the performance of NOMA-enabled NGWNs, it is essential to develop receivers equipped with improved detection algorithms that offer greater detection accuracy. Additionally, the reliance on instantaneous CSI poses a significant challenge for reliable communications, as SIC detection based on imperfect CSI can lead to decoding errors in SIC. The performance of NOMA can be significantly impacted by the inefficiencies in SIC. To enhance the performance of NOMA-enabled NGWNs and address SIC issues, improved detection algorithms are needed for receivers, ensuring higher accuracy to reduce error propagation. Moreover, When combining NOMA with MIMO, the complexity increases because the decoding order becomes challenging to design, given the matrix representation of channel conditions. In addition, ongoing research is focused on developing dynamic SIC receivers for downlink NOMA with power constraints and finding solutions for dealing with imperfect CSI. These are important areas that require additional research. We noted that researchers from both academia and industry are working on robust techniques to improve the efficacy of SIC to ensure its effective performance in massive connectivity scenarios. Such techniques include low-complexity SIC [440], approximation algorithms [441], hybrid SIC methods [442], parallel interference cancellation (PIC) [443], and machine learning-based techniques [444].

### D. High-Mobility Users

Due to the user's mobility, the channel conditions that exist between the transmitter and the receiver might change quickly. This can lead to fading, interference, and other channel impairments, which can have an effect on NOMA's performance. The impact of mobility on the operation of NOMA is contingent upon various variables, including the speed and direction of movement, the distance between users, the number of users, and the transmission power levels. In general, increased user mobility can result in more severe channel impairments and decreased NOMA performance. Using adaptive power allocation and decoding order techniques is one method that can be utilized to reduce the negative impact that mobility has on NOMA. NOMA is able to adapt to the varying channel impairments and maintain a high performance level because it can dynamically adjust the power levels and decoding orders based on the characteristics of the channel. In the literature, there are also works that used advanced multi-antenna techniques [414], [445] to improve coverage and connectivity in high mobility scenarios to enhance the performance of NOMA. By utilizing advanced multi-antenna techniques, the signal quality can be improved with reduced interference in high-mobility scenarios. Furthermore, in such environments, using multiple antennas at the transmitter and receiver can assist NOMA by taking into account the spatial diversity of communications channels [446].

### *E. Signal Superposition: Security and Privacy Risks*

NOMA is a promising wireless communications system technology that allows multiple users to share the same channel resources simultaneously. Yet, security and privacy considerations can impede NOMA's effectiveness and applicability in multiple ways. NOMA's superposition of signals also poses security and privacy risks. For example, unauthorized users or attackers may attempt to eavesdrop on the signals transmitted by legitimate users or inject their own signals into the channel. These attacks can jeopardize the communication's privacy and secrecy or disrupt the entire system. In order to overcome these issues, NOMA includes a variety of security techniques to safeguard the transmitted signals. Encryption, which scrambles the signal and renders it unintelligible to unauthorized users, is one of the most often employed security measures. The user's signal can be encrypted using methods such as Advanced Encryption Standard (AES) [447] or Secure Hash Algorithm (SHA) [448] to ensure that only the intended receiver can decode it. Authentication, which verifies the identity of the signal's source and receiver, is an additional crucial NOMA security mechanism. Authentication can prohibit intruders from injecting their signals into the channel and prevent attacks.

Recently, Blockchain has surfaced as a viable security and privacy technology for the next generation of wireless communications. Blockchain is a distributed ledger technology that enables secure, transparent transactions eliminating the necessity for intermediaries. In the context of NOMA, blockchain can be utilized to structure a decentralized and secure framework to handle user identities and authorizations. Identity management, secure transactions[14], privacy and decentralization are some ways in which blockchain can assist the security and privacy of NOMA. Authors in [449] proposed a blockchain-based FD-NOMA framework for vehicular networks with physical layer security.

Furthermore, NOMA can benefit greatly from applying ML techniques to enhance security and privacy in the next generation of wireless communications. Machine learning algorithms can be utilized to identify anomalous network behaviour, such as abnormal traffic patterns, anticipated access attempts, and abnormal resource utilization. These algorithms can identify possible security issues by examining historical data and learning from normal behaviour patterns. By anonymizing or obscuring sensitive information, machine learning algorithms can be utilized to protect the privacy of users' data. Differential privacy techniques, for instance, can add noise to data to prevent individual users from being identified while still maintaining the integrity of aggregate statistics.

[14]In wireless communications, a blockchain transaction may refer to data exchange between connected devices or networks. Such transactions are recorded on a blockchain to assure wireless communication's transparency, security, and accountability.

## VI. Design Recommendations and Insights

Recent years have seen a flurry of endeavors related to the study of NOMA's potential role in 5G and 6G wireless networks. These studies have adopted a number of cutting-edge techniques and protocols to boost the communications efficiency of wireless networks. NOMA, as a contestant of the NGMA, constitutes notable advantages in comparison to conventional MA schemes. The benefits encompass enhanced spectral efficiency, increased system capacity, and improved fairness and quality of service. Moreover, NOMA exhibits flexibility, rendering it appropriate for diverse applications and deployment scenarios. The advancements in ICT have given rise to novel prospects for the efficient implementation of NOMA in NGWNs. Nevertheless, in order to fully actualize the prospective advantages of NOMA, various factors must be taken into account. These factors encompass the demands of future applications, the characteristics of wireless communications systems, the available resources, the performance goals, and the operational complexity. With a thorough examination of these variables, it is feasible to devise and implement NOMA solutions that are optimized to particular use cases and capable of fulfilling the demands of NGWNs. This section aims to cover design aspects and lessons learned by answering different questions for readers.

**What can machine learning offer to NOMA?** Given the heterogeneous, dynamic, and complex nature of scenarios pertaining to 6G networks, it is imperative to develop novel approaches in order to effectively address the increasingly challenging problems that arise within these scenarios. The majority of conventional approaches utilized in the present research contributions necessitate stringent preconditions that have been formulated based on robust assumptions. In contrast to conventional algorithms, AI-based algorithms possess the capability to tackle problems within a broader and more comprehensive framework. In addition, ML enables NOMA-based networks to achieve the interaction between BS and the dynamic environment. Furthermore, it is worth noting that in traditional approaches, the control policy is designed with the primary goal of promptly attaining immediate advantages for networks without taking into account the broader, long-term evolution of the network. This, however, is precisely the objective of machine learning algorithms. ML is anticipated to hold significant prominence in the advancement and execution of future wireless networks. Several potential impacts of ML on the development of next-generation wireless networks have been identified in the literature [450]. ML techniques have become increasingly popular for the purpose of resource allocation in heterogeneous wireless networks. This is due to their ability to learn, particularly in network environments that are time-varying and unpredictable [27]. Recently, researchers have used ML techniques to fulfil the needs of NOMA based 5G and 6G networks, such as user pairing [451], power allocation [210], joint user pairing and power allocation [452], interference management [453], beam management [454], relay selection [455], channel estimation [456] and so on. ML is ideal to handle the complex requirements of future wireless networks because it is capable of learning patterns and making predictions based on massive amounts of data. The performance of NOMA depends on optimal resource allocation and user clustering. It is worth noting that the optimization of user clustering is NP-hard. Therefore, for such types of problems, the authors in [185], [457] emphasized using ML instead of conventional mathematical optimization techniques.

*1) Deep Learning for NOMA:* The progress in deep learning algorithms has revolutionized it into a potent technology for improving the performance of the NGWNs. The transmission of data in advanced NGWNs is expected to exhibit remarkable diversity and demonstrate complicated correlations. The applications of deep learning in NOMA are widely researched. They offer significant benefits for solving complex data processing challenges, resource allocation, and obtaining the CSI [27], [458]. We have noted that a significant amount of research is available in which researchers have deployed deep learning-based solutions to optimize the performance of the NOMA-driven wireless networks and to solve the problems, such as acquiring the CSI [92], signal detection [459], user pairing [460], and power allocation [461]. The fundamental concept behind the application of deep learning techniques in NOMA is that the deep learning model employs a DNN to determine a suitable data representation at each layer. The authors in [185] proposed a DL-based resource management framework to efficiently address the challenges associated with power allocation, subchannel allocation, and user association in NOMA. This work primarily centered around the enhancement of the system's energy efficiency, taking into consideration various constraints such as interference, service quality, and power. The problem associated with the user association was successfully resolved by implementing the Lagrange double decomposition scheme. Furthermore, the challenges related to power allocation and subchannel allocation were effectively addressed using the DNN and semi-supervised learning techniques. Results showed that the DL-based resource management framework achieved higher energy efficiency while reducing the overall complexity. Also in [462] an active user identification in grant-free enabled NOMA systems was proposed through the utilization of a DNN. By integrating the training information into a DNN, the methodology that was devised examined the irregular mapping between active devices and obtained NOMA signals to identify active users. The simulation outputs demonstrated that the proposed method outperformed the conventional approaches in terms of both computational complexity and success probability of active user identification.

*2) Reinforcement Learning for NOMA:* Reinforcement learning focuses on the idea of an agent performing actions in an environment, guided by its observations of that environment. The agent generally performs actions according to a set of guidelines that define its conduct at any specific time. The agent receives feedback from the environment, which defines the overarching goal of the algorithm and reflects the agent's performance at the time the feedback is provided. The agent aims to enhance its total reward by observing its surroundings and the feedback received, subsequently performing actions based on these observations. The enhancement of the total reward is typically expressed through a value function. The value function differs from the reward signal in that the reward signifies a preferred immediate outcome, whereas the value function reflects the possible future rewards that the agent may obtain depending on its present state. Additionally, challenges in reinforcement learning typically involve creating a representation of the environment. The agent evaluates the model to understand the dynamics of the environment, which is then used to develop a course of action [463].

Reinforcement learning has been widely studied in the literature to optimize the performance of the NOMA networks. In this regard, the authors in [464] proposed a power control scheme based on reinforcement learning for downlink NOMA transmission, which was operated without being aware of the jamming and radio channel parameters. The Dyna architecture, which developed a learned world model based on real antijamming transmission experiences, along with the hot-booting technique that leverages experiences from similar scenarios to set initial quality values, was employed to enhance the learning speed of Q-learning-based power allocation. This technique ultimately improved the communication efficiency in NOMA transmission amidst the challenges posed by smart jammers. Furthermore, in [465], in order to supplement terrestrial infrastructures, the authors suggested a NOMA-enhanced network that deploys several UAVs in three dimensions. By combining the optimization of the multi-UAV's dynamic trajectory with the power allocation policy based on user channel state information, the authors developed the sum rate maximization problem based on the proposed model. In order to maximize the overall throughput, a multi-agent mutual deep Q-network was proposed to jointly optimize the 3D trajectory and power allocation policy of UAVs. The proposed MDQN algorithm markedly enhanced the training efficiency of the agents, demonstrating a superior convergence rate compared to the independent DQN scheme, while the real-time optimal trajectory design surpassed benchmarks, including the congeneric 2D trajectory and circular deployment. In [195], the authors proposed two reinforcement learning techniques, namely SARSA and DRL, to solve the sum rate maximization problem in the uplink NOMA-IoT system. The authors proved that the sparse activations improve the performance of the DNNs when compared to the traditional mechanisms. In [466], the authors proposed a novel and effective DRL-assisted joint resource management approach for multicarrier NOMA systems. Their extensive experiments verify that the proposed scheme provides close-to-optimal results in the case of small-scale problems. In the case of large-scale problems, the proposed scheme is superior to existing benchmarks DUCPA [467], JPCA [468], CFJBA [469], and JPCA-DRL [470] in terms of the weighted sum rate and resistance to interference.

*3) Federated Learning for NOMA:* Distributed machine learning (DML) techniques, such as FL, partitioned learning, and distributed reinforcement learning, have been increasingly applied to wireless communications. This is due to improved capabilities of terminal devices, explosively growing data volume, congestion in the radio interfaces, and increasing concern for data privacy. The unique features of wireless systems, such as large scale, geographically dispersed deployment, user mobility, and the massive amount of data, give rise to new challenges in the design of DML techniques [471]. In [201], authors proposed a scheduling policy and power allocation scheme using NOMA to maximize the weighted sum data rate during the learning process. However, it is still an open problem regarding how to jointly optimize the radio resource allocation for NOMA transmission, the computation-resource allocation for FL training and aggregating, as well as the training accuracy of FL, with the objective of optimizing the overall performance of the FL. In [202], authors studied the NOMA-assisted FL powered by the WPT by formulating a joint optimization of the WPT durations for the UEs, the UE's NOMA transmission, the BS's broadcasting transmission, the UE's and BS's processing rates, as well as the training-accuracy of FL, with the objective of minimizing the system-wise cost that includes the total energy consumption and the overall latency for FL convergence.

*4) Transfer Learning for NOMA:* In contrast to traditional machine learning methods that focus on addressing a specific problem, transfer learning utilizes important insights from related tasks and past experiences to greatly improve the performance of conventional machine learning approaches [472]. Consequently, transfer learning offers a variety of benefits compared to conventional machine learning methods, including improved quality and quantity of training data, accelerated learning processes, reduced computing requirements, minimized communication overhead, and protecting data privacy [473]. In [474], the authors developed a detection algorithm rooted in transfer learning, in addition to three deep transfer learning strategies tailored for the learned preconditioned conjugate gradient descent network featuring the SIC detector. The proposed algorithm aimed to implement a suitable deep transfer learning approach in response to variations in environmental factors like channel model, modulation scheme, or power allocation, enabling the rapid acquisition of an effective model with minimal dataset costs. In [475], The authors examined the joint problem of user-cell association and beam selection to enhance the overall capacity of the network. A transfer Q-learning algorithm was introduced to address the optimization problem. The results indicated that the proposed transfer Q-learning algorithm surpassed the benchmarks in dynamic scenarios and demonstrated faster convergence compared to the traditional Q-learning algorithm.

*5) Meta-Learning for NOMA:* Meta-learning embodies a novel methodology that facilitates the extraction of hyper-parameters from samples of related tasks, setting it apart from conventional learning techniques regarding the level of "learning" involved. Conventional learning techniques employ samples gathered from one or several tasks to train a deep neural network with defined hyper-parameters, allowing it to acquire a policy. This policy functions as a framework that connects environmental parameters, such as channel gains, to refined decisions, including power allocation. Meta-learning employs samples gathered from diverse tasks to discern the common hyperparameters relevant to these tasks. As a result, meta-learning is frequently referred to as "learning to learn," and the resulting inductive bias is typically known as "meta-knowledge" [476]. In [477], by simultaneously optimizing the phase shift of the RIS and the power allocation at the BS, the authors developed a sum rate maximization issue and proposed a QoS-based clustering technique that sought to increase the resource efficiency of RIS-NOMA network. An approach based on meta-learning was proposed to tackle the joint optimization problem with a low model complexity and a faster convergence rate. Furthermore, in [478], the authors investigated an uplink transmission system enhanced by multiple STAR-RISs, where the diverse reflections among the various STAR-RISs supported the communication from single-antenna users to a multi-antenna base station. The problem of maximizing the overall sum rate was tackled using a technique baed on the joint optimization of active beamforming, power distribution, transmission, and reflection beamforming at the STAR-RIS, along with the allocation of users to the STAR-RIS. A new deep reinforcement learning algorithm was introduced to tackle the non-convex optimization problem, combining meta-learning with DDPG. The numerical results demonstrated that the proposed Meta-DDPG algorithm shows a notable performance enhancement compared to the traditional DDPG algorithm, achieving a 53% improvement. Furthermore, integrating multi-order reflections across various STAR-RISs leads to a 14.1% enhancement in the total data rate.

**What can researchers do to apply ML techniques for NOMA?** According to the literature, optimizing resource allocation is an essential component in strengthening the functioning of NOMA networks. ML approaches, such as DL, DRL, etc., show great potential for bettering the functionality of NOMA networks. DRL has the ability to analyze data collected from the network environment, enabling it to make decisions and optimize resource allocation in NOMA networks. Moreover, FL lets multiple users or devices train a shared model collaboratively without compromising their private data. This kind of decentralized data processing can minimize the necessity of transferring data to a central server, thereby decreasing communications costs and bandwidth requirements. Furthermore, ML techniques can enhance the functionality of NOMA networks by optimizing user pairing and clustering by taking into account contextual data, network dynamics, and user requirements. Also, the utilization of ML techniques can enhance the security and privacy of NOMA networks. The use of machine learning techniques can be helpful in detecting and preventing security threats, such as eavesdropping and jamming attacks. ML techniques can significantly improve the sustainability of NOMA networks by optimizing power allo-cation and transmission strategies, leading to reduced energy consumption. In summary, ML techniques hold great promise for upgrading the performance of NOMA networks through low-complexity resource allocation optimization, improved security and privacy, and reduced energy consumption.

To achieve this, researchers need to focus on designing algorithms that are well-suited for the unique requirements of NOMA networks and collecting realistic datasets to train and validate the models. Additionally, integrating ML with other optimization and control techniques can further improve the performance of NOMA networks. To improve the performance of future NOMA networks using ML, researchers can focus on the following areas:

1) Dataset collection: In order to enhance the efficacy of ML-driven NOMA networks, it is imperative to gather real-world and representative datasets. The utilization of real-world data for training ML models has the potential to enhance their precision and performance, thereby enabling researchers to achieve better outcomes.
2) Algorithm design: ML algorithms need to be intelligently tailored to NOMA networks. This requires a deep understanding of the network characteristics and challenges. Algorithms that are well-suited for the unique requirements of NOMA networks can significantly improve their performance.
3) Model validation: ML models need to be validated using real-world experiments to ensure their robustness and accuracy. This will help researchers to identify potential issues and improve the performance of the models.
4) Integration with other techniques: ML techniques can be integrated with other optimization and control techniques to improve the performance of NOMA networks further. For example, RL can be combined with traditional mathematical optimization techniques to achieve better performance.

**Why do we need a unified NOMA framework?** A unified NOMA framework is crucial to the standardization of NOMA in NGWN and ought to maintain the exceptional performance promised by various 5G-NOMA variants, such as supporting massive connectivity, satisfying user fairness, and striking a balanced tradeoff between energy efficiency and spectral efficiency. It is important to note that the majority of existing 5G-NOMA variants were developed for specific purposes. SCMA, for instance, was developed primarily to support immense connectivity for mMTC, whereas PD-NOMA is recognized for its ability to increase system throughput for eMBB. Therefore, a general framework is necessary for the deployment of NOMA towards NGMA, and sophisticated tools, such as multi-objective optimization techniques and multi-task learning, will be advantageous. The unified NOMA framework must be resistant to dynamic wireless propagation environments and complex mobility profiles of users. Instead of relying on users' instantaneous channel conditions, which can swiftly change in practice, the use of users' heterogeneous QoS requirements may be more reliable for achieving a robust NOMA transmission because users' QoS requirements are changing in a much slower manner. Particularly, users with low QoS demands can be grouped with users with stringent QoS demands for spectrum sharing, which can notably enhance spectral efficiency. It has been demonstrated that the use of the delay-Doppler plane yields more degrees of freedom for system design when there are users with heterogeneous mobility profiles than that of the conventional time-frequency plane [414], [479], [480].

**What is the role of efficient and robust power allocation techniques in achieving fair resource distribution among users in NOMA-assisted NGWNs?**
When using NOMA, the transmitter allocates the available power among multiple users sharing the same RB. Power allocation in NOMA is aimed at improving system performance by fairly allocating available power among users. Inefficient power allocation strategies may result in substantial interference, user unfairness, system outage, and energy inefficiency, all of which can have an impact on system performance. The power allocation approach employed in NOMA is characterized by the channel conditions, the presence of CSI, QoS specifications, and the total available system power. The power allocation strategy in NOMA considers the subsequent factors:

1) Channel conditions: The power allocation approach utilized by NOMA necessitates the consideration of user's channel conditions, including channel gains, fading, and path loss. The power allocation scheme is designed to allocate less power to users with stronger channel conditions and more power to users with weaker channel conditions. The rationale behind this phenomenon is that users with favourable channel conditions necessitate a lesser power to achieve their desired QoS, while users with poor channel conditions necessitate a higher power.
2) Availability of CSI: The power allocation scheme utilized in NOMA necessitates the consideration of CSI, which is employed to gauge the channel conditions of the users.
3) User's QoS requirements: The power allocation scheme utilized in NOMA necessitates consideration of the QoS requirements of the users. It is imperative to establish an optimal power distribution to ensure the meeting of QoS requirements for users while maximizing the overall efficacy of the network. This can be facilitated by the selection of an objective function that accurately reflects the QoS demands and system performance metrics. Examples of such metrics include the sum rate, proportional fairness, and energy efficiency.
4) Total available system power: The power allocation technique utilized in NOMA necessitates consideration of the total available system power, which is restricted by various factors such as hardware constraints, interference, and regulatory obligations. It is important to guarantee optimal power allocation across users while concurrently fulfilling QoS requirements and minimizing interference.

User fairness is one of NOMA's most important features. The NOMA scheme enables a more flexible management of the users' achievable rates and is an efficient way to enhance user fairness. User fairness in the NOMA refers to the distri-bution of the resources among users to ensure a certain level of QoS. Several methods have been used in the literature to measure the user's fairness, such as Max-Min fairness (MMF) [481], proportional fairness (PF) [482] and Jain's fairness index [483], [484]. Moreover, the authors in [485], defined the user fairness in terms of ergodic capacity, outage probability and bit error probability. In NOMA users with weaker channel conditions are given more power so they can be served as users with better channel conditions. This is achieved by using SIC, whereby the users with weaker channel conditions are decoded first, and their interference is cancelled before decoding the users with stronger channel conditions. Various optimization techniques, such as linear programming, convex optimization, or non-convex optimization, can be employed to implement the power allocation strategy in NOMA in practical scenarios. Based on the objective function and constraints, the optimization problem can be formulated. The optimal power allocation coefficients for each user are provided by the solution to the optimization problem and are utilized by the transmitter to allocate power to the users. In a nutshell, effective power allocation is critical for improving the performance of NOMA systems. By optimizing power allocation, NOMA systems can achieve increased capacity, efficiency, and fairness while minimizing interference and energy consumption.

**What is the significance of efficient user clustering techniques in NOMA-assisted NGWNs?** NOMA allows multiple users to share the same resource block in the power domain. This means that users from different channels can simultaneously transmit and receive signals within the same RB without the need for orthogonalization in terms of time, frequency, or code. The power allocation scheme implemented in this system prioritizes users with weaker channel gains by allocating them higher power, while users with stronger channel gains receive lower power allocations. This approach ensures a fair distribution of resources among all users. The decoding process involves the weak user treating the strong user's signals as interference or noise while the strong user applies SIC in the DL to decode the signal intended for it. The effectiveness of SIC and the increase in total capacity are directly proportional to the difference in channel gains between paired users. In a cellular network, when cell-center users with high gain are matched with cell-edge users with low gain in a specific cell, the mid-gain users will either be paired with each other or remain unpaired. The pairing of mid-users can result in a decrease in capacity due to the degradation of SIC performance. In the scenario where these users are not paired, they do not experience the SIC performance issue due to their utilization of OMA. Consequently, they are unable to take advantage of the capacity improvements achieved through NOMA. An additional concern within the context of user pairing pertains to the maximum number of users that can be effectively multiplexed together. Even so, it is important to note that as the number of users increases beyond two, the complexity of the situation also increases.

The performance of user pairing in NOMA depends on several factors, such as the difference between the channel gains of the paired users, the number of users that can be multiplexed together, and the complexity of the SIC process. Ideally, the paired users should have a large difference in their channel gains, which leads to better SIC performance and higher overall system capacity. The complexity of the system, however, rises as the number of users exceeds two, which can result in system complexity and performance degradation. To deal with these problems, researchers came up with a number of advanced user pairing methods and algorithms, such as ML-based approaches, joint optimization of user pairing and power allocation, and C-NOMA systems. Based on the channel conditions, QoS requirements, and other factors, machine learning-based approaches use advanced algorithms to predict the optimum user pairing and power-sharing coefficients.

One of the challenges in user pairing is the signalling overhead, which refers to the amount of overhead signalling required to exchange information between the base station and the users. In NOMA, user pairing involves exchanging information about the channel gains, QoS requirements, and other criteria of the users to select the best pair of users. This information exchange can lead to significant signalling overhead, particularly in scenarios with a large number of users. Therefore, advanced user pairing techniques, such as machine learning-based approaches, can be used to reduce the signalling overhead by predicting the optimal user pairing based on historical data or other criteria.

User pairing poses a challenge due to the system's complexity, which pertains to the computational resources needed for user pairing and power allocation. The user pairing in NOMA involves selecting the most suitable pair of users by considering factors such as their channel gains, QoS requirements, and other relevant criteria. To determine the best user pairing and power allocation coefficients, complex computations and algorithms need to be performed. Furthermore, with an increase in the number of users, the system's complexity also increases, potentially resulting in a decline in performance and a decrease in system capacity. Advanced techniques for user pairing, such as joint optimization of user pairing and power allocation, can be employed to streamline system complexity by optimizing user pairing and power allocation jointly, thereby achieving optimal overall system performance and spectral efficiency.

**How does the joint optimization of user pairing and power allocation affect the complexity of the NOMA network?** Joint user pairing and power allocation is an advanced approach in NOMA that optimizes user pairing and power allocation jointly in order to ensure significant performance enhancement and spectral efficiency. This technique maximizes system capacity and user fairness by picking the best possible pair of users to share the same RB based on their channel conditions and QoS needs and then assigning power levels based on their channel gains. In NOMA, the joint optimization of user pairing and power sharing is a hard, non-convex optimization problem that needs advanced methods and techniques to solve. Numerous algorithms and techniques have been proposed in the literature to address this optimization problem, including convex relaxation, successive convex approximation, and gradient-based approaches. The joint optimization of user pairing and power allocation in NOMA presents a number of challenges that must be overcome in order to fully realize its potential. The complexity of the system increases with

an increase in the number of users and RBS, which affects the network performance and capacity. Accurate CSI is vital for power allocation and user pairing in joint user pairing and power allocation. CSI is prone to several impairments in practical scenarios, including feedback delay, limited feedback bandwidth, and channel estimation error. Hence, the utilization of sophisticated CSI acquisition techniques, including pilot contamination mitigation, compressed sensing, and machine learning-based techniques, can be employed to enhance the accuracy and swiftness of CSI in NOMA. Meanwhile, joint optimization of user pairing and power allocation also causes signalling overhead, which also degrades the system's performance. Hence, designing more sophisticated tools and techniques is required to ensure the smooth optimization of joint user pairing and power assignment.

**In what ways can NOMA be used to minimize the average age of information (AoI) in wireless networks, and what are the benefits of doing so?** Recently, researchers have shown an enormous interest [486], [487] in a new performance metric, termed Age of Information (AoI), due to its advantages in characterizing the timeliness of data transmission in status update systems. The timeliness of status updates is of great importance, especially in real-time monitoring applications, in which the dynamics of the monitored processes need to be well grasped for further actions. The AoI is defined as the time elapsed since the generation time of the latest received status update at the destination. According to the definition, the AoI is jointly determined by the transmission interval and the transmission delay.

Researchers have recently shown interest in exploring the potential of using NOMA to reduce the average AoI in wireless networks [41], [42], [488]. The idea of NOMA is to leverage the power domain to enable multiple clients to be served on the same time-frequency resource block. This can lead to possible AoI drops for more than one user without requiring more radio resources. While in OMA, only the served client may have an AoI drop, and the AoI of other clients all increase. For instance, in [42], a low-complexity form of NOMA, termed NOMA-assisted random access, is applied to grant-free transmission in order to illustrate the two benefits of NOMA for AoI reduction, namely increasing channel access and reducing user collisions. In [489], the authors presented a study on a RIS-assisted NOMA network for collecting packets of IoT devices. The proposed approach utilizes a novel AoI model and adopts the DDPG to jointly optimize the phase-shift matrix of RIS and service time of packets, with the aim of minimizing the average peak AoI. Furthermore. in [251], the authors optimized the AoI and packet drops of the entire network by switching between NOMA and OMA according to the packet existence in each NOMA pair and outage probability requirements.

## VII. Future Perspectives

Although there are still challenges to be addressed in optimizing the performance of NOMA networks, recent research has identified several promising future perspectives that could potentially overcome these challenges and advance the technology. This section covers some of the significant future perspectives in NOMA research, such as novel applications, emerging technologies, and potential use cases.

### A. *Connected Autonomous Vehicles*

Vehicular technologies are evolving rapidly, from connected vehicles called cellular vehicle-to-everything (C-V2X) to autonomous vehicles to the combination of the two, the networks of connected autonomous vehicles (CAVs) [490], [491]. As a fundamental component of CAVs, C-V2X is rapidly developing with the advent of intelligent networked vehicles, the integration of vehicular sensors, and the collaboration between vehicles, road infrastructures, and the cloud. C-V2X enables the CAVs to communicate efficiently with each other and the BS through other CAVs, roadside units (RSU), cellular users, and others. In cities such as Paris, Las Vegas, Tokyo, and Dubai, similar multi-passenger autonomous vehicles[15] have been developed and are undergoing testing. In Singapore[16], they've gone a step farther, constructing a whole village to test and enhance robot bus technology.

Considering the expected growing number of devices connecting to the C-V2X network in the future, the low efficiency of orthogonal multiple access in the conventional OFDM-based LTE network will cause congestion, leading to significant access delay and pose a great challenge, especially for safety-critical applications. For the forthcoming 6G, which will enable limitless connectivity, the NOMA technology has been widely regarded as a practical solution as multiple users can share the same resource blocks, which improves system capacity, spectral efficiency, and energy efficiency. Multiple users with distinct traffic requirements can transmit on the same channel to increase spectrum efficiency and reduce data traffic congestion, reducing latency. Authors have already worked on designing the C-V2X frameworks using NOMA. For example, in [275], authors designed a framework based on a NOMA-assisted C-V2X network, which guarantees ultra-reliable and low-latency communications. In [492], the authors proposed an energy-efficient power allocation scheme for the RSU-assisted NOMA multicasting for B5G C-V2X networks.

### B. *Internet of Robotic Things for Smart and Sustainable Cities*

The Internet of Things focuses on enabling services for ubiquitous monitoring, sensing, and tracking, whereas autonomous robots are more involved in action, interaction, and autonomous behaviour. The IoRT system consists of several components that are connected to the Internet, such as intelligent robots, sensors, and clouds. A system of systems is created when all of the individual components are connected via communications protocols. These robots, in general, need to be able to do three things: *sensing, intelligence, and motion*. The *Seoul Metropolitan Government* is designing the world's first pioneer robot science museum (RSM) to promote public education in robotics and improve general knowledge and interest in robots. Not only will robots be on display at RSM, which promotes and advances science, technology, and innovation throughout society, but robots will also be in control of all aspects of the event, from its planning to its construction to its provision of services. RSM, constructed by onsite robots, will start its 'first exhibition' in July 2023 [17].

[15]https://knowhow.distrelec.com/transportation/how-robots-are-changing-transport/

[16]*Singapore Self-Driving Buses*

NOMA techniques have been considered to be well-suited for machine-type communication or IoRT due to their better spectrum efficiency and massive connectivity. NOMA's usability can assure integration of *Communications, Computing, and Control (3C)* into 6G IoRT for satisfying ultra-reliable low-latency connectivity requirements in a smart city, where applications of IoRT vary from intelligent transportation systems to remote surgeries and beyond, ensuring effective communications, computing, and control of these intelligent robots can truly determine the future of smart and sustainable cities. In addition, the design flexibility of NOMA-based networks is enhanced by the fact that robots are operated frequently in dynamic scenarios, meaning that the channel conditions of robots are different at each time slot. In order to improve the received SINR for both robots and other mobile users, it is possible to assign them to the same cluster dynamically. That's why NOMA can improve the effectiveness and security of independently operating robots. The authors in [493] proposed a deep learning-based model for path designing and resource management of indoor intelligent robots. Efficient resource allocation and communications-aware NOMA-enhanced IoRT designs can result in improved spectrum efficiency. For example, the authors in [494] proposed a scheme for communications aware path planning NOMA enhanced indoor robots system.

### C. Industry 5.0

The fifth industrial revolution, often known as Industry 5.0, is completely altering the ways in which corporations manufacture, innovate, and distribute their goods. Manufacturers are integrating technologies like IIoT, cloud computing, big data analytics, AI, and machine learning into their manufacturing facilities and across their operations. These "factories of the future" are equipped with intelligent sensors, embedded software incorporated into hardware, and robotics, all of which collaborate to collect data, analyze it, and facilitate improved decision-making. The IIoT has entered the era of big data as a result of the rapid growth of industrial data, which requires a larger bandwidth to send massive amounts of data. However, limited wireless resources and even wired connections have severely hampered IIoT development. Furthermore, due to the increasing number of industrial devices to be connected, spectrum scarcity is a constraint that will affect the QoS of each device.

NOMA, as a leading NGMA contender, can effectively enhance spectrum utilization and alleviate the spectrum shortage problem by allocating the same spectrum resource to multiple users. By incorporating NOMA into Industry 5.0, the IIoT can boost overall transmission capacity by connecting more sensing nodes while utilizing limited resources. The authors in [495] proposed NOMA-based multiple access for IIoT to simultaneously improve spectral efficiency, system capacity, and energy efficiency. The authors in [329] presented NOMA-inspired mobile edge computing enabled IIoT to increase the capacity and performance of the network.

### D. Adaptability of NOMA in e-Health

The materialization of e-Health applications, such as telemedicine, telemonitoring, e-prescriptions, health informatics, and others, has been made possible by combining cutting-edge communications technologies with intelligent sensors. In 2021, the size of the e-Health market was estimated at US$ 246.01 billion, and it has been predicted that from 2021 to 2029, the market would generate a total of US$ 701.7 billion, growing at a rate of 14%. The primary reasons propelling the market expansion of e-Health are an increase in the need for remote monitoring services, an increase in the prevalence of chronic diseases, and a technological boom, [496], [497]. HTx[18] an EU project, creates a framework for next-generation health technology assessment (HTA), and supports patient-centered, society-oriented, and real-time decision-making for integrated healthcare throughout Europe.

The incorporation of NOMA in e-Health allows multiple patients using wearable sensors or IoMT-based Wireless Body Area Networks (WBAN) to transmit data to central healthcare providers over the same spectrum resources, thereby optimizing spectral efficiency and providing high-speed, low-latency connectivity for remote e-Health systems. [498]. A significant amount of literature is available online which incorporates NOMA in e-Health [499], [500].

### E. Sustainability

In order to improve spectral efficiency and reduce energy consumption, NOMA can provide a smarter and more efficient use of the available spectrum. NOMA can make the use of spectrum more sustainable by allowing different users to share the same RBs. Moreover, NOMA has the capability to cater to a greater number of applications and use cases by offering enhanced and reliable connectivity, which leads to more sustainable and efficient operations. NOMA has the potential to be a more sustainable option for wireless communications systems by minimizing energy consumption through efficient resource allocation and power management techniques. NOMA technology enables us to build a more sustainable and connected world. By leveraging this technology, wireless communications systems can achieve greater efficiency and performance while also minimizing their impact on the environment.

NOMA, as the frontrunner of NGMA, has the potential to bring significant improvements to wireless communications systems. There are several challenges and limitations that must be addressed to fully realize its potential, such as interference management, receiver complexity, security and privacy, compatibility, implementation complexity, and so on. By developing efficient and intelligent algorithms, addressing the challenges of NGWN, ensuring compatibility with existing and new wireless communications technologies, working with industry stakeholders and regulatory bodies, and adapting to changing channel conditions, researchers can overcome these challenges and create a more efficient and reliable wireless communications system for the future.

[17] *Robot Science MuseumRobots developed by six domestic enterprises delivered contributions in every moment of the event by practising disease control and prevention protocol, providing guidance, moderating the event, breaking ground, and giving performances.*

[18] HTx–Next Generation Health Technology

## VIII. Conclusion

This survey highlighted NOMA as a suitable candidate for the NGMA contest to meet the requirements of 6G. It emphasized the viability of NOMA for various wireless communications technologies, such as multiantenna systems, CoMP, MEC, backscatter communications, UAV communications, IoT, and more, by reviewing state-of-the-art literature. Several NOMA variants, along with their fundamental operations, were presented along with a set of challenges NOMA encounters in massive connectivity scenarios. Due to the benefits that NOMA has offered to 5G and B5G, this study has elucidated the adaptability and flexibility of NOMA as an attractive multiple access technique for NGMA in the wide variety of 6G applications and technologies. NOMA emerges as a prominent option for NGMA, with the potential to assume a pivotal role in facilitating the smooth and effective functioning of forthcoming 6G and beyond. Moreover, it holds the key to unlocking new prospects for innovation and collaboration across various applications and technologies.

## IX. Acknowledgments

This paper is supported by the Innovation Team and Talents Cultivation Program of the National Administration of Traditional Chinese Medicine (No. ZYYCXTD-D-202208).

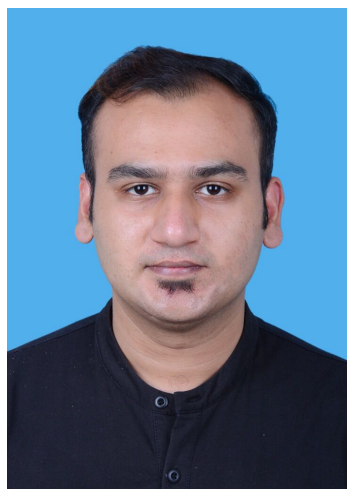

**Adeel Ahmed** (Graduate Student Member, IEEE) received B.S. degree in Telecommunications and Networking from COMSATS University, Pakistan, in 2015 and an M.S. degree in Computer Science and Technology from the School of Computer Sciences and Technology at the University of Science and Technology of China, Hefei, China, in 2021. Currently, he is pursuing his Ph.D. from the School of Computer Sciences and Technology at the University of Science and Technology of China. His research interests include Next Generation Multiple Access, Machine Learning for Wireless Communication, Internet of Things, and Software-Defined Networking.

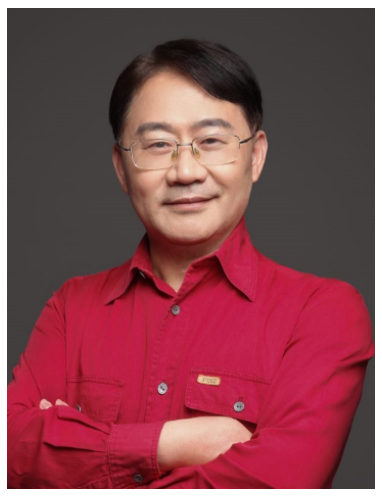

**Xingfu Wang** (Member, IEEE) received a B.S. degree in electronic and information engineering from the Beijing Normal University of China, Beijing, China, in 1988, and an M.S. degree in computer science from the University of Science and Technology of China, Hefei, China, in 1997. He is an Associate Professor with the School of Computer Science and Technology, University of Science and Technology of China. His current research interests include information security, data management, and IoT.

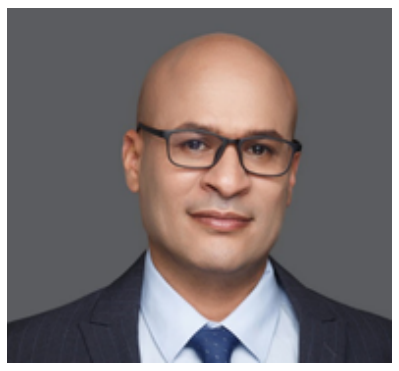

**Ammar Hawbani** (Member, IEEE) received a B.S. degree in computer software and theory and the M.S. and Ph.D. degrees from the University of Science and Technology of China (USTC), Hefei, China, in 2009, 2012, and 2016. Following his Ph.D. completion, he served as a Postdoctoral Researcher at the School of Computer Science and Technology, USTC, from 2016 to 2019. Subsequently, he worked as an Associate Researcher at the School of Computer Science and Technology, USTC, from 2019 to 2023. Currently, he is a Full Professor at the School of Computer Science, Shenyang Aerospace University. His research interests include IoT, WSNs, WBANs, VANETs, and SDN.

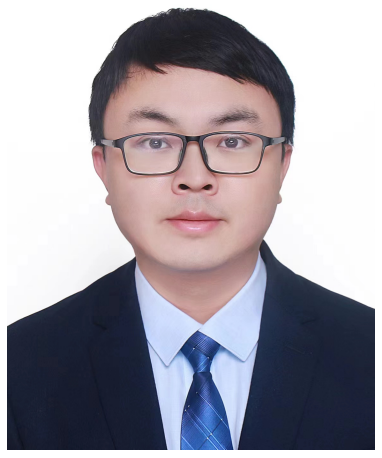

**Weijie Yuan** (Senior Member, IEEE) received the Ph.D. degree from the University of Technology Sydney, Australia, in 2019. In 2016, he was a Visiting Ph.D. Student with the Institute of Telecommunications, Vienna University of Technology, Austria. He was a Research Assistant with the University of Sydney, a Visiting Associate Fellow with the University of Wollongong, and a Visiting Fellow with the University of Southampton, from 2017 to 2019. From 2019 to 2021, he was a Research Associate with the University of New South Wales. He is now an Assistant Professor with the Southern University of Science and Technology. He currently serves as an Associate Editor for the IEEE Transactions on Wireless Communications, IEEE Communications Standards Magazine, IEEE Transactions on Green Communications and Networking, IEEE Communications Letters, IEEE Open Journal of Communications Society, and EURASIP Journal on Advances in Signal Processing. He is the Lead Editor for the series on ISAC in IEEE Communications Magazine. He was an Organizer/the Chair of several workshops and special sessions on OTFS and ISAC in flagship IEEE and ACM conferences, including IEEE ICC, IEEE/CIC ICCC, IEEE SPAWC, IEEE VTC, IEEE WCNC, IEEE ICASSP, and ACM MobiCom. He is the Founding Chair of the IEEE ComSoc Special Interest Group on OTFS (OTFS-SIG). He has been listed in the World's Top 2% Scientists by Stanford University for citation impact since 2021. He was a recipient of the Best Editor from IEEE CommL, the Exemplary Reviewer from IEEE TCOM/WCL, as well as the Best Paper Award from IEEE ICC 2023, IEEE/CIC ICCC 2023, and IEEE GlobeCom 2024.

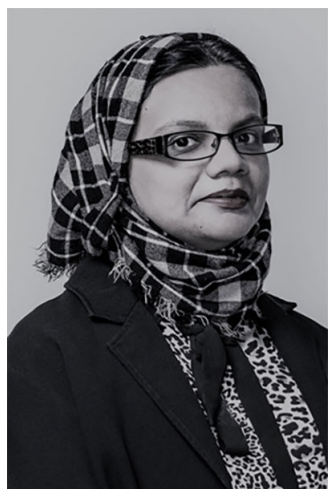

**Hina Tabassum** (Senior Member, IEEE) received the Ph.D. degree from the King Abdullah University of Science and Technology (KAUST). She is currently an Associate Professor with the Lassonde School of Engineering, York University, Canada, where she joined as an Assistant Professor, in 2018. Prior to that, she was a postdoctoral research associate at University of Manitoba, Canada. She is also appointed as a Visiting Faculty at University of Toronto in 2024 and the York Research Chair of 5G/6G-enabled mobility and sensing applications in 2023, for five years. She is also appointed as the IEEE ComSoc Distinguished Lecturer for the term 2025-2026. She is listed in the Stanford's list of the World's Top Two-Percent Researchers in 2021-2024. She received the Lassonde Innovation Early-Career Researcher Award in 2023 and the N2Women: Rising Stars in Computer Networking and Communications in 2022. She has published over 100 refereed articles in well-reputed IEEE journals, magazines, and conferences. Her publications thus far have garnered 6400+ citations with an h-index of 38 (according to Google Scholar). She has been recognized as an Exemplary Editor by the IEEE Communications Letters (2020), IEEE Open Journal of the Communications Society (IEEE OJCOMS) (2023 - 2024), and IEEE Transactions on Green Communications and Networking (2023). She was recognized as an Exemplary Reviewer (Top 2% of all reviewers) by IEEE Transactions on Communications in 2015, 2016, 2017, 2019, and 2020. She is the Founding Chair of the Special Interest Group on THz communications in IEEE Communications Society (ComSoc)-Radio Communications Committee (RCC). She served as an Associate Editor for IEEE Communications Letters (2019–2023), IEEE OJCOMS (2019–2023), and IEEE Transactions on Green Communications and Networking (2020–2023). Currently, she is also serving as an Area Editor for IEEE OJCOMS and an Associate Editor for IEEE Transactions on Communications, IEEE Transactions on Wireless Communications, and IEEE Communications Surveys and Tutorials.

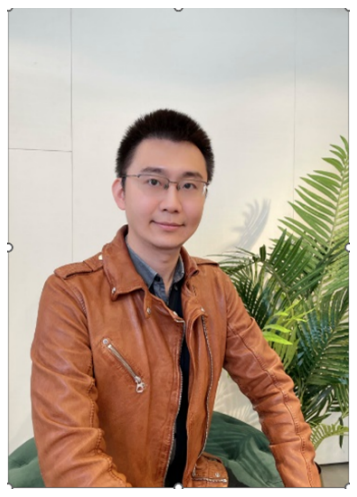

**Yuanwei Liu** (S'13-M'16-SM'19-F'24, https://www.eee.hku.hk/~yuanwei/) is a tenured full Professor in Department of Electrical and Electronic Engineering (EEE) at The University of Hong Kong (HKU). Prior to that, he was a Senior Lecturer (Associate Professor) (2021-2024) and a Lecturer (Assistant Professor) (2017- 2021) at Queen Mary University of London (QMUL), London, U.K, and a Postdoctoral Research Fellow (2016-2017) at King's College London (KCL), London, U.K. He received the Ph.D. degree from QMUL in 2016. His research interests include non-orthogonal multiple access, reconfigurable intelligent surface, near field communications, integrated sensing and communications, and machine learning.

Yuanwei Liu is a Fellow of the IEEE, a Fellow of AAIA, a Web of Science Highly Cited Researcher, an IEEE Communication Society Distinguished Lecturer, an IEEE Vehicular Technology Society Distinguished Lecturer, the rapporteur of ETSI Industry Specification Group on Reconfigurable Intelligent Surfaces on work item of "Multi-functional Reconfigurable Intelligent Surfaces (RIS): Modelling, Optimisation, and Operation", and the UK representative for the URSI Commission C on "Radio communication Systems and Signal Processing". He was listed as one of 35 Innovators Under 35 China in 2022 by MIT Technology Review. He received IEEE ComSoc Outstanding Young Researcher Award for EMEA in 2020. He received the 2020 IEEE Signal Processing and Computing for Communications (SPCC) Technical Committee Early Achievement Award, IEEE Communication Theory Technical Committee (CTTC) 2021 Early Achievement Award. He received IEEE ComSoc Outstanding Nominee for Best Young Professionals Award in 2021. He is the co-recipient of the 2024 IEEE Communications Society Heinrich Hertz Award, the Best Student Paper Award in IEEE VTC2022-Fall, the Best Paper Award in ISWCS 2022, the 2022 IEEE SPCC-TC Best Paper Award, the 2023 IEEE ICCT Best Paper Award, and the 2023 IEEE ISAP Best Emerging Technologies Paper Award. He serves as the Co-Editor-in-Chief of IEEE ComSoc TC Newsletter, an Area Editor of IEEE Transactions on Communications and IEEE Communications Letters, an Editor of IEEE Communications Surveys & Tutorials, IEEE Transactions on Wireless Communications, IEEE Transactions on Vehicular Technology, IEEE Transactions on Network Science and Engineering, and IEEE Transactions on Cognitive Communications and Networking. He serves as the (leading) Guest Editor for Proceedings of the IEEE on Next Generation Multiple Access, IEEE JSAC on Next Generation Multiple Access, IEEE JSTSP on Intelligent Signal Processing and Learning for Next Generation Multiple Access, and IEEE Network on Next Generation Multiple Access for 6G. He serves as the Publicity Co-Chair for IEEE VTC 2019-Fall, the Panel Co-Chair for IEEE WCNC 2024, Symposium Co-Chair for several flagship conferences such as IEEE GLOBECOM, ICC and VTC. He serves the academic Chair for the Next Generation Multiple Access Emerging Technology Initiative, vice chair of SPCC and Technical Committee on Cognitive Networks (TCCN).

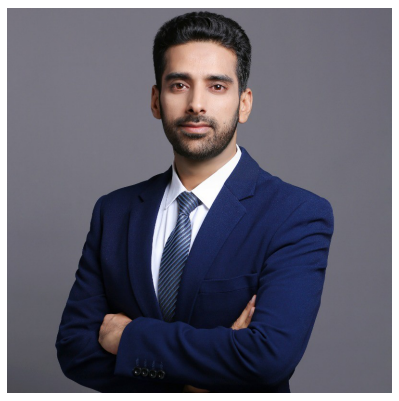

**Muhammad Umar Farooq Qaisar** (Member, IEEE) received his B.S. degree from the International Islamic University, Islamabad, Pakistan, in 2012, and his M.S. degree in Computer Science and Technology from the University of Science and Technology of China in 2017. He earned his Ph.D. degree in Computer Science and Technology from the University of Science and Technology of China in 2022. He was a postdoctoral fellow with the School of System Design and Intelligent Manufacturing at the Southern University of Science and Technology. Currently, he is an Associate Professor in the School of Computer Science at Northwestern Polytechnical University. His main research interests include IoT, WSN, SDN, VANETs, ISAC, UAVs, and communication security

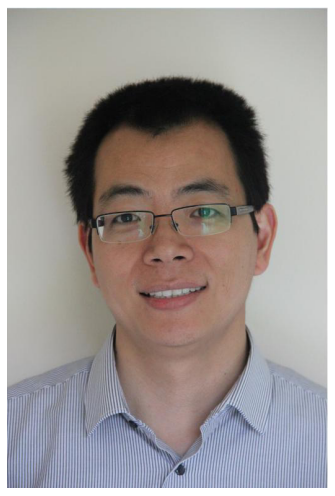

**Zhiguo Ding** (Fellow, IEEE) received his B.Eng from the Beijing University of Posts and Telecommunications in 2000, and the Ph.D degree from Imperial College London in 2005. He is currently a Professor in Communications at Khalifa University. Dr Ding' research interests are 6G networks, cooperative and energy harvesting networks and statistical signal processing. He is serving as an Area Editor for the IEEE Transactions on Wireless Communications and IEEE Open Journal of the Communications Society, an Editor for IEEE Transactions on Vehicular Technology, SCIENCE CHINA Information Sciences and IEEE Communications Surveys & Tutorials, and was an Editor for IEEE Wireless Communication Letters, IEEE Transactions on Communications, IEEE Communication Letters. He recently received the EU Marie Curie Fellowship 2012-2014, the Top IEEE TVT Editor 2017, IEEE Heinrich Hertz Award 2018, IEEE Jack Neubauer Memorial Award 2018, IEEE Best Signal Processing Letter Award 2018, Friedrich Wilhelm Bessel Research Award 2020, and IEEE SPCC Technical Recognition Award 2021. He is a Fellow of the IEEE, a Distinguished Lecturer of IEEE ComSoc, and a Web of Science Highly Cited Researcher in two categories 2022.

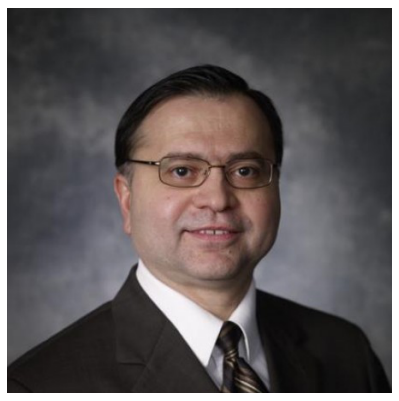

**Naofal Al-Dhahir** (Fellow, IEEE) is Erik Jonsson Distinguished Professor and ECE Associate Head at UT-Dallas. He earned his PhD degree from Stanford University and was a principal member of technical staff at GE Research Center and AT&T Shannon Laboratory from 1994 to 2003. He is co-inventor of 43 issued patents, co-author of over 600 papers and co-recipient of 8 IEEE best paper awards. He is an IEEE Fellow, AAIA Fellow, received 2019 IEEE COMSOC SPCC technical recognition award, 2021 Qualcomm faculty award, and 2022 IEEE COMSOC RCC technical recognition award. He served as Editor-in-Chief of IEEE Transactions on Communications from Jan. 2016 to Dec. 2019. He is a Fellow of the US National Academy of Inventors, a Member of the European Academy of Sciences and Arts, and a Web of Science Clarivate Highly Cited Researcher.

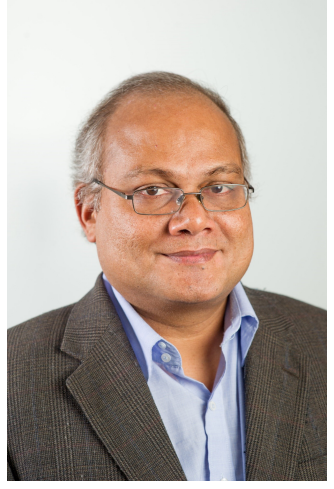

**Arumugam Nallanathan** (S'97-M'00-SM'05-F'17) is Professor of Wireless Communications and Head of the Communication Systems Research (CSR) group in the School of Electronic Engineering and Computer Science at Queen Mary University of London since September 2017. He was with the Department of Informatics at King's College London from December 2007 to August 2017, where he was Professor of Wireless Communications from April 2013 to August 2017 and a Visiting Professor from September 2017 till August 2020. He was an Assistant Professor in the Department of Electrical and Computer Engineering, National University of Singapore from August 2000 to December 2007. His research interests include Artificial Intelligence for Wireless Systems, Beyond 5G Wireless Networks and Internet of Things (IoT). He published more than 700 technical papers in scientific journals and international conferences. He is a co-recipient of the Best Paper Awards presented at the IEEE International Conference on Communications 2016 (ICC'2016), IEEE Global Communications Conference 2017 (GLOBECOM'2017) and IEEE Vehicular Technology Conference 2018 (VTC'2018). He is also a co-recipient of IEEE Communications Society Leonard G. Abraham Prize in 2022. He is an IEEE Distinguished Lecturer. He has been selected as a Web of Science Highly Cited Researcher in 2016, and 2022-2024.

He was a Senior Editor for IEEE Wireless Communications Letters, an Editor for IEEE Transactions on Wireless Communications, IEEE Transactions on Communications, IEEE Transactions on Vehicular Technology and IEEE Signal Processing Letters. He served as the Chair for the Signal Processing and Communication Electronics Technical Committee of IEEE Communications Society and Technical Program Chair and member of Technical Program Committees in numerous IEEE conferences. He received the IEEE Communications Society SPCE outstanding service award 2012 and IEEE Communications Society RCC outstanding service award 2014.

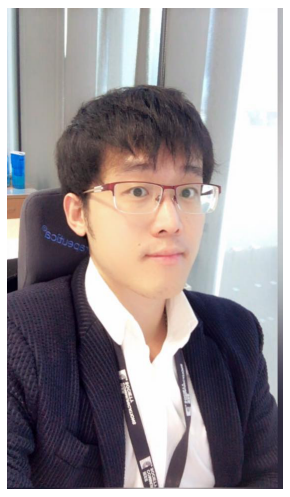

**Derrick Wing Kwan Ng** (Fellow, IEEE) received the bachelor's (Hons.) and M.Phil. degrees in electronic engineering from The Hong Kong University of Science and Technology (HKUST), Hong Kong, in 2006 and 2008, respectively, and the Ph.D. degree from The University of British Columbia (UBC) in 2012. He was a Senior Post-Doctoral Fellow at the Institute for Digital Communications, Friedrich– Alexander-University Erlangen–Nürnberg (FAU), Germany. He is currently working as a Scientia Associate Professor at the University of New South Wales, Sydney, Australia. His research interests include convex and non-convex optimization, physical layer security, IRS-assisted communication, UAV-assisted communication, wireless information and power transfer, and green (energy-efficient) wireless communications. He received the Australian Research Council (ARC) Discovery Early Career Researcher Award in 2017, the Best Paper Award from the WCSP 2020 and 2021, the IEEE TCGCC Best Journal Paper Award in 2018, INISCOM 2018, the IEEE International Conference on Communications (ICC) in 2018 and 2021, the IEEE International Conference on Computing, Networking and Communications (ICNC) in 2016, the IEEE Wireless Communications and Networking Conference (WCNC) in 2012, the IEEE Global Telecommunication Conference (Globecom) in 2011 and 2021, and the IEEE Third International Conference on Communications and Networking in China in 2008. He has been serving as an Editorial Assistant to the Editor-in-Chief for the IEEE TRANSACTIONS ON COMMUNICATIONS from January 2012 to December 2019. He is currently serving as an Editor for the IEEE TRANSACTIONS ON COMMUNICATIONS and the IEEE TRANSACTIONS ON WIRELESS COMMUNICATIONS and an Area Editor for the IEEE OPEN JOURNAL OF THE COMMUNICATIONS SOCIETY. He has been listed as a Highly Cited Researcher by Clarivate Analytics (Web of Science) since 2018.